\documentclass[]{aa}
\usepackage{graphicx}
\usepackage{txfonts}
\usepackage[export]{adjustbox}
\usepackage[utf8]{inputenc}
\usepackage{geometry}
\usepackage{longtable}
\usepackage{epstopdf}
\usepackage{multirow}
\usepackage{comment}
\usepackage{arydshln}
\usepackage{verbatim}
\usepackage{url}
\usepackage{rotating}
\usepackage{color}
\usepackage{amssymb}
\titlerunning{CPz}
\authorrunning{Fotopoulou \& Paltani}

\begin{document}

    \title{CPz: Classification-aided photometric-redshift estimation\thanks{The accompanying table is available in electronic form at the CDS via anonymous ftp to  to cdsarc.u-strasbg.fr (130.79.128.5) or via http://cdsweb.u-strasbg.fr/cgi-bin/qcat?J/A+A/}}

    \author{S. Fotopoulou\inst{\ref{Geneva}},\inst{\ref{Durham}}
            \and S. Paltani\inst{\ref{Geneva}}
        }

    \institute{Department of Astronomy, University of Geneva, ch. d'Ecogia 16, 1290 Versoix, Switzerland\label{Geneva} % S. Fotopoulou, S. Paltani
	\and Center for Extragalactic Astronomy, Department of Physics, Durham University, South Road, Durham, DH1 3LE, United Kingdom\label{Durham} \email{sotiria.fotopoulou@durham.ac.uk}}

    \abstract
    {Broadband photometry offers a time and cost effective method to reconstruct the continuum emission of celestial objects.  Thus, photometric redshift estimation has supported the scientific exploitation of extragalactic multiwavelength surveys for more than twenty years. Deep fields have been the backbone of galaxy evolution studies and have brought forward a collection of various approaches in determining photometric redshifts. In the era of precision cosmology, with the upcoming Euclid and LSST surveys, very tight constraints are put on the expected performance of photometric redshift estimation using broadband photometry, thus new methods have to be developed in order to reach the required performance.     
    We present a novel automatic method of optimizing photometric redshift performance, the classification-aided photometric redshift estimation (CPz). The main feature of CPz is the unified treatment of all classes of objects detected in extragalactic surveys: galaxies of any type (passive, starforming and starbursts), active galactic nuclei (AGN), quasi-stellar objects (QSO), stars and also includes the identification of potential photometric redshift catastrophic outliers. The method operates in three stages. First, the photometric catalog is confronted with star, galaxy and QSO model templates by means of spectral energy distribution fitting. Second, three machine-learning classifiers are used to identify 1) the probability of each source to be a star, 2) the optimal photometric redshift model library set-up for each source and 3) the probability to be a photometric redshift catastrophic outlier. Lastly, the final sample is assembled by identifying the probability thresholds to be applied on the outcome of each of the three classifiers. Hence, with the final stage we can create a sample appropriate for a given science case, for example favoring purity over completeness. We apply our method to the near-infrared VISTA public surveys, matched with optical photometry from CFHTLS, KIDS and SDSS, mid-infrared WISE photometry and ultra-violet photometry from the Galaxy Evolution Explorer (GALEX). We show that CPz offers improved photometric redshift performance for both normal galaxies and AGN without the need for extra X-ray information.}

    \keywords{Methods: data analysis, Galaxies: distances and redshifts -- surveys -- general -- active -- quasars}

    \maketitle

\section{Introduction}

Looking at a single broadband image, obtained for example in the $r$ band (6000\AA), we are able to distinguish distinct sources and classify them according to their morphologies, ranging from round and smooth elliptical galaxies to the impressive grand design spiral galaxies \citep{Hubble1926}. We now know that the morphologies are linked to the type of stellar populations and the gas and dust content of the galaxy. For example, passive galaxies are known to consist of mainly old star populations, while star-forming galaxies consist of younger, bluer stars populating the galaxy's spiral arms.  However, from a single image we can not say with certainty for example if a point-like source is really a star or a quasi-stellar object (QSO), or what is the cosmological distance of the source. Source classification has been subsequently enhanced by attributes both from photometric (e.g., colors) and spectroscopic (e.g., line widths) measurements. Plotting pairs of attributes has proven to be a powerful tool in identifying the parameter space occupied by each galaxy class for example separating between starforming and passive galaxies using the bimodality cloud, \citep{Bell2004}, or separating between starforming galaxies and active galactic nuclei (AGN) through the BPT diagram, \citep{Baldwin1981}. The main limitation of this approach is the dimensionality reduction of a wealthy parameter space to usually only two to four attributes and the introduction of hard limits to separate between the classes. 

Another approach in identifying the nature of astronomical sources is the comparison of stellar population synthesis models (SSP) to observations through spectral energy distribution (SED) fitting. SSPs demonstrate that the age and metallicity of the star population will create a distinct galaxy continuum emission which can be probed with broadband photometry which however suffers from degeneracies in color space \citep[e.g.,][]{Bruzual2003,Maraston2005}.
A selection of models deemed representative of the sample at hand can be used to identify stars vs galaxies and to estimate physical properties such as the stellar mass, star-formation rate, amongst others, judged by least $\chi^2$ \citep[e.g.,][]{Bolzonella2000,Robin2007,Ilbert2009, Fotopoulou2012, Dahlen2013}. Lastly, machine-learning algorithms are very well applicable to an astronomical context and they have been embraced since the early 90's. supervised and unsupervised methods are able to identify correlations, groupings and even outliers in vast datasets that would otherwise be impossible to visualize by a human eye. Many works have explored classification in astronomy using machine-learning techniques aiming towards separating stars from galaxies \citep{Odewahn1993, Soumagnac2015}, QSO identification \citep{Brescia2015}, estimating physical parameters \citep{Ucci2017}, finding peculiar objects \citep{Meusinger2012} etc. 

In addition to the class of the object, multiwavelength information are useful in providing an estimate of the redshift as proposed by \citet{Baum1962}. All modern extragalactic surveys make extensive use of photometric redshift estimation as it is a cost-efficient method to determine distances of galaxies (COMBO-17, CFHTLS, CDFS and ECDFS, Lockman Hole, AEGIS, COSMOS, XXL, to name a few). However, most surveys are mainly focused on the normal galaxy population. In the presence of deep X-ray flux measurements, variability and morphology information, \citet{Salvato2009,Salvato2011} showed that the optimal photometric redshift solution can be achieved by dissecting the galaxy population in three categories containing 1) normal galaxies, 2) normal galaxy and AGN emission, 3) AGN dominated emission, which we refer to as QSO. This method has been successfully applied to other extragalactic fields such as the Lockman Hole \citep{Fotopoulou2012}, the Chandra Deep Field South \citep{Hsu2014} and AEGIS-X \citep{Nandra2015} and it is the current state of the art, used when X-ray data are available for the whole field in consideration. The core of the method is the use of independent information such as X-ray flux, morphology and variability to pin-point the SED models that will give the optimal photometric redshift solution for each population. By doing so, the degeneracies between models are reduced, thus achieving higher photometric redshift performance. However, the correct implementation of the method requires splitting the sample into point-like and varying sources and also having an estimate of X-ray flux. This information will not be available to the desired sensitivity for the majority of the source population detected by Euclid and LSST. 

In this paper, we generalize the idea of population-specific libraries designed for application on surveys for which X-ray fluxes and variability information is not available and the morphology is estimated by the half light radius. The classification-aided photometric-redshift estimation (CPz) utilizes machine-learning classification and spectral energy distribution (SED) fitting for photometric redshift. 
We show that this classification step allows the production of optimized photometric redshift for galaxies, AGN and QSO while at the same time identifying stars and catastrophic outliers.

\section{Background}

Before we present the CPz method, we briefly describe the concepts and nomenclature of machine-learning and template fitting methods used in the rest of the paper. We use as an example a multiwavelength extragalactic survey. From the photometric imaging we retrieve information such as fluxes, colors, shapes, and positions. From the spectroscopic follow up on the same area of the sky, we might associate to the same sources additional information such as redshift, emission line widths, star vs galaxy classification etc. The input quantities (photometry, colors, shapes, emission lines widths, etc) are called attributes in the machine learning nomenclature and we will use this term hereafter. The label, ``star'' - ``not star'' for our example, is also called the target, while the procedure of assigning a source to a given class is called labeling. 

Suppose that we want to identify all stars within the survey. With the template fitting method, we might take all the photometric measurements, perform model fitting using star and galaxy templates and select the model that represents best our data. On the other hand, for supervised machine-learning we would select a subsample of sources that we have spectroscopically confirmed to be stars and a subsample that we have confirmed they are not stars. We would then give as input to the classifier the photometric measurements, the colors and shapes and attach a label ``star'', ``not star'' next to each source. We would then split the sample into training and test samples. The algorithm will then create a mapping between the input photometry, colors and shapes and their labels using the training sample and assess the quality of the mapping by predicting the labels of the test sample. If the algorithm is a neural network it will assign weights to all inputs and iteratively optimize the weights until the prediction of the mapping between the input quantities and assigned labels reaches a success threshold. One the other hand, if we choose to use decision trees, the algorithm will create automatically the split on the input quantities, until the prediction of the classification has reached a certain success threshold.

\subsection{Random forest}
A plethora of machine-learning algorithms is available, each with advantages and disadvantages. A detailed account is however outside the scope of this work\footnote{See for example \citet{Ivezic2014}}. We chose to use the random forest \citep[RF,][]{Breiman2001} because it gathers an attractive set of advantages. To name a few, it is robust against over-fitting, present when the algorithm learns artificial structure in the data (for example noise), robust against correlated input attributes, it is a fuzzy classifier providing a probability for a source to belong to each class and it is very fast to train.

These advantages are a direct consequence of the method. A RF is a collection of decision trees. Each tree is created with a random subsample of the input attribute set. For example, if we have available 100 colors, each tree will be created with e.g., 20 colors. As a result, each tree will learn only part of the input attribute pattern and will be a weak classifier. Since each tree only learns a subsample of the data the forest does not learn artificial structures or become affected by correlated attributes to the same degree as for example neural networks. The final answer of the RF is the average of all trained trees, which provides a probability for each object to belong in one of the specified classes.

\subsection{Classification quality}\label{sec:quality}

We will define here a few concepts used to quantify classification quality.

\paragraph{Classification performance}
In order to assess the quality of a classifier, a comparison between the input labels and predicted classes is needed. Taking as an example the question "Is this source a star?" we have i) true positive (TP) a real star, classified as a star, ii) true negative (TN) a non-stellar object, classified as not a star, iii) false positive (FP) a non-stellar object, classified as a star, and iv) false negative (FN) a star, classified as not a star.

This information is often summarized in a confusion matrix. We also define the following measures of quality for the classifiers:
\paragraph{Accuracy} Fraction of correct predictions:
\begin{equation}
\rm{ACC = \frac{TP+TN}{TP+TN+FP+FN}}.
\end{equation}
\paragraph{Precision} Fraction of correct positive predictions, also referred to as purity in astronomy which we will also adopt for the rest of the paper:
\begin{equation}
\rm{P = \frac{TP}{TP+FP}}.
\end{equation}
\paragraph{Recall} Fraction of truly positive predictions, also referred to as completeness in astronomy which we will also adopt for the rest of the paper:
\begin{equation}
\rm{R = \frac{TP}{TP+FN}}.
\end{equation}	
\paragraph{F1 measure} Harmonic mean of precision and recall:
\begin{equation}
\rm{F1 = 2\cdot\frac{P\cdot R}{P+R}}.
\end{equation}	
\paragraph{Fall-out} Fraction of FP over negative predictions:
\begin{equation}
\rm{F = \frac{FP}{TN+FP}}.
\end{equation}	

A good classifier will have high accuracy, precision, recall and F1 measure and low fall-out.

\subsection{Template fitting}

With template fitting methods a selection of theoretical or empirical models are confronted first with the response function of the telescope creating a flux library ($\rm{flux_{temp}}$) and then with the data ($\rm{flux_{obs}}$). Through a maximum likelihood approach the best model representing the data is selected. When the uncertainties in the data ($\rm{\sigma^2_{obs}}$) follow a Gaussian distribution, the maximum likelihood approach is equivalent to selecting the model with minimum $\chi^2$ defined as:

\begin{equation}
\chi^2=\sum_{i}^N\frac{(flux_{obs,i}-\alpha\cdot flux_{temp,i})^2}{\sigma^2_{obs,i}},
\end{equation}
where N is the number of data-points (filters) and $\alpha$ the scaling factor that minimizes the $\chi^2$ value.
The significant advantage of template fitting over machine learning methods for the particular application of photometric redshift estimation is that template fitting can recover redshift solutions that are outside the training sample used for machine learning (e.g., high redshift galaxies).

\subsection{Photometric redshift quality}

We introduce here also the two measures of quality of the photometric redshift estimation.

\paragraph{Accuracy, $\sigma$} The photometric redshift accuracy ($\sigma$) is usually defined in the literature as the normalized median absolute deviation (NMAD) \citep{Ilbert2009}:
\begin{equation}\label{eq:sigma}
\sigma_{NMAD} = 1.48\times median\Big(\frac{|z_{phot}-z_{spec}|}{1+z_{spec}}\Big).
\end{equation}
The usage of $\sigma_{NMAD}$ is preferred over the standard deviation because the median is less sensitive to extreme values, while the scale factor 1.48 is introduced to allow the interpretation of $\sigma_{NMAD}$ as the the standard deviation of normally distributed data.
\paragraph{Catastrophic outliers, $\eta$} For a number of sources the photometric redshift is wrong. Usually referred to as catastrophic outliers ($\eta$), they are quantified as the percentage of sources for which eq. (\ref{eq:eta}) holds.
\begin{equation}\label{eq:eta}
|z_{phot}-z_{spec}|>0.15\cdot(1+z_{spec}).
\end{equation}

\section{The CPz method}

\begin{figure*}
    \centering
    \begin{tabular}{c}
    \includegraphics[width=\textwidth]{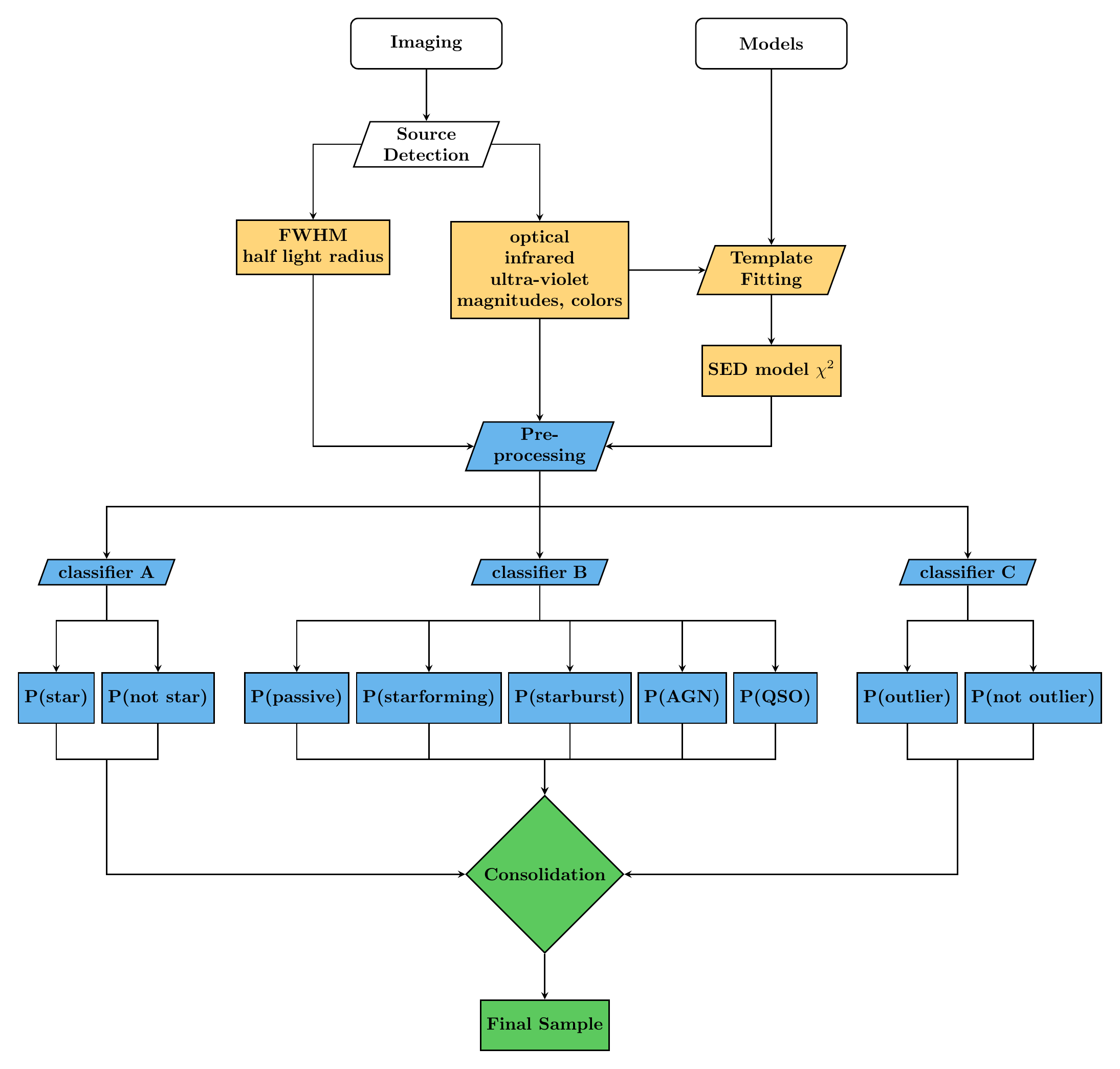} \\
	\end{tabular}
    \caption{Components of the classification aided photometric redshift estimation method (CPz). Photometric and morphometric parameters are extracted from the observed images. Magnitudes, colors and an estimation of the source shape (e.g., half light radius) make up the input attributes. The spectral energy distributions are fitted with model templates of star, galaxy and active galactic nuclei populations, the $\chi^2$ value of the best fitting model per library is added to the input attribute set. The pre-processing of the data consists of normalization and whitening of the attribute distributions (zero mean and variance of one) and the sample is split into training, testing and validation subsamples. The training sample is presented to three distinct classifiers, producing a probability for each source to A) be a star B) have an optimal photometric redshift from a collection of distinct libraries C) be a photometric redshift outlier. The final phase of CPz is the consolidation of the results, during which the test sample is used to decide on the thresholds adopted for each classifier and the assessment of the final accuracy on the classification and the photometric redshift performance using the validation sample.
    \label{fig:flowchart}}
    \end{figure*}

The CPz consists of three main stages shown in Fig. \ref{fig:flowchart}. Stage I is the collection of the information on the flux and morphology of the sources and the subsequent photometric redshift estimation using template fitting. Stage II is the classification using three RFs to identify the A) probability of being a star, B) optimal photometric redshift setup and C) probability of being a photometric redshift outlier, given the colors, $\chi^2$ values from the template fitting step and morphology estimates (in our case the half-light radius). Finally, Stage III consists of the consolidation phase, during which a source is assigned a photometric redshift solution and i) probability to be a star, ii) probability for the redshift to be wrong (eq. \ref{eq:eta}). We will describe the details behind each processing step by applying the method on the near-infrared VIKING and VIDEO Public VISTA Surveys cross matched with the CFHTLS, KiDS and SDSS optical surveys.

\subsection{Stage I: catalogs and template fitting}\label{sec:data}

Stage one of the CPz method consists of collecting information on the morphology (e.g., FWHM, half radius) and flux measurements and the spectral energy distribution (SED) model fitting. At this stage, all $\chi^2$ estimates of the best fitting model per library are saved and propagated into the pre-processing.

\subsubsection{Photometric surveys}\label{sec:photdata}

Modern extragalactic surveys profit from multiwavelength coverage from the ultra-violet to the mid-infrared, which allows for the estimation of good quality photometric redshifts. Future surveys for example with Euclid and LSST will provide photometry from the $u$-band up to the $H$-band with continuous coverage in wavelength. To test our method in comparable conditions of Euclid plus LSST, we use the ESO near-infrared Public VISTA surveys \footnote{\url{https://www.eso.org/sci/observing/PublicSurveys/sciencePublicSurveys.html}} \citep{Arnaboldi2007} using the z, Y, J, H, and K photometric filters. We are using both the VIKING ($\rm{J_{lim, AB}=22.1}$, PI W. Sutherland) and VIDEO ($\rm{J_{lim, AB}=24.5}$, PI M. Jarvis) surveys in order to benefit from the large area coverage and the depth, respectively. Euclid observations will not have a $K$-band coverage, therefore we will also discuss the impact of this particular filter on our results. The optical wavelength coverage (filters u, g, r, i, z) is from the SDSS survey \citep[DR12, $\rm{i_{lim, AB}=21.3}$,][]{Alam2015}, CFHTLS \citep[T0007, $\rm{i_{lim, AB}=24.8}$,][]{Hudelot2012} and KiDS \citep[DR2, $\rm{i_{lim, AB}=24.2}$,][]{Jong2015} surveys. We are also using mid-infrared observations in the W1 and W2 filters of the WISE satellite \citep[ALLWISE\footnote{\url{http://wise2.ipac.caltech.edu/docs/release/allwise/}}, $\rm{W1_{lim,AB}=20.3}$,][]{Wright2010,Mainzer2011} and ultra-violet (filters FUV, NUV) from the GALEX satellite \citep[GR6/7, $\rm{NUV_{lim,AB}=20.5}$,][]{Morrissey2007}. We corrected all photometric measurements according to the Schlegel maps of Galactic absorption \citep{Schlegel1998} and the Cardelli law for the Milky way \citep{Cardelli1989}.

\subsubsection{Spectroscopic surveys}

We have selected spectroscopic redshift surveys which combined span a large redshift range $\rm(z\in[0-4])$, represent all galaxy types and include spectroscopically confirmed stars. These are SDSS\footnote{\url{http://www.sdss.org/dr12/data\_access/bulk/}} \citep[DR12-- $\rm{7.5\times10^4}$ sources,][]{Alam2015}, GAMA\footnote{\url{http://www.gama-survey.org/dr2/data/cat/SpecCat/v08/}} \citep[DR2 -- $\rm{1.3\times10^4}$ sources,][]{Liske2015}, VIPERS\footnote{\url{http://vipers.inaf.it/rel-pdr1.html}} \citep[DR1 -- $\rm{3\times10^3}$ sources,][]{Garilli2014}, VVDS \citep[DR2-- $\rm{3\times10^3}$ sources,][]{LeFevre2013}, PRIMUS \citep[DR1 -- $\rm{2.7\times10^4}$ sources,][]{Coil2011,Cool2013}, 6df \citep[DR3 -- $\rm{3\times10^3}$ sources,][]{Jones2004,Jones2009}. Some of these surveys provide a label, or comment introduced by visual inspection classifying the objects into star, galaxy, AGN or QSO. However, this labeling is neither homogeneous across all surveys, nor objective. Thus, we keep this information for comparison with our machine-learning approach, but we do not use it during the training. 

\begin{figure}
\centering
\begin{tabular}{c}
\includegraphics[width=\columnwidth]{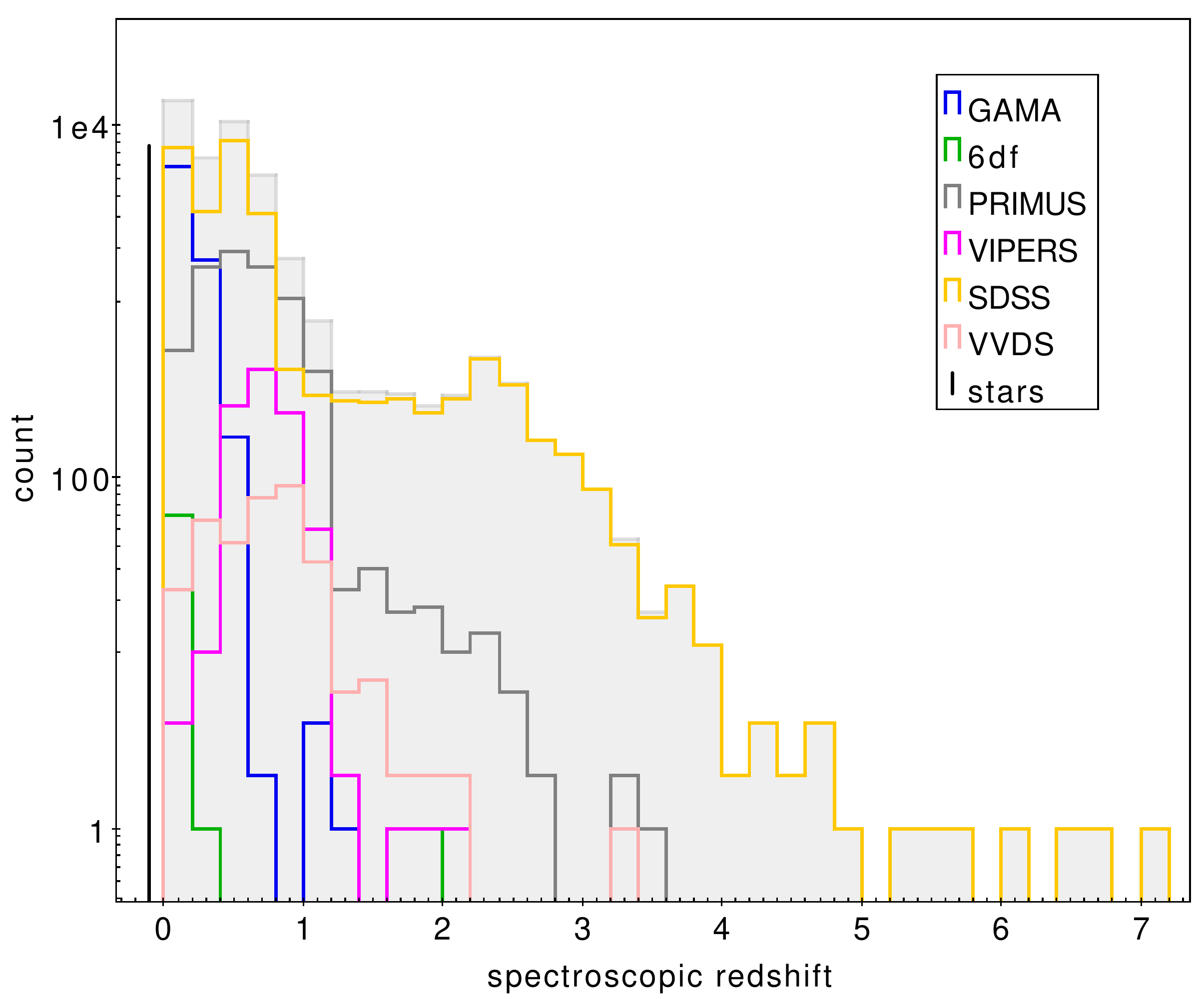}\\
(a)\\
\includegraphics[width=\columnwidth]{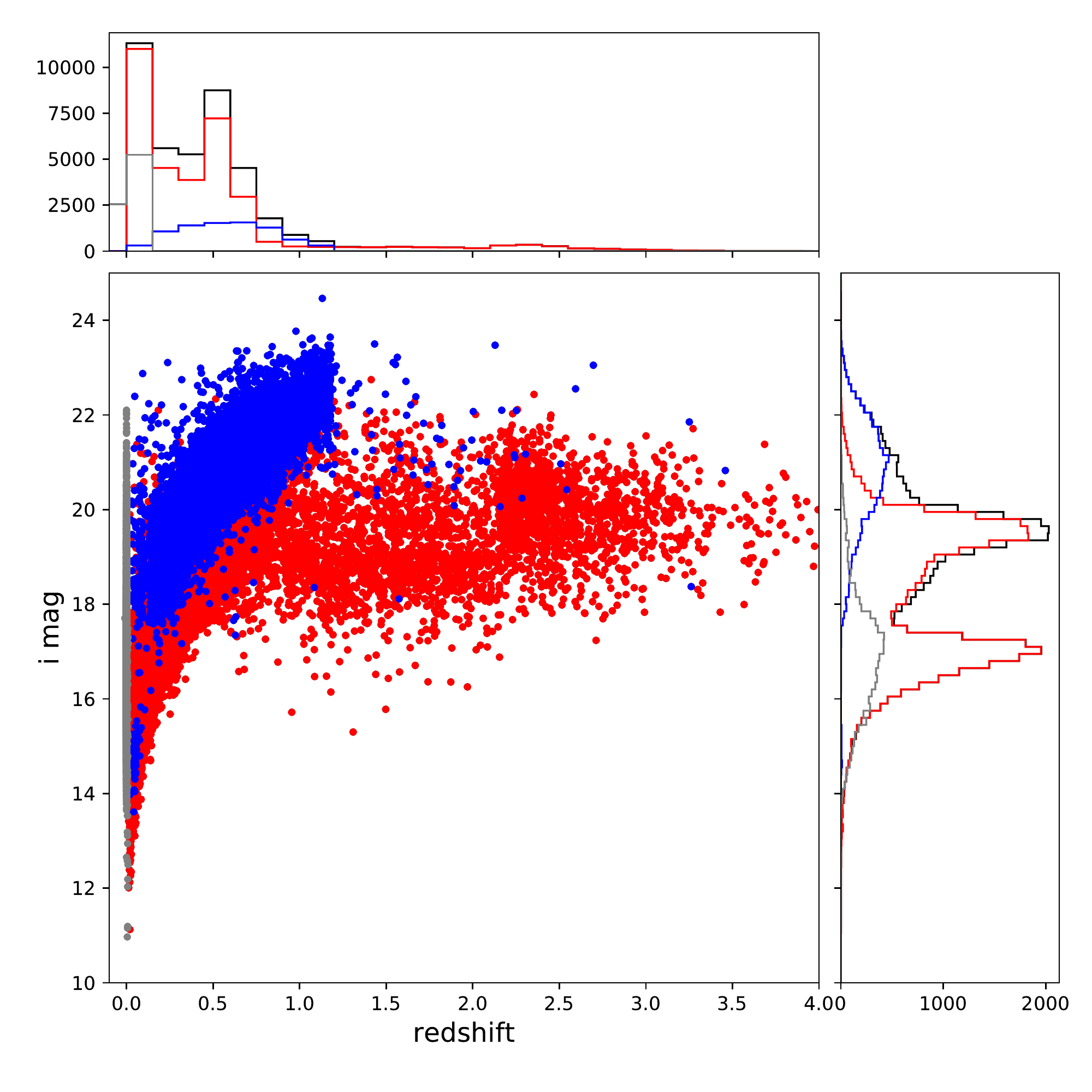} \\
(b)
\end{tabular}
\caption{Spectroscopic redshift distribution of the u-k -- IR sample used in this work. (a) The gray shaded area shows the total sample while the colored lines show the survey of origin. The black spike located for clarity at z=-0.1 shows the number of stars in the sample. (b) Magnitude distribution for SDSS (red) and CFHTLS (blue) surveys comprising 80\% and 20\% of the galaxy sample respectively. The stars are shown with grey colour on this plot.}
\label{fig:reddist}
\end{figure}

The sample used for this work consists exclusively of sources with the highest spectroscopic redshift quality\footnote{Sources denoted as stars and galaxies with zflag $\ge$ 3 for GAMA and 6dF, \textsc{ZWARNING=0} for SDSS, and zflag=\textsc{XX}3 or \textsc{XX}4, \textsc{X}=0,1,2 for VIPERS and VVDS.}, matched to photometric detections using a radius of $1''$. We use the following combination of wavelengths as quoted in the analysis and discussion, namely: UV (filters FUV, NUV from GALEX), optical (filters u-z from any of SDSS, CFHTLS, KIDS), near infrared survey (filters z-K from any of VIKING, VIDEO), mid infrared (filters W1, W2 from WISE). Fig. \ref{fig:reddist} (a) shows the redshift distribution per survey (colored lines) and the total galaxy sample (shaded gray area). The black spike located for demonstration purposes at -0.1 denotes the star sample. The panel (b) of the same figure shows the corresponding photometry for the spectroscopic sample. SDSS photometry corresponds to 80\% of the sample ($\sim$200 $\deg^2$), while CFHTLS due to the extensive spectroscopic follow of the W1 field ($\sim$30 $\deg^2$) extends our sample to fainter magnitudes compared to SDSS.

We performed four CPz Runs using the following filter combinations 1) u-K ($\sim7.8\times10^4$ sources), 2) u-K -- IR ($\sim5\times10^4$ sources), 3) UV -- u - K ($\sim1.5\times10^4$ sources) and 4) UV -- u-K -- IR ($10^4$ sources) filters. This work does not address the impact of missing values in the photometry, therefore each run is comprised of sources that have good quality spectroscopic redshifts and photometric measurements in all filters.
As the impact of the mid-infrared bands on the quality of the result is significant, we discuss in detail for the rest of the presentation of the method only Run u-K -- IR. We differ the discussion of Runs one, three and four to Section \ref{sec:depfilt}.

\subsubsection{Template fitting}

\begin{table*}
\begin{center}
\caption{Left hand side: normal galaxy models used for template fitting. The starformation increases from top to bottom (see \citet{Ilbert2009} for more details). Right hand side: normal, AGN-QSO and hybrid models. The AGN fraction varies from 0-100\% as noted in the name of the model, eg. S0\_90\_QSO2\_10 is a combination of 90\% flux from a disk galaxy and 10\% flux contribution from an obscured QSO. The prefix ``pl\_'' denotes that the model has been extended to the ultra-violet assuming a power-law. (see \citet{Salvato2009} for a detailed description).}
\label{tab:SEDlib}
\begin{tabular}{lcc|lcc} \hline \hline
model &		description	 & filename  & model & description & filename\\ \hline
1  & elliptical & Ell1\_A\_0 	 & 32  & spiral &  CB1 \\
2  & '' &  Ell2\_A\_0 	 & 33  & disk &  S0 \\
3  & '' &  Ell3\_A\_0 	 & 34  & spiral &  Sb \\
4  & '' &  Ell4\_A\_0 	 & 35  & spiral &  Spi4 \\
5  & '' &  Ell5\_A\_0 	 & 36  & AGN &  M82 \\
6  & '' &  Ell6\_A\_0 	 & 37  & starburst/AGN &  I22491 \\
7  & '' &  Ell7\_A\_0 	 & 38  & Seyfert 1.8 &  Sey18 \\
8  & disk &  S0\_A\_0 	 & 39  & Seyfert 2.0 &  Sey2\\
9  & spiral &  Sa\_A\_0 	 & 40  & Disk-obscured QSO Hybrid &  S0\_90\_QSO2\_10 \\
10  & '' &  Sa\_A\_1 	 & 41  & '' &  S0\_80\_QSO2\_20 \\
11  & '' &  Sb\_A\_0 	 & 42  & '' &  S0\_70\_QSO2\_30 \\
12  & '' &  Sb\_A\_1 	 & 43  & '' &  S0\_60\_QSO2\_40 \\
13  & '' &  Sc\_A\_0 	 & 44  & '' &  S0\_50\_QSO2\_50 \\
14  & '' &  Sc\_A\_1 	 & 45  & '' &  S0\_40\_QSO2\_60 \\
15  & '' &  Sc\_A\_2 	 & 46  & '' &  S0\_30\_QSO2\_70 \\
16  & '' &  Sd\_A\_0 	 & 47  & '' &  S0\_20\_QSO2\_80 \\
17  & '' &  Sd\_A\_1 	 & 48  & '' &  S0\_10\_QSO2\_90 \\
18  & '' &  Sd\_A\_2 	 & 49  & AGN &  Mrk231\\
19  & irregular &  Sdm\_A\_0 	 & 50  & Starforming-unobsc. QSO Hybrid &  I22491\_90\_TQSO1\_10 \\
20  & extreme starforming&  SB0\_A\_0 	 & 51  & '' &  I22491\_80\_TQSO1\_20 \\
21  & ''&  SB1\_A\_0 	 & 52  & '' &  I22491\_70\_TQSO1\_30 \\
22  & ''&  SB2\_A\_0 	 & 53  & '' &  I22491\_60\_TQSO1\_40 \\
23  & ''&  SB3\_A\_0 	 & 54  & '' &  I22491\_50\_TQSO1\_50 \\
24  & ''&  SB4\_A\_0 	 & 55  & '' &  I22491\_40\_TQSO1\_60 \\
25  & ''&  SB5\_A\_0 	 & 56  & '' &  pl\_I22491\_30\_TQSO1\_70 \\
26  & ''&  SB6\_A\_0 	 & 57  & '' &  pl\_I22491\_20\_TQSO1\_80 \\
27  & ''&  SB7\_A\_0 	 & 58  & '' &  pl\_I22491\_10\_TQSO1\_90 \\
28  & ''&  SB8\_A\_0 	 & 59  & QSO &  pl\_QSOH \\
29  & ''&  SB9\_A\_0 	 & 60  & QSO &  pl\_QSO \\
30  & ''&  SB10\_A\_0  & 61  & QSO &  pl\_TQSO1\\
31  & ''&  SB11\_A\_0  &  &  \\ \hline\hline
\end{tabular}
\end{center}
\end{table*}

We performed SED fitting using the code LePhare\footnote{\url{www.cfht.hawaii.edu/~arnouts/lephare.html}}. We used the template set selected for the COSMOS Survey \citep{Scoville2007}. The normal galaxy templates were introduced in \citet{Ilbert2009} and the AGN hybrid and QSO templates in \citet{Salvato2009}. We used a redshift step of 0.01 from z=0 to z=6 for all models. The E(B-V) values used are 0.,0.05,0.1,0.15,0.2,0.3. The inclusion of emission lines, dust attenuation, absolute magnitude prior per library is reported in Table \ref{tab:SEDfit}.

We also fit a model library of 154 stars (1150\AA-25000\AA) containing normal spectral types, F-K dwarfs and G-K giant components from the Pickles Atlas \citep{Pickles1998}, white dwarfs \citep{Bohlin1995} and subdwarf O and B stars \citep{Bixler1991}.

\subsection{Stage II: classification}

Stage two of CPz is the machine-learning classification, starting with the normalization transformations applied to the input attributes (pre-processing) created in Stage I. In the following section we discuss the setup of the classifiers. We are using the \verb scikit-learn  implementation in Python \citep{Pedregosa2011} to pre-process and classify the data.

\subsubsection{Input attributes and class definition}\label{sec:classdef}

The attributes used for the classification are all color combinations and magnitudes of the photometric bands described in \S \ref{sec:photdata}. For filters $u$ to $K$ we use both total (auto) magnitudes and 3'' aperture magnitudes corrected to total, to account for flux lost due to the fixed size of the aperture. The correction is estimated on point-like sources and applied to the entire catalog. Particularly for the W1 and W2 WISE bands, we are using only the total magnitudes since the PSF of WISE is much larger compared to the optical and near-infrared bands ($\sim$6'' compared to 0.8''-1.3'' respectively). As a proxy for the morphology, we are using the half light radius estimated for the bands $g$ up to $K$, defined as the radius up to which 50\% of the total flux is enclosed. For point-like sources (stars and QSO) this radius will be very close to the FWHM of the PSF, while for extended objects it is significantly larger. Additionally, we are using the values of the $\chi^2$ for the star models in classifier A (star classifier) and the values of the $\chi^2$ of galaxies, AGN, QSO one for each corresponding library setup for classifier B (galaxy classifier). The total number of input attributes is 263.

The definition of the target classes is of paramount importance for supervised machine-learning.  With our method we aim to train three classifiers, thus we need three labels for each object. Namely, we create a star classifier (classifier A), a galaxy classifier tuned to return the class for which the photometric redshift solution is optimal (classifier B) and a classifier to identify photometric redshift outliers (classifier C). Even though technically possible, we do not combine all categories into one classifier. We opt for a flexible scheme within which we have the best photometric redshift estimation for all sources and impose only during the consolidation phase probability thresholds to identify stars and outliers. With this approach, we can tune during the consolidation phase the completeness and purity as desired for each specific science application.

The star - no star label is assigned using a sample of spectroscopically confirmed stars, used to train classifier A. The galaxy class label is tuned to identify the optimal photometric redshift class. It is assigned according to the minimum value of $\Delta z_i =\rm{|z_{phot_i}-z_{spec}|}$, where $i$ corresponds to the library configurations given in Table \ref{tab:SEDfit}. Thus, Classifier B will recognize, for example for Case III, which of the five classes (passive, starforming, starburst, AGN, QSO) will provide the best photometric redshift estimate compared to the available spectroscopic value. Finally, if $\rm{|z_{phot_i}-z_{spec}|/(1+z_{spec})>0.15}$ for all photometric redshift classes, the solution is considered a catastrophic outlier and is used as training in Classifier C. The catastrophic outliers will thus contain stars, QSOs - notoriously difficult to model due to their featureless SEDs - and rare galaxies (e.g., extreme obscured or extremely starforming) that are not represented by our template selection.

\subsubsection{Pre-processing: whitening and normalization}

There are two main operations that must take place before the data can be presented to the classifier, collectively called pre-processing. These operations are the treatment of missing values, also called data imputation and the whitening and normalization of the dataset. 

A typical approach in machine-learning when it comes to the treatment of missing values is the substitution of the value with the mean of the distribution. The mean of the distribution is the preferred substitution in the case of lacking observational data and this is the approach we adopt for this work. However, in an astronomical context other data imputation methods could be considered. For example, it can be that a source has an upper limit (non-detection), lower limit (saturation), or missing value (not observed). The impact of the data imputation depends on the specifics of each dataset, taking into account the sky coverage of the observations, including the tiling of the observations and any masking due to bright stars and the depth of the photometry. 

Whitening and normalization are transformations that center the distribution of the input attributes around zero and make the distribution of the values have a standard deviation of one. Machine-learning algorithms require whitening and normalization of the data to avoid recognition of artificial structures in the data due to difference in the order of magnitude of the attribute values, for example by mixing flux and magnitude estimates. It is critical that the same transformation applied to the training sample must be applied to the test and validation samples.

\subsubsection{Train - test - validation samples}

We split our sample of about 50 000 sources into three subsamples with ratio 1:1:1 for training, testing and validation. It is important that all three samples are representative of each other, therefore we first sort according to redshift and then split the sample taking every third source for each of the classifiers under consideration. During the training phase, the algorithm creates the Random Forest that maps best the input attributes to the target classes. After the training is complete and the forest is fully grown, we use the test sample that was not part of the training sample to estimate the accuracy of the classifier and identify the appropriate thresholds to adopt to separate stars and outliers. Finally, after all optimizations are performed, we estimate the final quality of CPz method using the validation sample which is never seen by the classifier during the training and testing phases. If not enough data are available for a separate validation set, it is common practice to perform cross-validation. During this procedure a handful of data are purposely left aside during the training and the classifier accuracy is tested on them. If this procedure is performed $n$-times, we refer to $n$-fold cross-validation. We have verified that with three-fold cross-validation we obtain similar accuracy results to the ones estimated with a dedicated testing sample.

A number of parameters are available to tune the performance of the Random Forest algorithm. We performed a grid search using three-fold cross validation in the training phase to select the best parameters for our work given in Table \ref{tab:rfsetup}. The parameter n\_estimators is the number of trees created. As a rule of thumb, the higher the number the more accurate the classification, however there is a limit after which the accuracy does not increase noticeably with an increasing expense on the computing time. In our case we found that 200 trees was a good trade-off between accuracy and computing time. The rest of the parameters in Table \ref{tab:rfsetup} control the split of the input attributes and the creation of the trees.

\begin{table}
\centering
\caption{Random Forest set-up parameters. The parameter n\_estimators is the number of trees created.}\label{tab:rfsetup}
\begin{tabular}{lc}\hline\hline
RF parameter & value\\\hline
n\_estimators & 200 \\
criterion & entropy\\
max\_leaf\_nodes & None\\ 
min\_samples\_leaf & 1\\
min\_samples\_split & 10\\
min\_weight\_fraction\_leaf&0.0\\
max\_features & 20\\
max\_depth&None\\
bootstrap & True \\ 
\hline\hline
\end{tabular}
\end{table}

\subsubsection{Classifier A: is it a star or not?}\label{sec:classifierA}

\begin{figure*}
\centering
\begin{tabular}{ccc}
\includegraphics[width=0.3\linewidth]{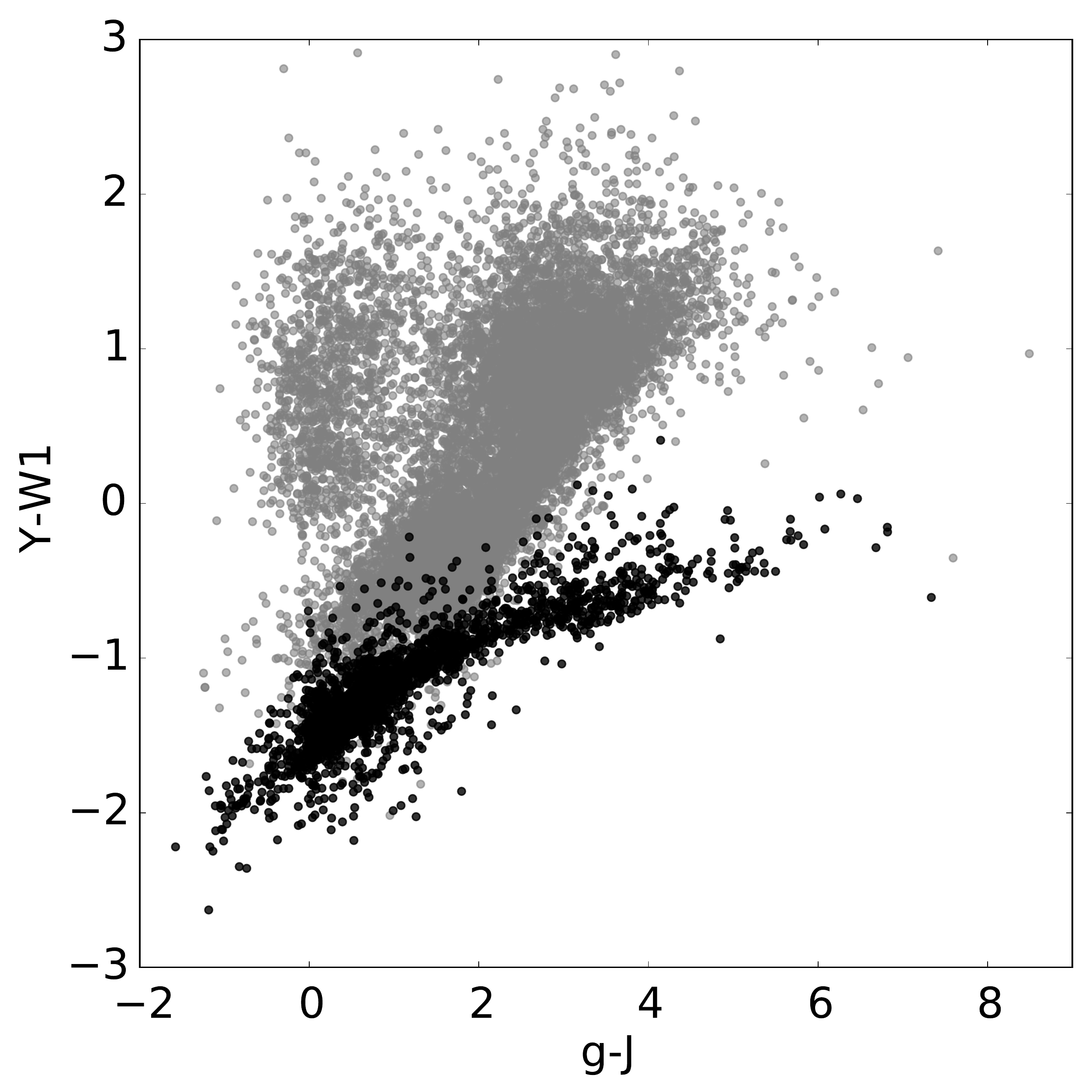} &
\includegraphics[width=0.3\linewidth]{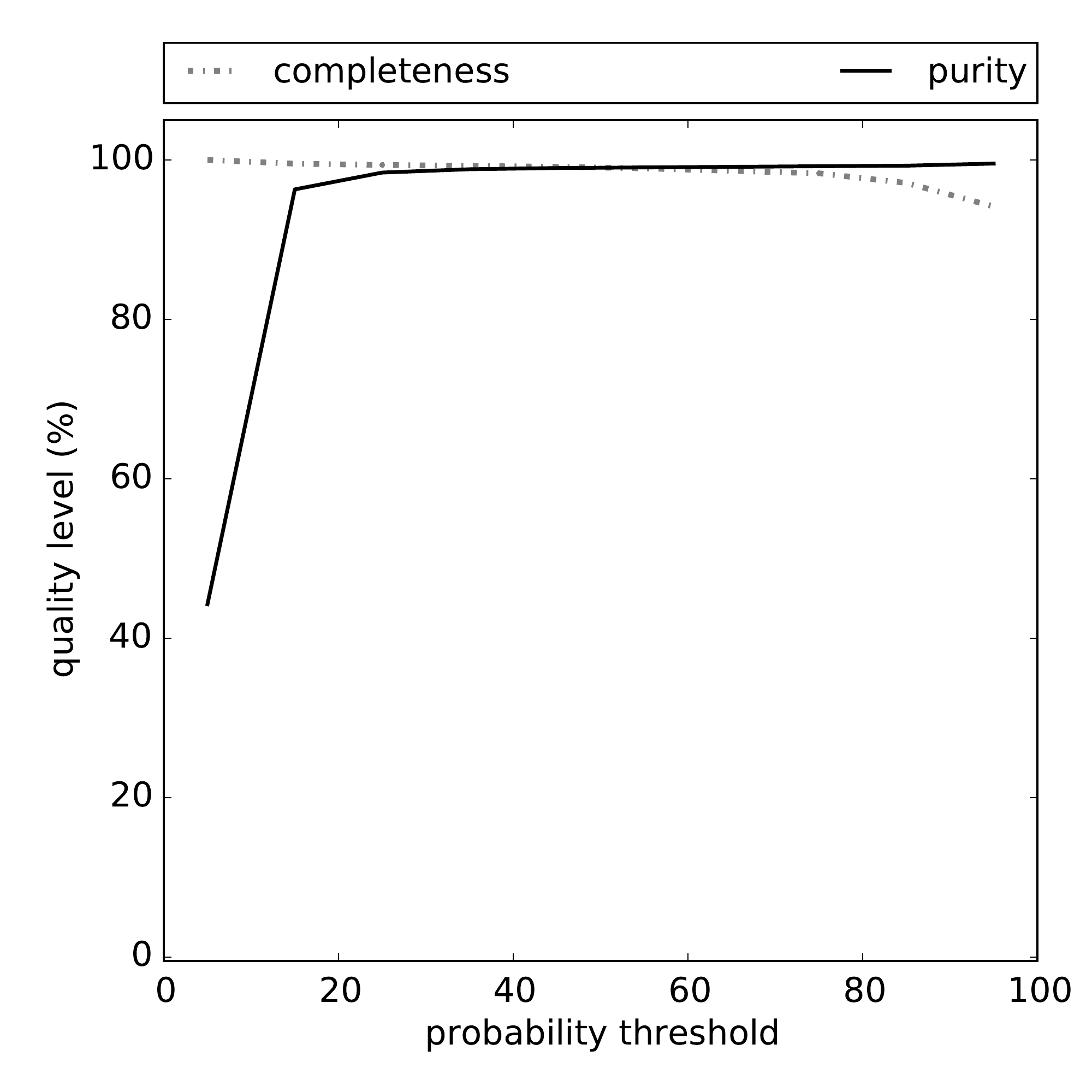} & 
\includegraphics[height=0.29\linewidth]{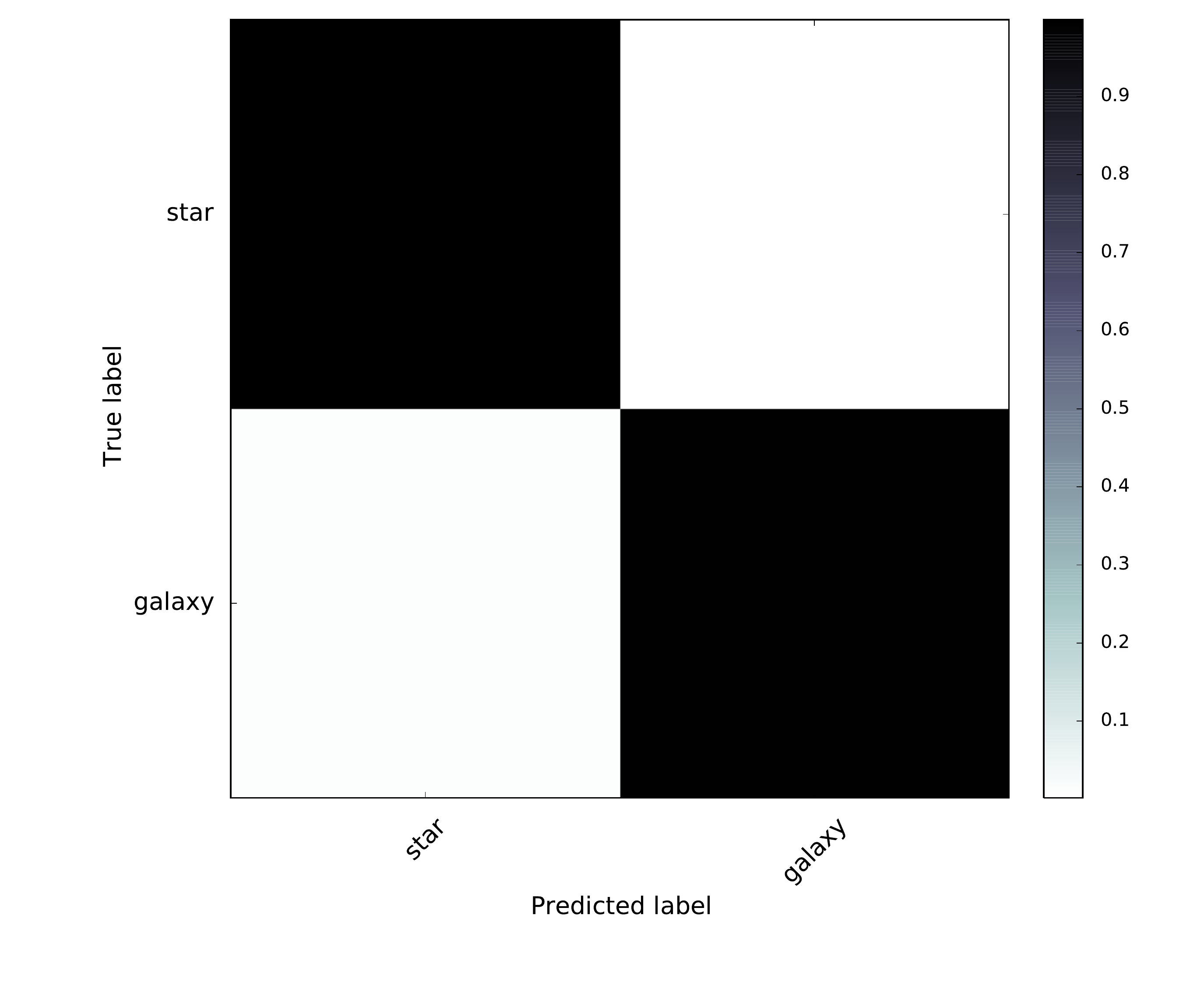} \\
(a) & (b) & (c) \\
\end{tabular}
\caption{Stars identified in the test sample. Panel (a) shows the color locus occupied by the stars (black points) compared to the color locus occupied by the non-stars, i.e. galaxies and AGN (gray points) adopting a probability threshold of 50\%. Panel (b) shows the completeness (gray dot-dashed line) and purity (black solid line) of the identified star sample for each given probability threshold given on the $x$-axis. Panel (c) summarizes the performance of the classifier in a confusion matrix.}
\label{fig:test_star}
\end{figure*}

First we examine the separation between stars and galaxies. We achieve accuracy of ACC=99.7\%, precision P=99.1\%, recall R=99\% and fall-out of only F=0.2\%. As the locus occupied by stars in color space is narrow and almost disjoint to that of galaxies, it is an easier classification problem especially if infrared data are available. In Figure \ref{fig:test_star} (a) we plot the results of the test sample. The color plot (Y-W1) vs (g-J) shows clearly the separation in color space between stars (black), selected as sources with Pr[star]>50\% and galaxies (gray). Similar plots using IRAC colors have been used previously in the literature \citep[e.g.,][]{Ilbert2009}. 

By default, the Random Forest implementation in Python will assign each object to the class with the highest probability. In the case of two categories the threshold is set at 50\%. Nevertheless, given the science interest, a selection of pure or complete sample might be of interest. Since the Random Forest provides also as output the probability for each class, it is possible to select the sample according to the desired specifications. Panel (b) of Fig. \ref{fig:test_star} shows the level of completeness (gray dot-dashed line) and purity (black line) as a function of probability threshold. For example, a threshold P[star]>80\% would lead to roughly 100\% purity and 95\% completeness.

The last panel of Fig. \ref{fig:test_star} visualizes the performance of classifier A in a confusion matrix. The diagonal elements represent the percentage of true positive classifications, while the off-diagonal elements show the percentage of false positive classifications. A perfect classifier will have black diagonal elements and white off-diagonal elements, classifier A with ACC=99.7\% shows excellent performance.

\subsubsection{Classifier B: which is the best photo-z library?}\label{sec:classifierB}

\begin{figure*}
\centering
\begin{tabular}{ccc}
\includegraphics[width=0.3\linewidth]{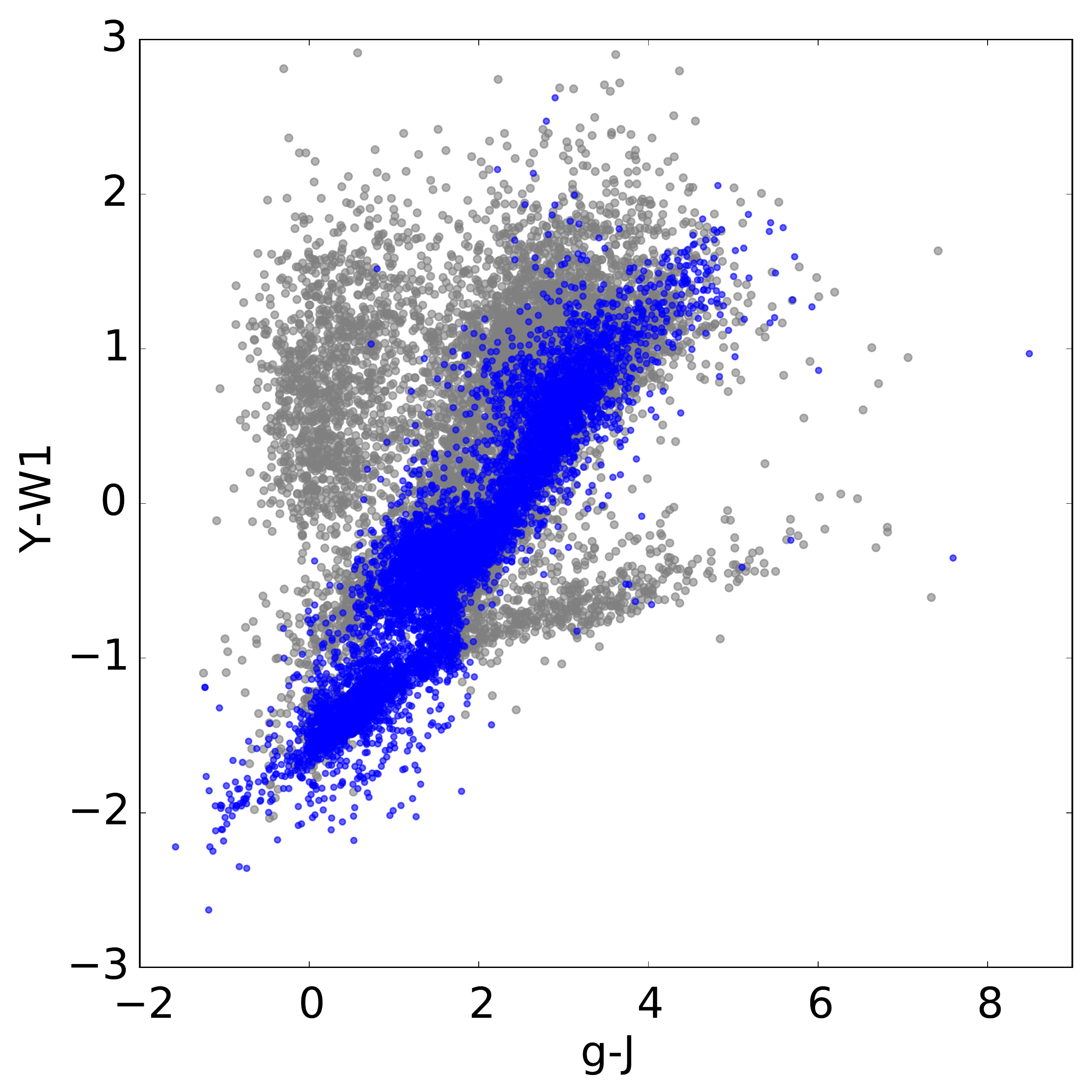} &
\includegraphics[width=0.3\linewidth]{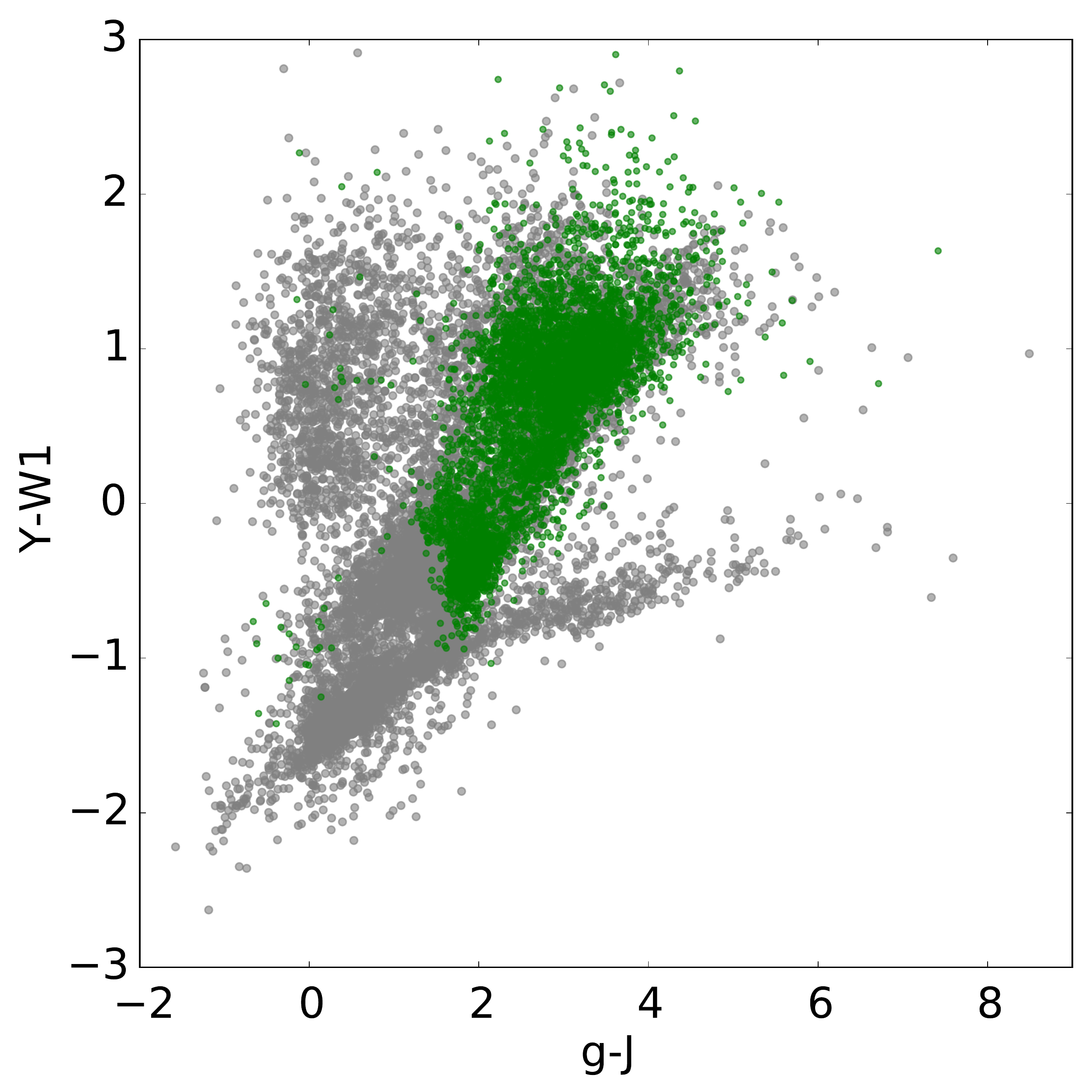} &
\includegraphics[width=0.3\linewidth]{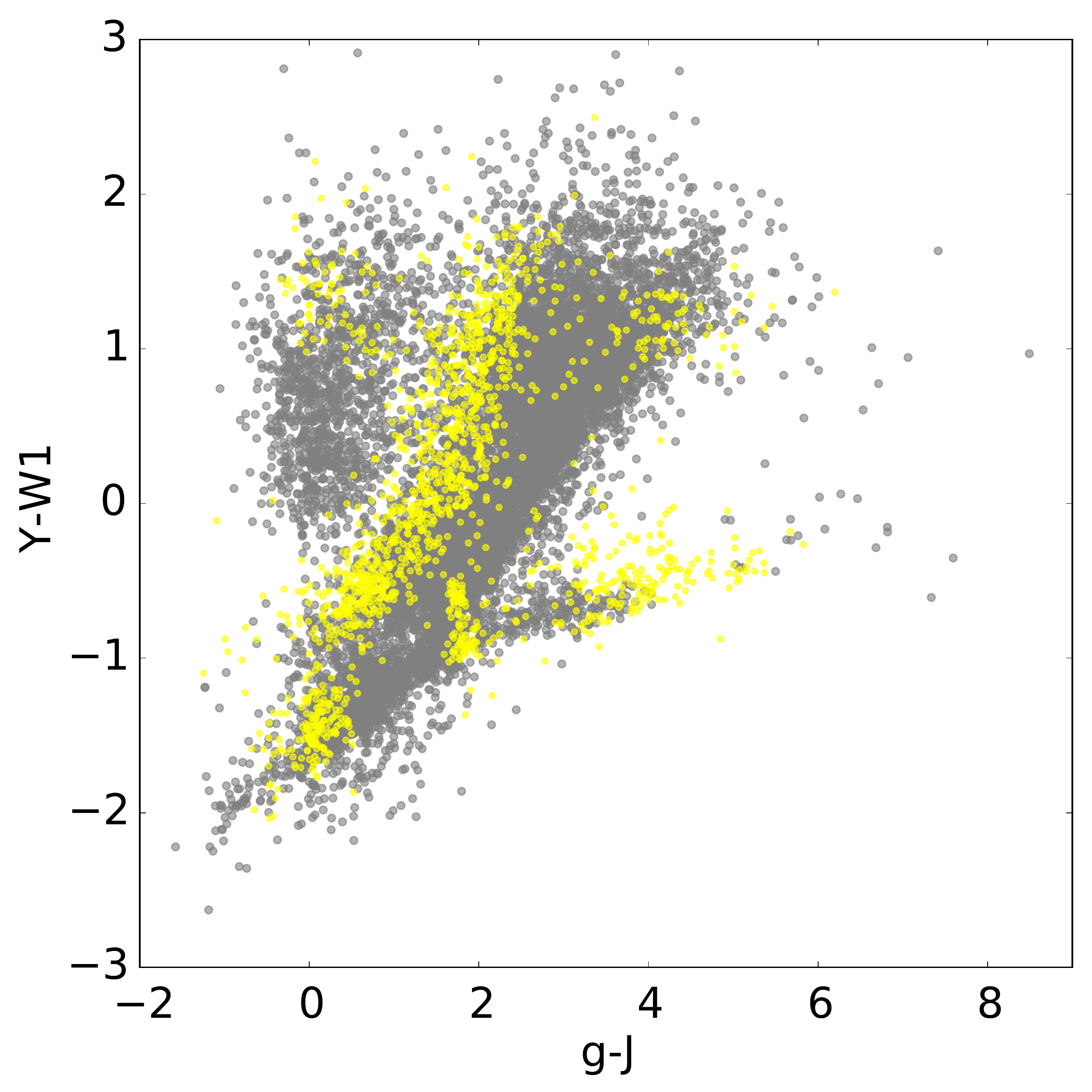} \\
 (a) passive & (b) starforming & (c) starburst\\
\includegraphics[width=0.3\linewidth]{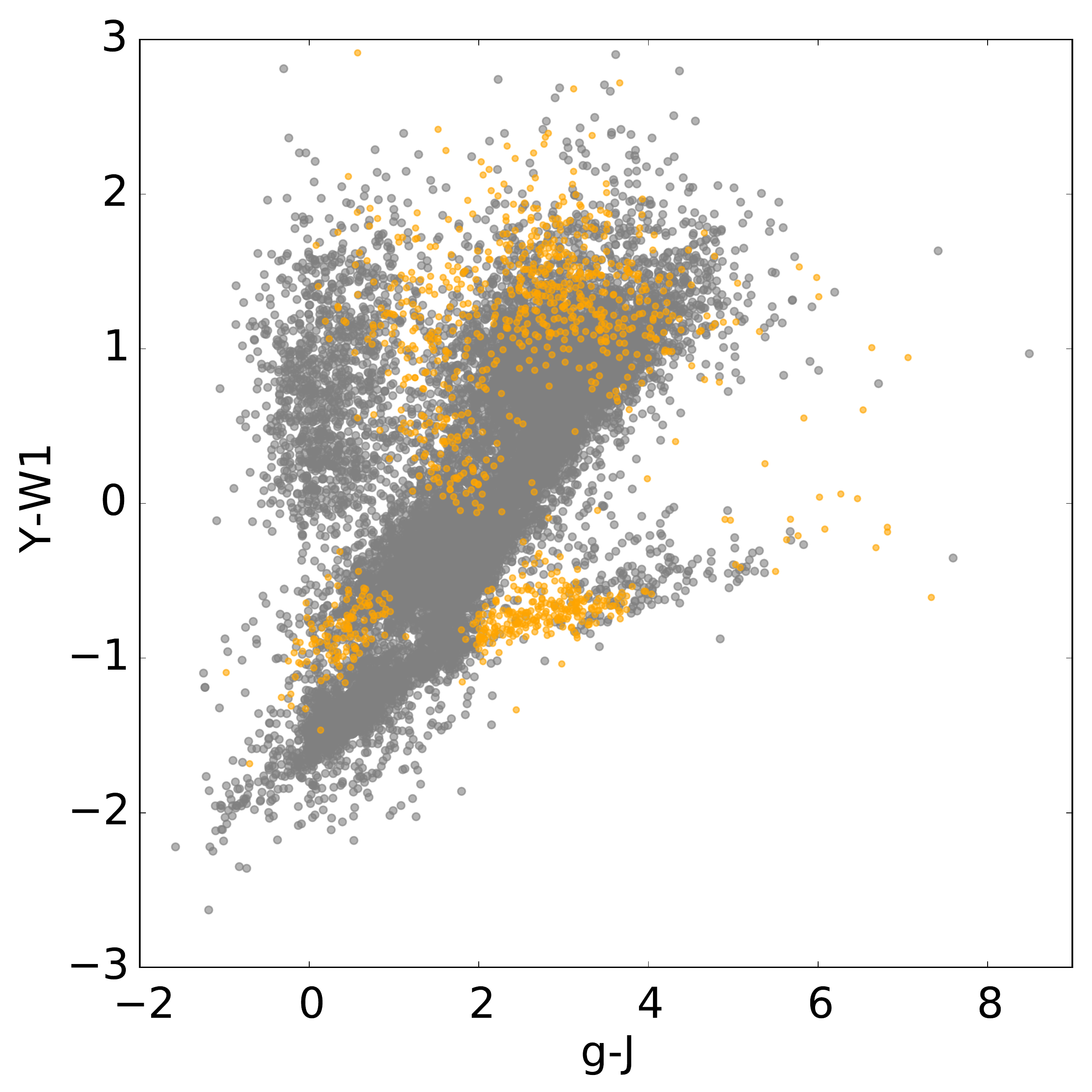} &
\includegraphics[width=0.3\linewidth]{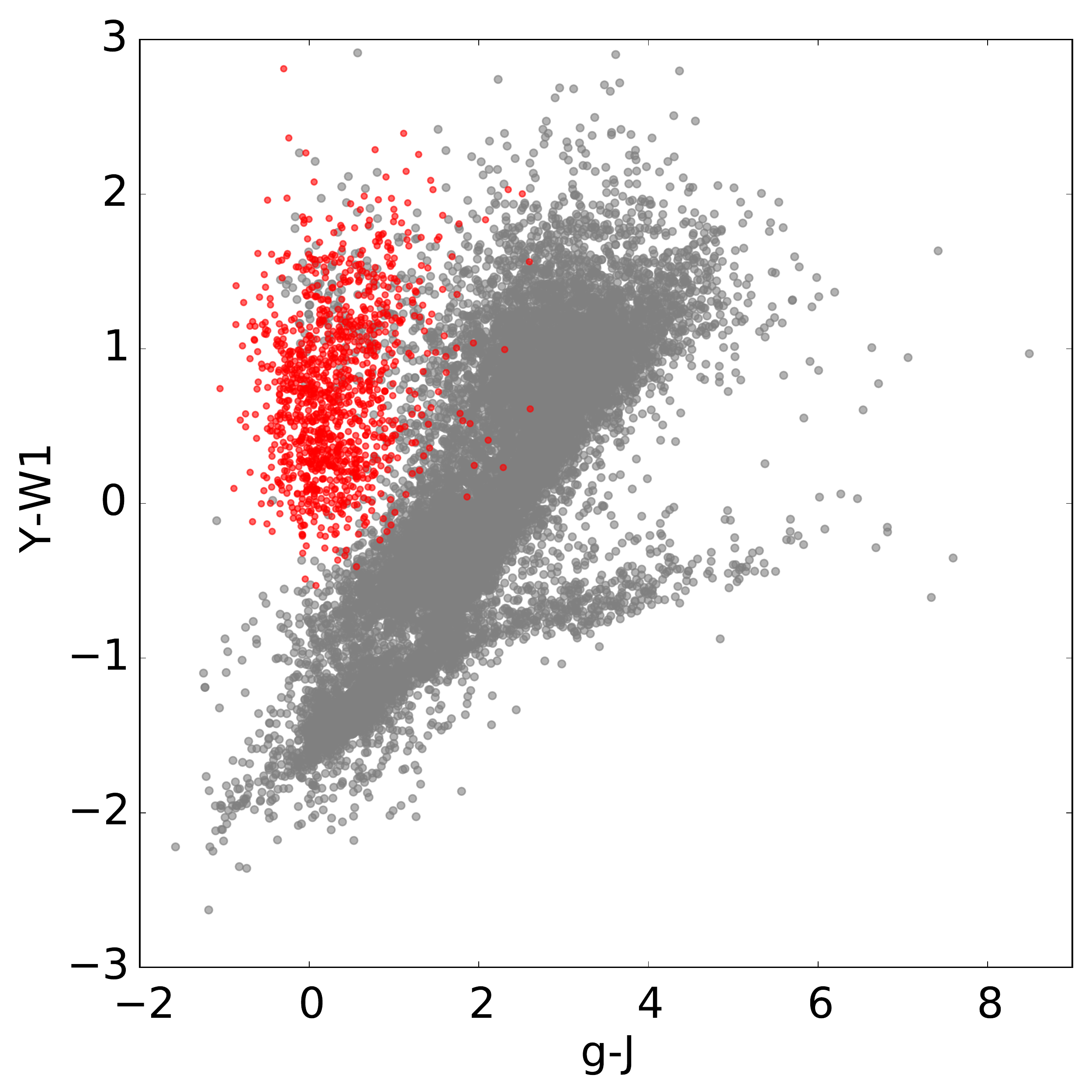} &
\\
 (d) AGN & (e) QSO &  \\
\end{tabular}
\caption{Output of the galaxy classification (colored points) compared to the total input population (gray points). The galaxy class is given below each plot. We note that stars have not been excluded a priori.}
\label{fig:test_B_class}
\end{figure*}

Classifier B is trained to identify the model library that will provide the optimal photometric redshift solution judged by the best photometric redshift. As discussed in Sec. \S \ref{sec:classdef}, we grouped the galaxy models roughly according to their star-formation into passive, starforming and extreme starforming which for short we will refer to as starburst. We have also included two additional libraries: AGN and QSO (see Table \ref{tab:SEDlib} for the full list of templates used and \S \ref{sec:galclass} for the explored galaxy class combinations). In Fig. \ref{fig:test_B_class} we show the objects identified in the test sample in each category using the same color-color plot as in Fig. \ref{fig:test_star}. The gray points show all of the sample for comparison. We note that classifier B is applied to the full sample without any a priori knowledge of stars. Passive and starforming galaxies trace a somewhat different region in the color space, albeit with a large overlap in this two-dimensional representation. QSOs on the other hand shown in panel (e) are very well isolated with respect to the normal galaxy population. Finally, as expected, the AGN population is located in the overlap region between the normal galaxies and the QSO, since by construction their SEDs are a mixture of the two components.

\begin{figure*}
\centering
\begin{tabular}{ccc}
\includegraphics[width=0.3\linewidth]{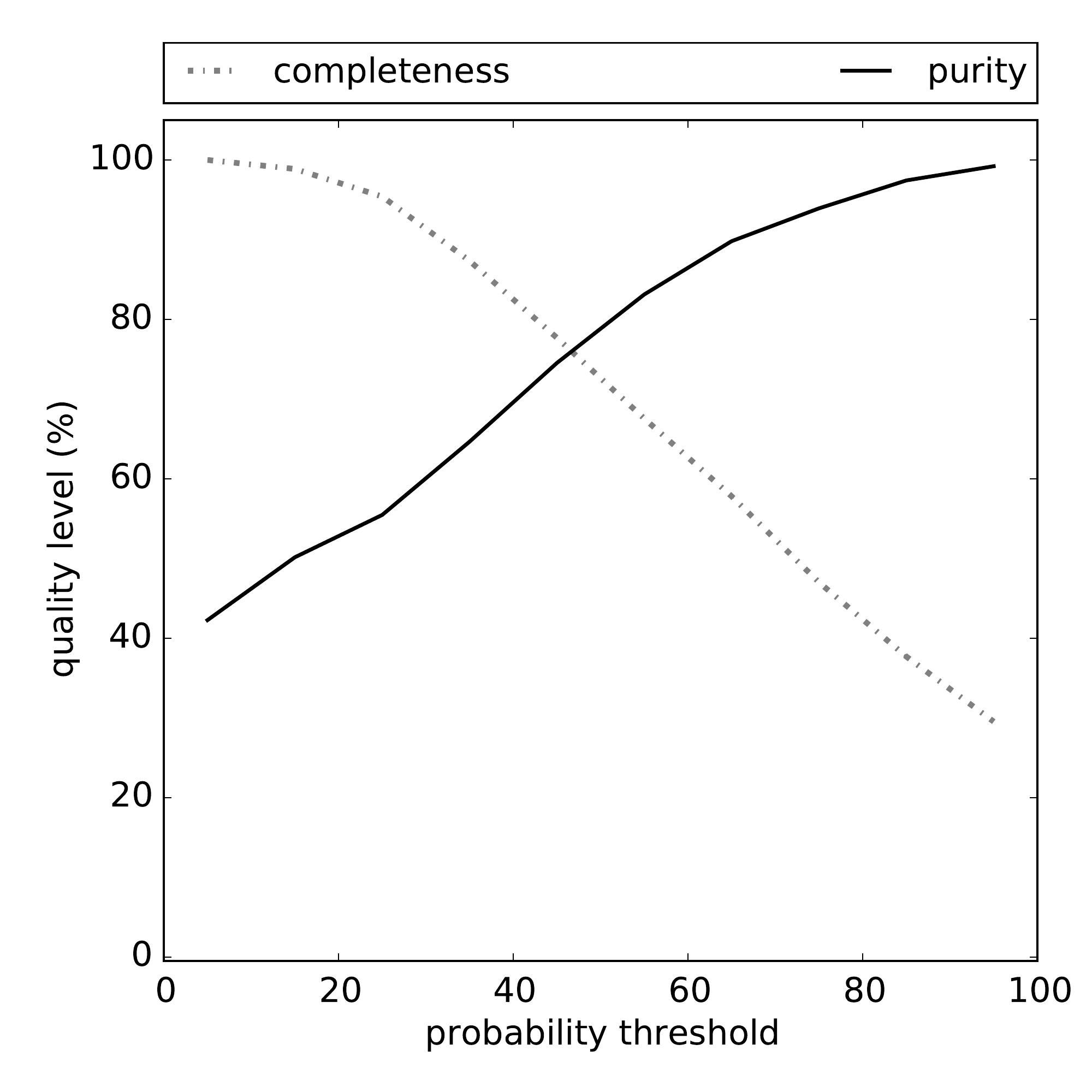} &
\includegraphics[width=0.3\linewidth]{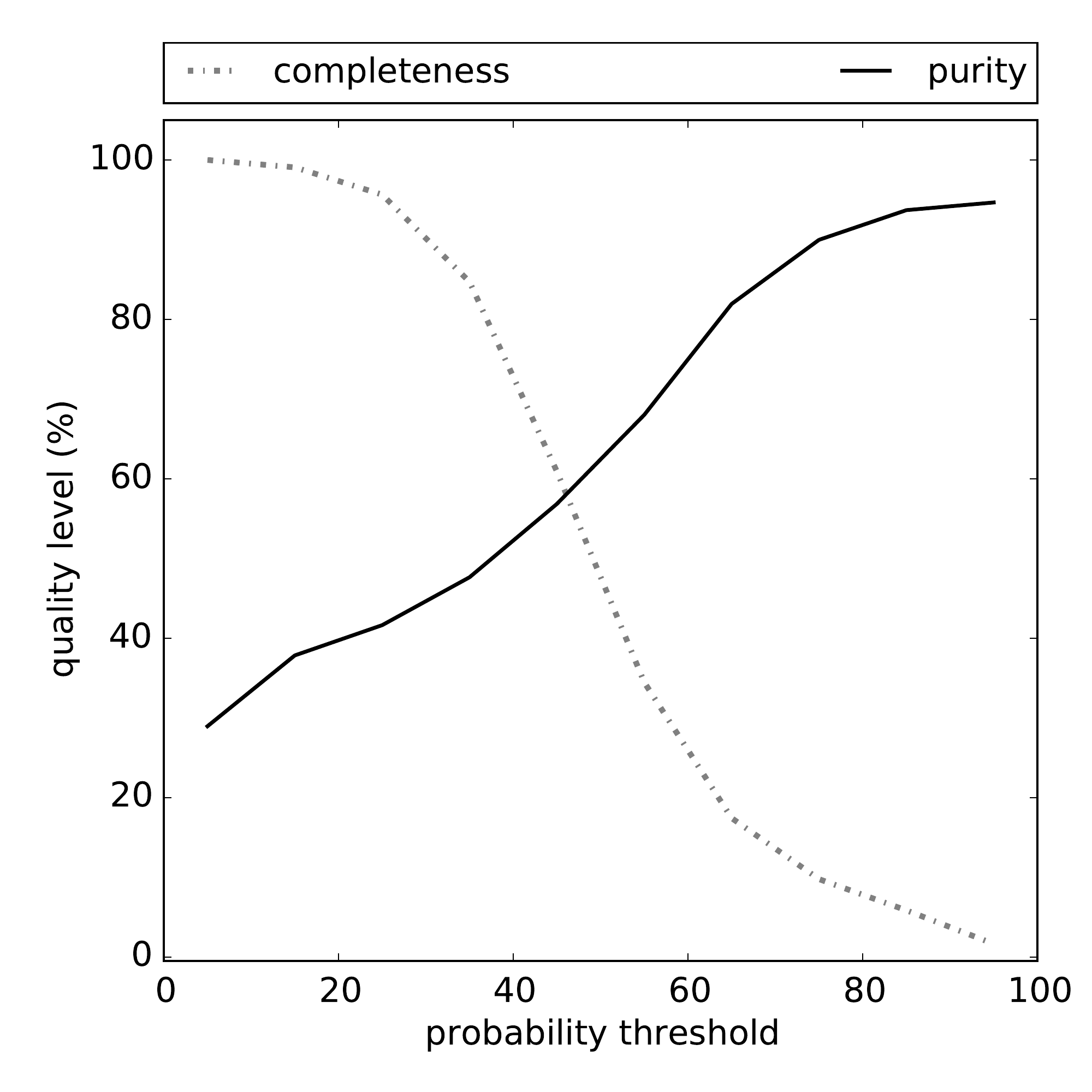} &
\includegraphics[width=0.3\linewidth]{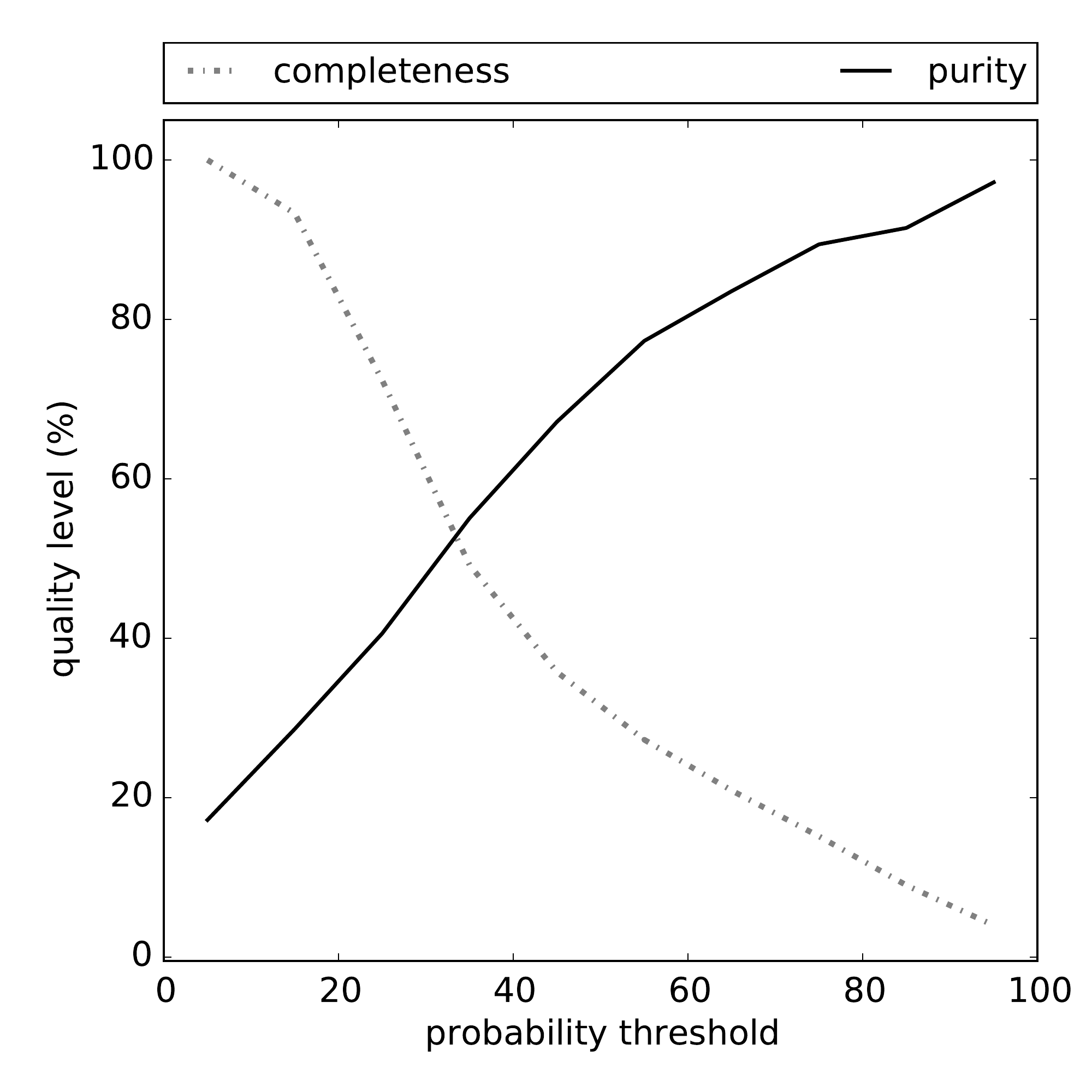} \\
 (a) passive & (b) starforming & (c) starburst\\
\includegraphics[width=0.3\linewidth]{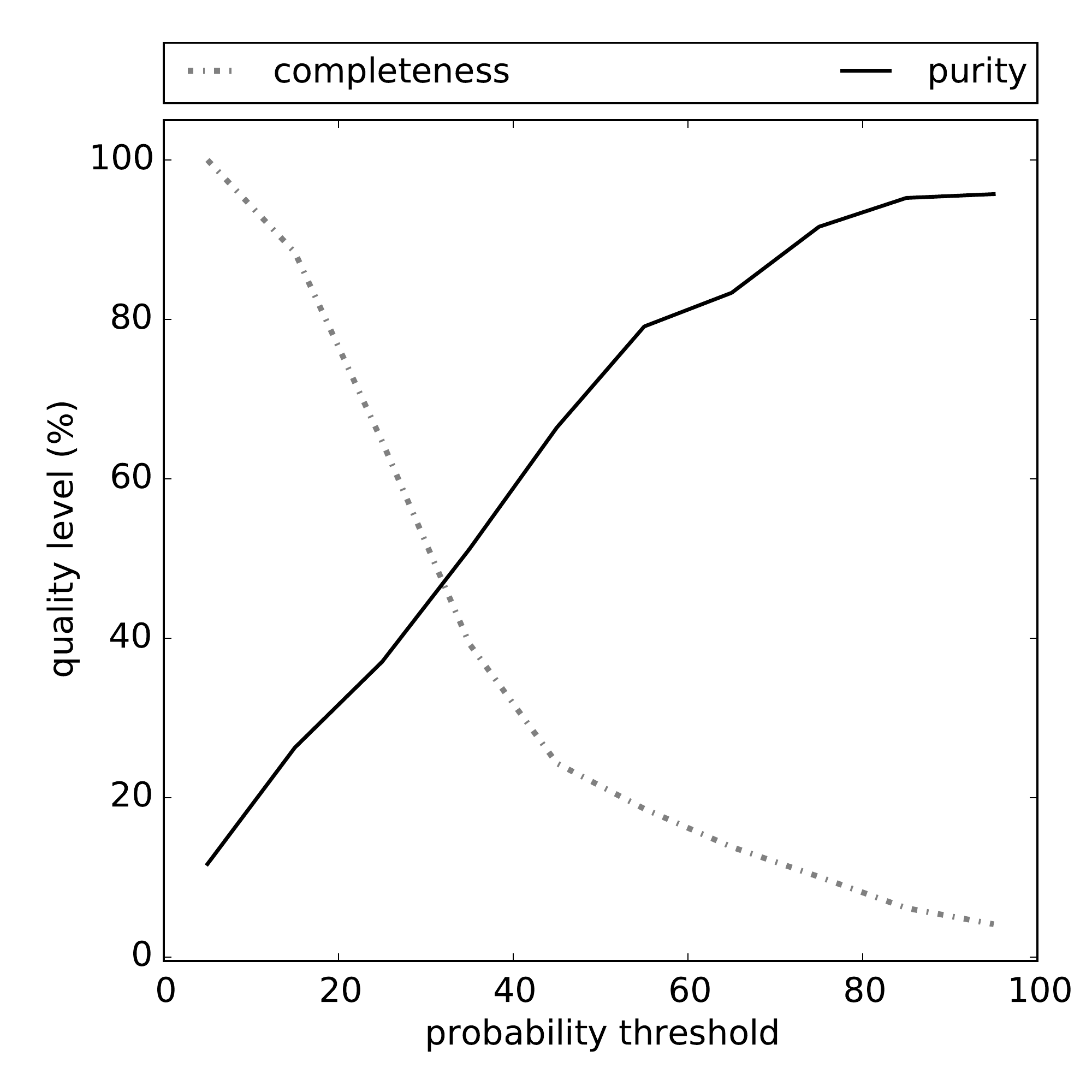} &
\includegraphics[width=0.3\linewidth]{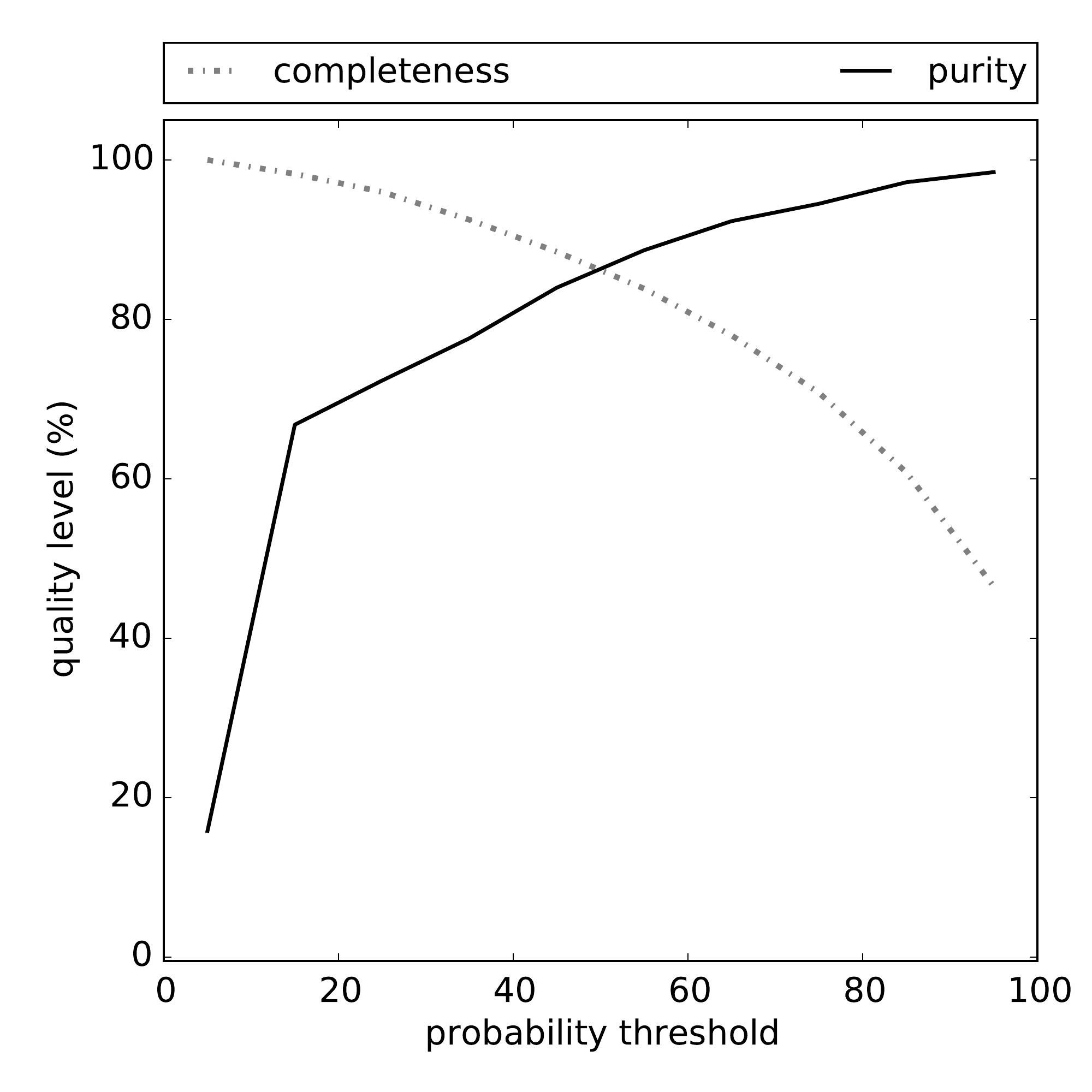} &
\includegraphics[height=0.29\linewidth]{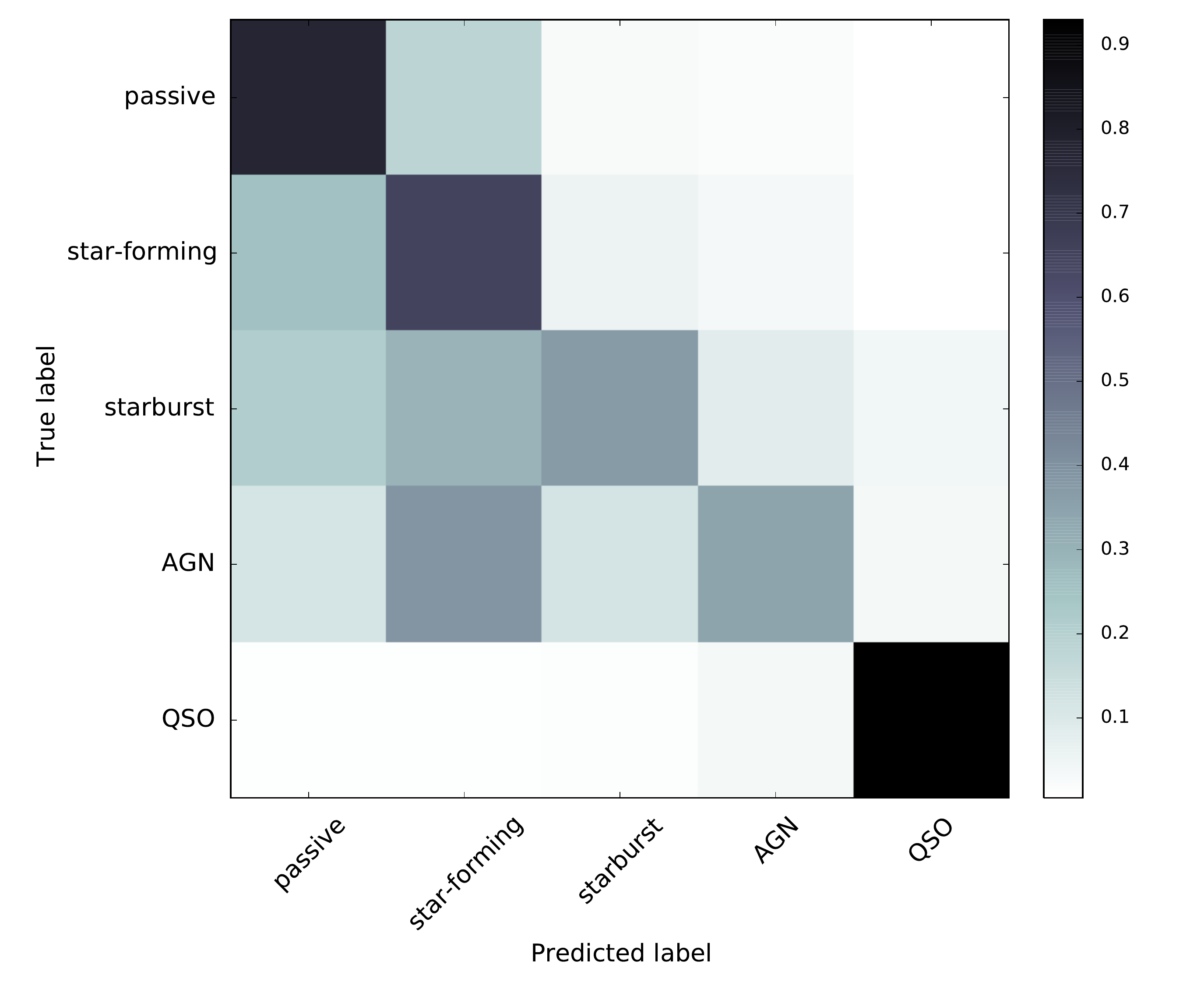} \\
 (d) AGN & (e) QSO & (f) \\
\end{tabular}
\caption{Completeness (dashed line) and purity (solid line) of each galaxy class as a function of probability threshold. Last panel shows the confusion matrix.}
\label{fig:test_B_purity}
\end{figure*}

The weighted average accuracy of classifier B over all classes is about 64\%. Since in this case we have a multiclass classification, the average accuracy is not as informative. In Fig. \ref{fig:test_B_purity} we show the completeness and purity per galaxy category as a function of probability threshold, similarly to Fig. \ref{fig:test_star}. The minimum threshold for class assignment is 20\%. As seen from panels (a)-(e) such a threshold would correspond to complete (70\%-90\%) but rather impure samples (30\%-60\% purity). We note that in the final galaxy sample the performance will be slightly better since for now the stars are still included in the sample. The confusion matrix in the last panel of Fig. \ref{fig:test_star} shows the relative mixing of the classes adopting the default class assignment. The QSOs are well separated from all other classes and with little false positive detections. The same is true, but with lesser quality for the passive galaxies. The largest mixing is present between the starforming and the AGN population.

Since in our sample we know the true classification between stars and galaxies, it is interesting to explore the confusion between these two classes as revealed in Fig. \ref{fig:test_B_class}. The sample of spectroscopically confirmed stars ($\sim$ 2500 sources) is distributed in four galaxy classes: passive (71\%), starforming (3\%), starburst (16\%), and AGN (10\%). No stars were classified as QSO. Nevertheless, 74\% of the star sample is placed at z=0 from the SED fitting. 
Stars can mimic passive galaxies are redshift zero (99\% of passive sample, 57\% of total star sample). Stars classified as starforming and starburst galaxies are placed either at redshift zero (64\%) with varying amounts of absorption, or at higher redshifts ($\mathrm{0.01<z<1.0}$) with the SED fitting selecting the template with the highest allowed absorption (E(B-V)=0.3). Similarly, 51\% of the stars classified as AGN are placed at z=0, with varying amounts of E(B-V) depending on the template SED. In the AGN case, the dominant template that is the best fit to stars at non-zero redshifts is template 48 consisting of 10\% S0 non-starforming disk galaxy and 90\% of QSO2, which is a template with intrinsic heavy obscuration in the UV-optical part of the spectrum.
If we use the results of classifier A to remove the stars from the galaxy sample the weighted classification accuracy is not affected significantly (worsens about 4\%), since the stars are classified with high accuracy (weighted average 85\%)  even using in Classifier B.\footnote{This means that 85\% of the star sample is distributed accurately to the input classes that originally have ``star-like'' SEDs. Incidentally, 60\% of the star sample is correctly placed at z=0 from fitting their SEDs with galaxy templates only.}

\subsubsection{Classifier C: is the photo-z solution acceptable?}\label{sec:classifierC}

Outlier identification is a difficult task, because there are many reasons that can lead to a failed estimation of photometric redshift. The reasons include intrinsic physical properties such as i) variability, as shown in \citet{Salvato2009}, ii) rare or not represented SED in the template library and iii) true degeneracies in color space (Richards 2006). External factors can also lead to a failed photometric redshift estimation i) bad photometry, for example saturation ii) source misassociation, for example blending due to large PSF or astrometric offsets iii) wrong spectroscopic redshift.

\begin{figure*}
\centering
\begin{tabular}{ccc}
\includegraphics[width=0.3\linewidth]{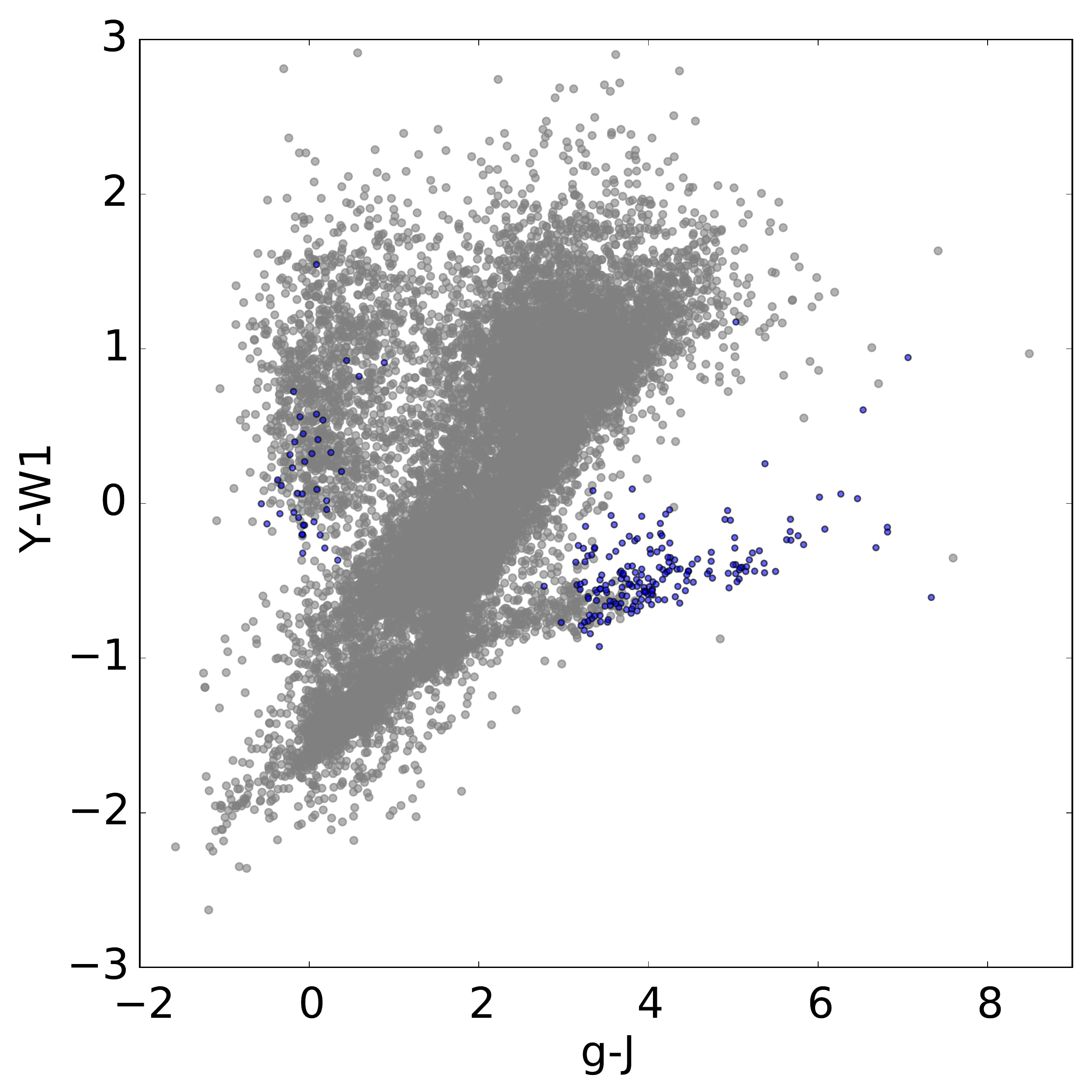} &
\includegraphics[width=0.3\linewidth]{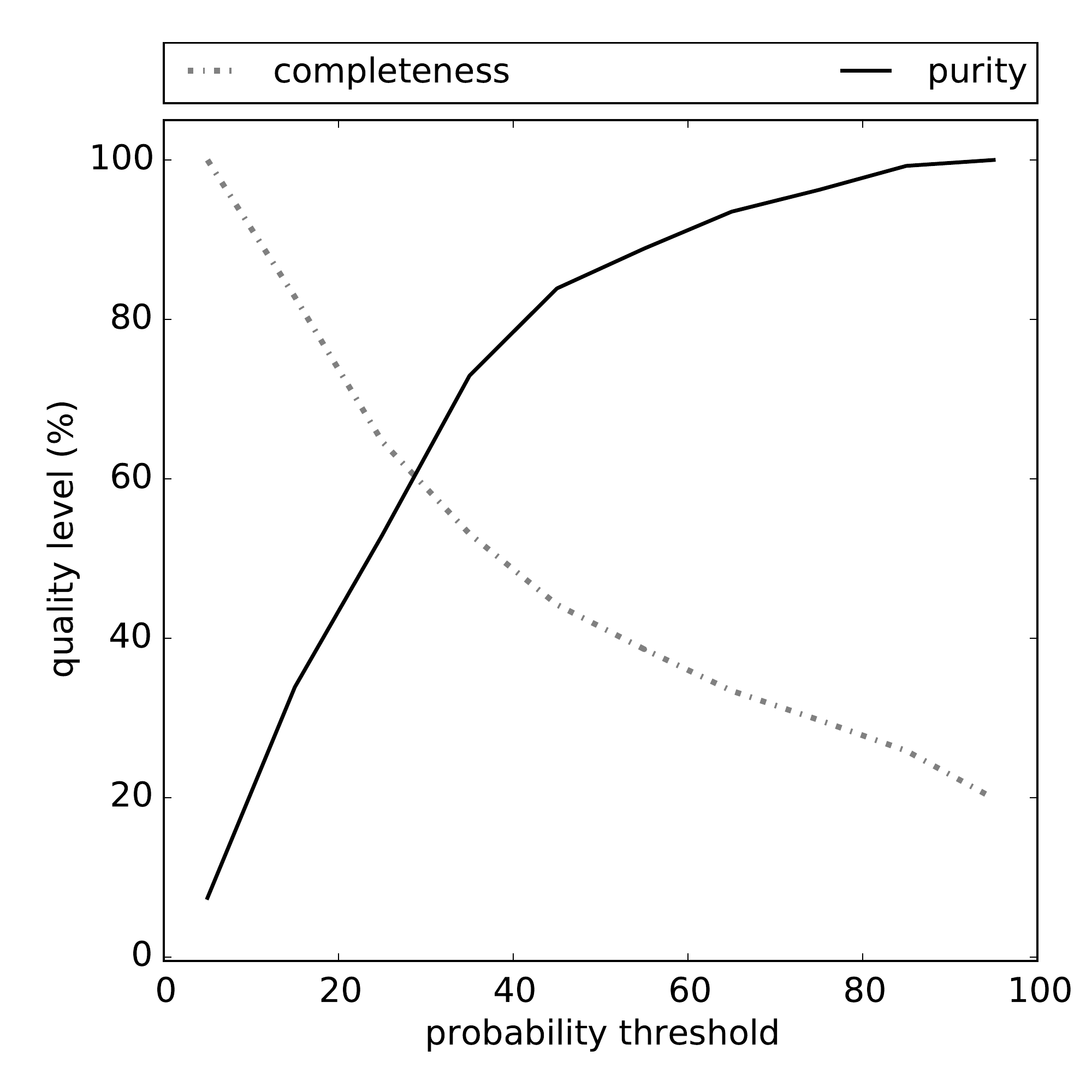} & 
\includegraphics[height=0.29\linewidth]{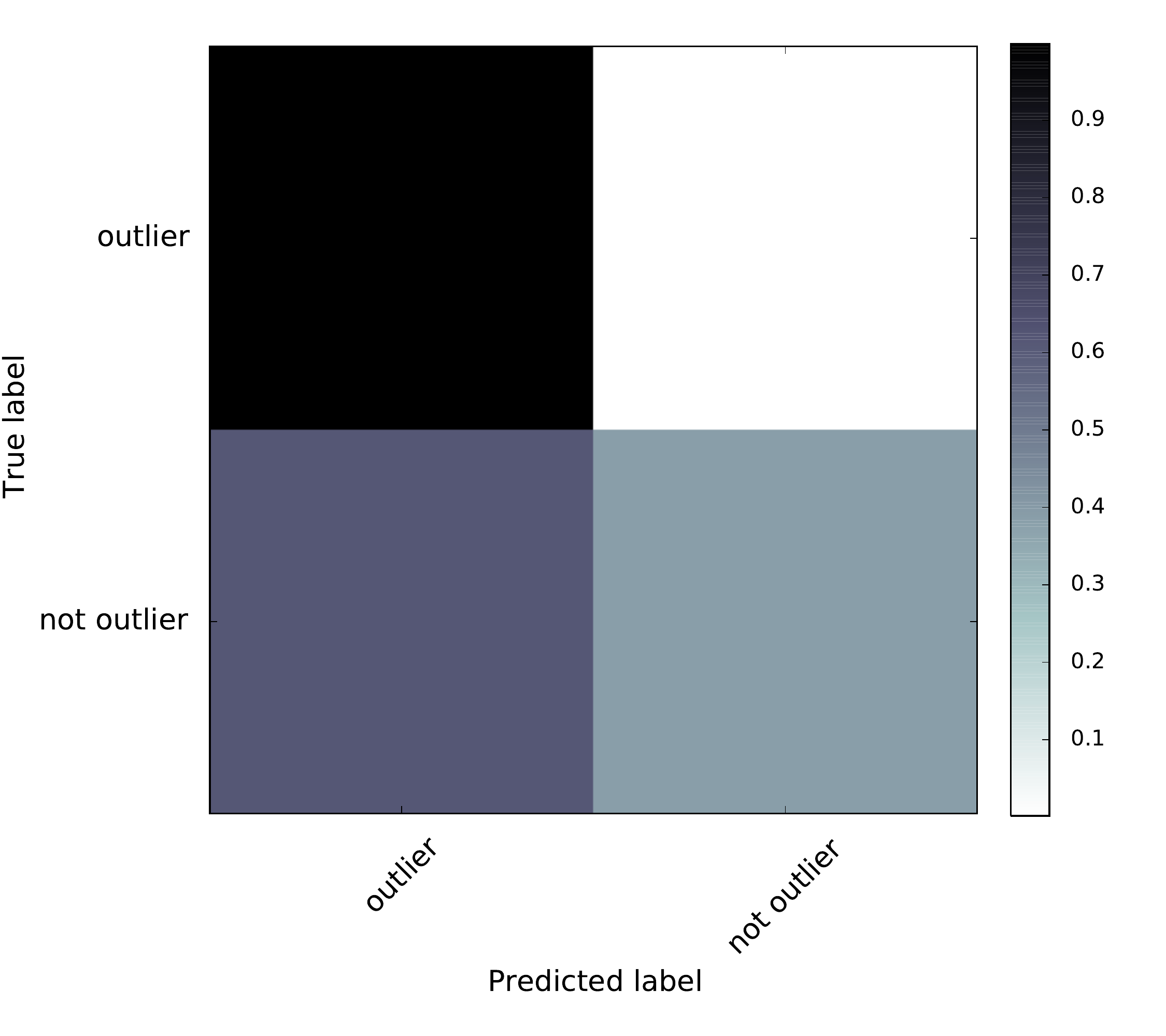} \\
(a) & (b) & (c) \\
\end{tabular}
\caption{Same as Fig. \ref{fig:test_star} but for classifier C.}
\label{fig:test_outlier}
\end{figure*}

We explored the possibility to identify outliers using a random forest classifier. The sample used contains the sources that follow eq. \ref{eq:eta}. In Figure \ref{fig:test_outlier} we show, similarly to the stars, the output classification of the test sample. The black points correspond to the sources with P[outlier]>50\%. We see that the majority of the outliers is gathered in the locus occupied by QSO and stars.The average accuracy of Classifier C is 97.9\%, this score is also immune to the presence of stars (see \S \ref{sec:classifierB}) as the stars are classified as outliers with very high accuracy (99\%).

Similarly to classifiers A and B, adopting a more conservative outlier threshold, will allow for a more complete sample, P[outlier]>20\% leads to 80\% complete sample in the expense of purity (40\%). However, the selection threshold can be adjusted according to the science case, or even ignored altogether. Examining the confusion matrix in the last panel of Fig. \ref{fig:test_outlier} we see that there are many false positive identifications of good photometric redshift estimations as outliers (60\%) at P[outlier]>50\%, however most outliers are indeed classified as outliers.

\subsection{Stage III: consolidation}\label{sec:consolidation}
By creating three classifiers to provide a star, galaxy and outlier classification we have flexibility in the consolidation phase. Here, the scientific goal of the survey can be taken into account in adopting the appropriate probability threshold selection for each classifier, using the results of the test sample presented in \S \ref{sec:classifierA}-\ref{sec:classifierC}.

\subsubsection{Threshold selection}
For classifier B, we have performed multiclass classification and the sum of the probabilities of the five classes is equal to unity. In this case, the default RF behavior assigns as best class the one with the highest probability (Consolidation A). However, we may also choose to make an extra threshold cut for the galaxy classes (Consolidation B). Here we adopt the threshold of 40\% for all normal galaxy classes and AGN, corresponding to at least 50\% pure samples and the default 20\% threshold for QSO. In Table \ref{tab:consolidation} we show the photometric redshift quality for the two cases. We notice a slight improvement in the overall photometric performance, both in terms of accuracy ($\sigma_A$=0.035, $\sigma_B$=0.032) and percentage of outliers ($\eta_A$=4.9\%-$\eta_B$=4.5\%) to the expense of the numbers of available sources with 84\% of the original sample retained after consolidation B.

\subsubsection{Rejection of stars and outliers}\label{sec:rejections}
In addition, we can select the threshold of the star and/or outlier identification. For the star labeling we adopt P[star]>50\% corresponding to about 99\% completeness and accuracy. Respectively for the outlier rejection we opt for completeness over purity adopting a threshold of P[outlier]>20\% which corresponds to about 70\% completeness and 40\% purity.
Table \ref{tab:consolidation} shows the performance of the photometric redshift estimation and the corresponding number of sources remaining the sample for each rejection step. We note that in general, rejection of stars and outliers will lead to higher accuracy and less outliers both when adopting the highest galaxy class probability (consolidation A) and when imposing an absolute threshold (consolidation B) compared to no rejection at all. Seemingly the accuracy of photo-z worsens when stars are rejected (from $\sigma_{no\;rej}=0.035$ to $\sigma_{star\;rej}=0.042$). This is due to the fact that the majority of the stars are placed correctly at redshift zero during SED fitting, even when using galaxy templates (60\% of the star sample, see \S \ref{sec:classifierB}).

\begin{table*}
\centering
\begin{tabular}{c|cccc|cccc}\hline\hline
Rej.    & \multicolumn{4}{c|}{Consolidation A} & \multicolumn{4}{c}{Consolidation B}\\
type &    N & \% & $\sigma$ & $\eta$(\%) & N & \% & $\sigma$ & $\eta$(\%)\\\hline
no rejection & 16\,394 & 100.0 & 0.035 & 4.9 & 13\,773 & 84.0 & 0.032 & 4.5 \\
star & 13\,859 & 84.5 & 0.042 & 4.2 & 11\,294 & 68.9 & 0.040 & 3.7\\
outlier & 15\,752 & 96.1 & 0.033 & 2.6 & 13\,212 & 80.6 & 0.030 & 2.1\\
star \& outlier & 13\,436 & 82.0 & 0.041 & 2.9 & 10\,937 & 66.7 & 0.039 & 2.3\\\hline\hline
\end{tabular}
\caption{Number of sources and photometric redshift performance after (A) class assignment according to the highest probability and/or star and outlier rejection (B) class assignment according to the highest probability but imposing a probability threshold for all galaxy classes.\label{tab:consolidation}}
\end{table*}

\subsubsection{Rejection of uninformative PDFs}

\begin{figure*}
\centering
\begin{tabular}{cc}
\includegraphics[width=\columnwidth]{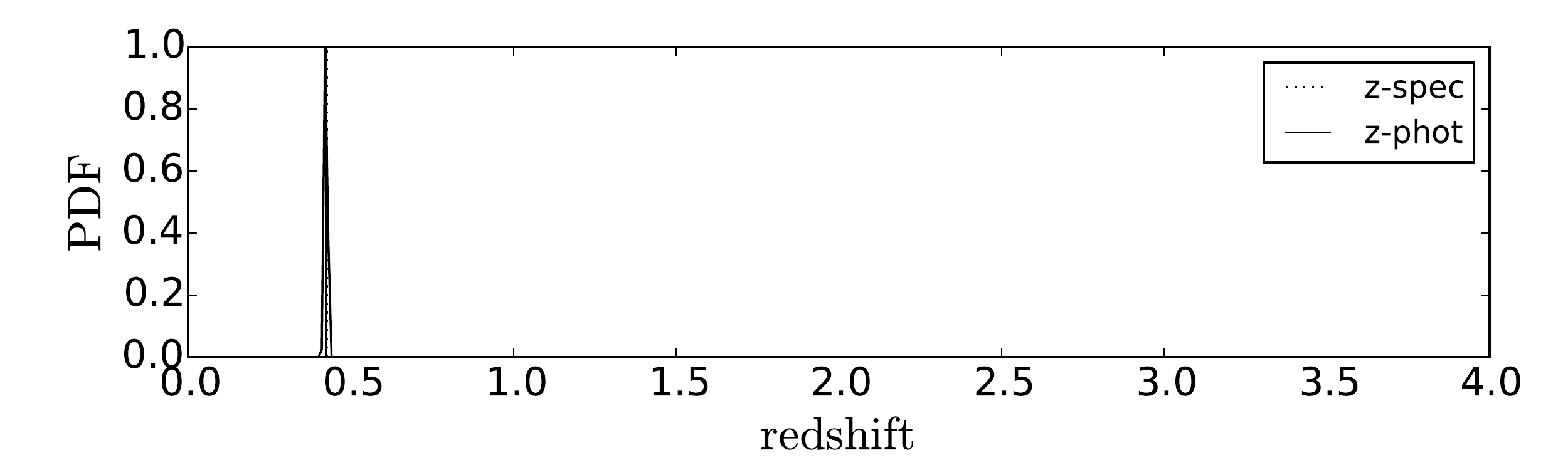}& \includegraphics[width=\columnwidth]{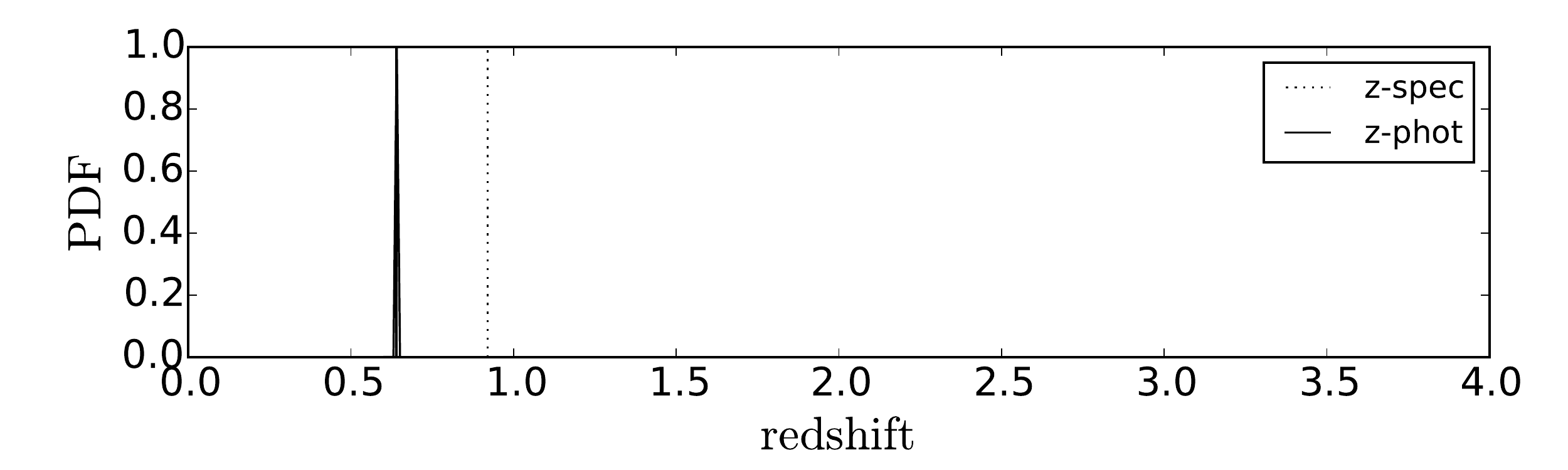}\\
\includegraphics[width=\columnwidth]{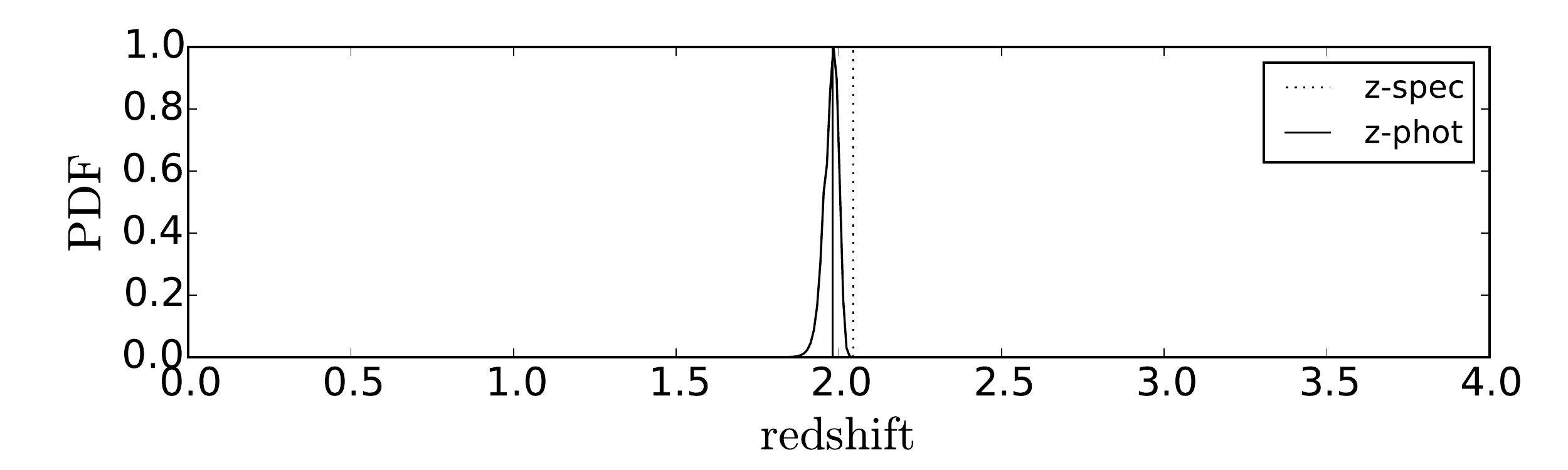}&\includegraphics[width=\columnwidth]{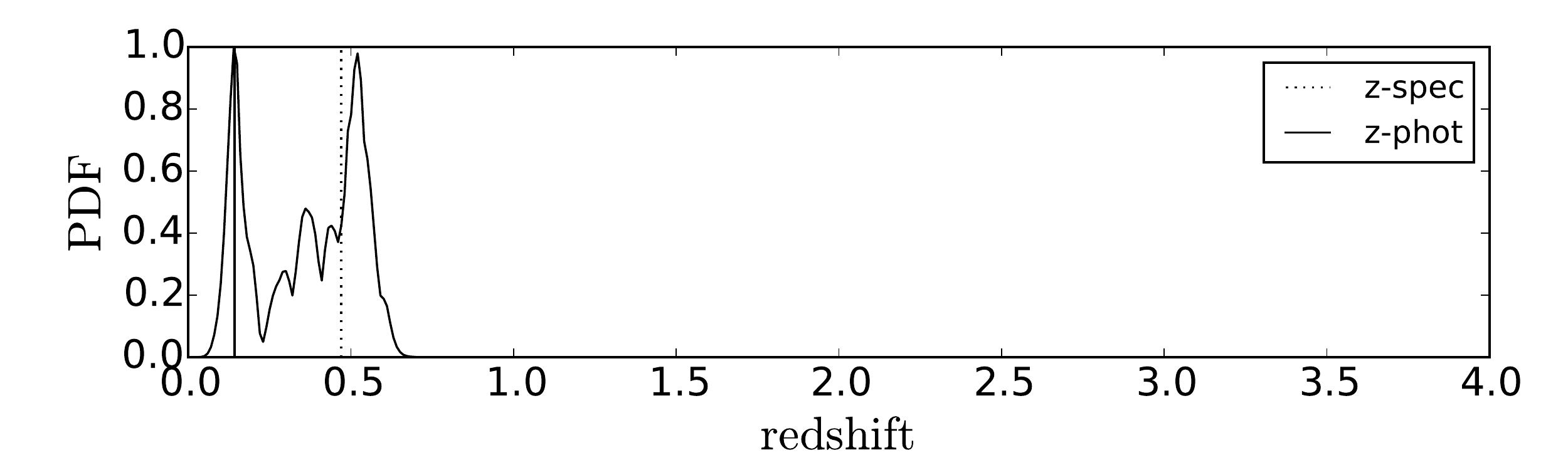}\\
\includegraphics[width=\columnwidth]{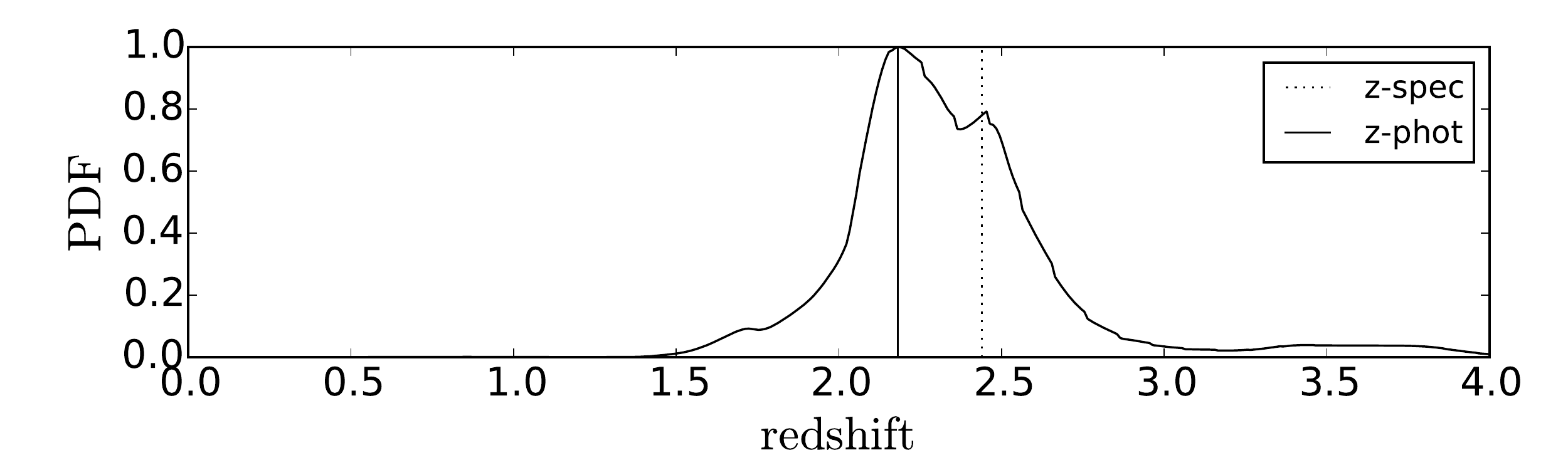}&\includegraphics[width=\columnwidth]{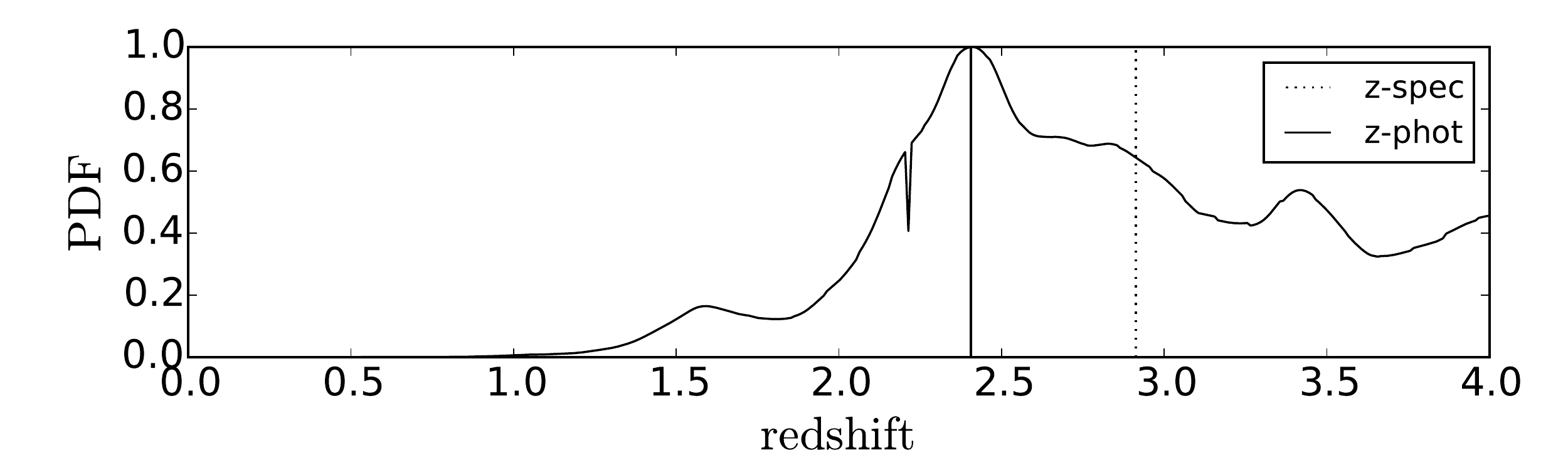}\\
\end{tabular}
 \caption{Example probability distribution functions (PDF) from the consolidated sample. The first row shows two sources with high information content, $D_{KL}\sim7$. The second shows broader and multimodal PDFs with $D_{KL}\sim4$, while the last row shows the PDFs with the least information content $D_{KL}<2$.}
\label{fig:PDF}
\end{figure*}

\begin{figure*}
\centering
\begin{tabular}{cc}
\includegraphics[width=\columnwidth]{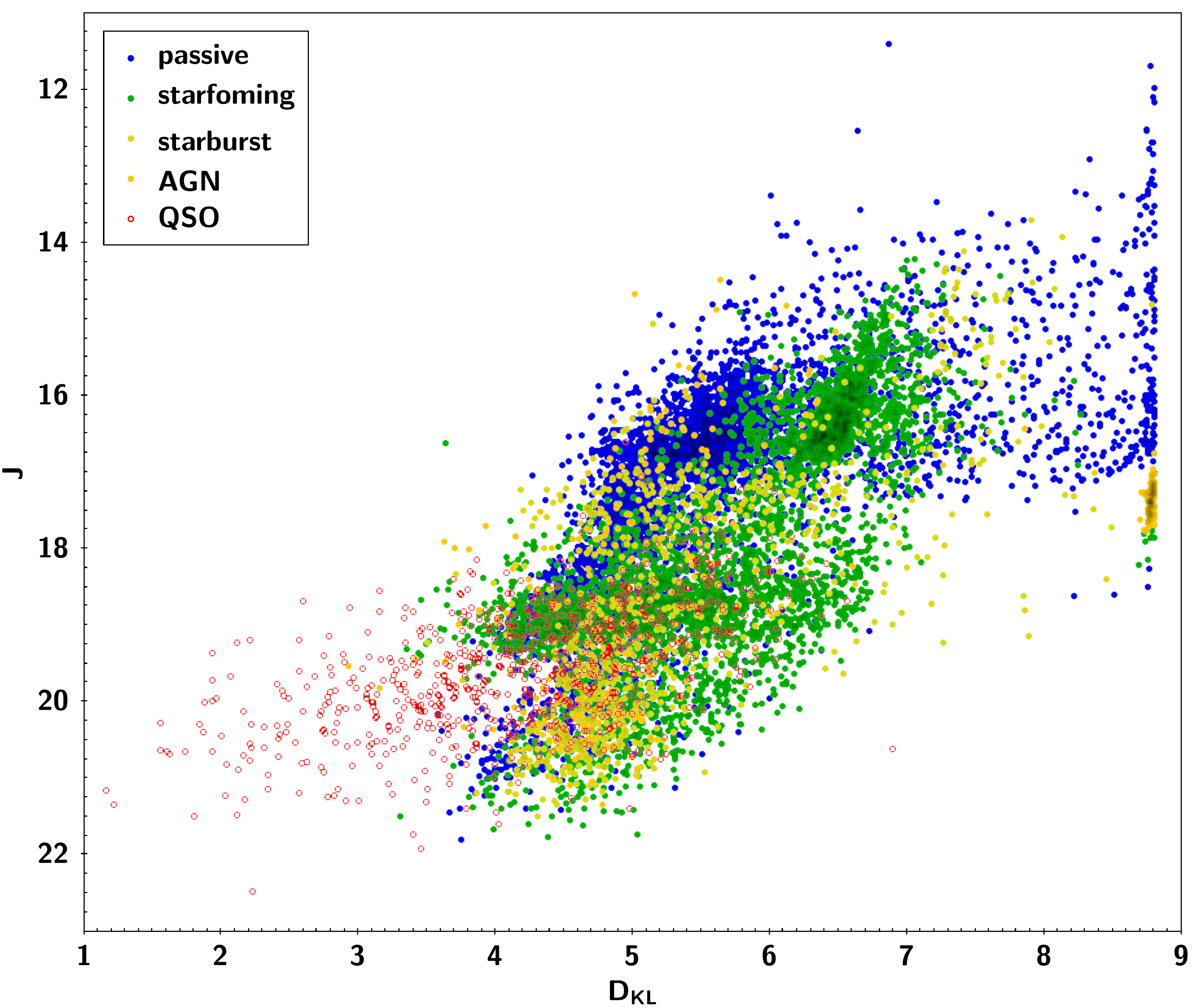}&\includegraphics[width=\columnwidth]{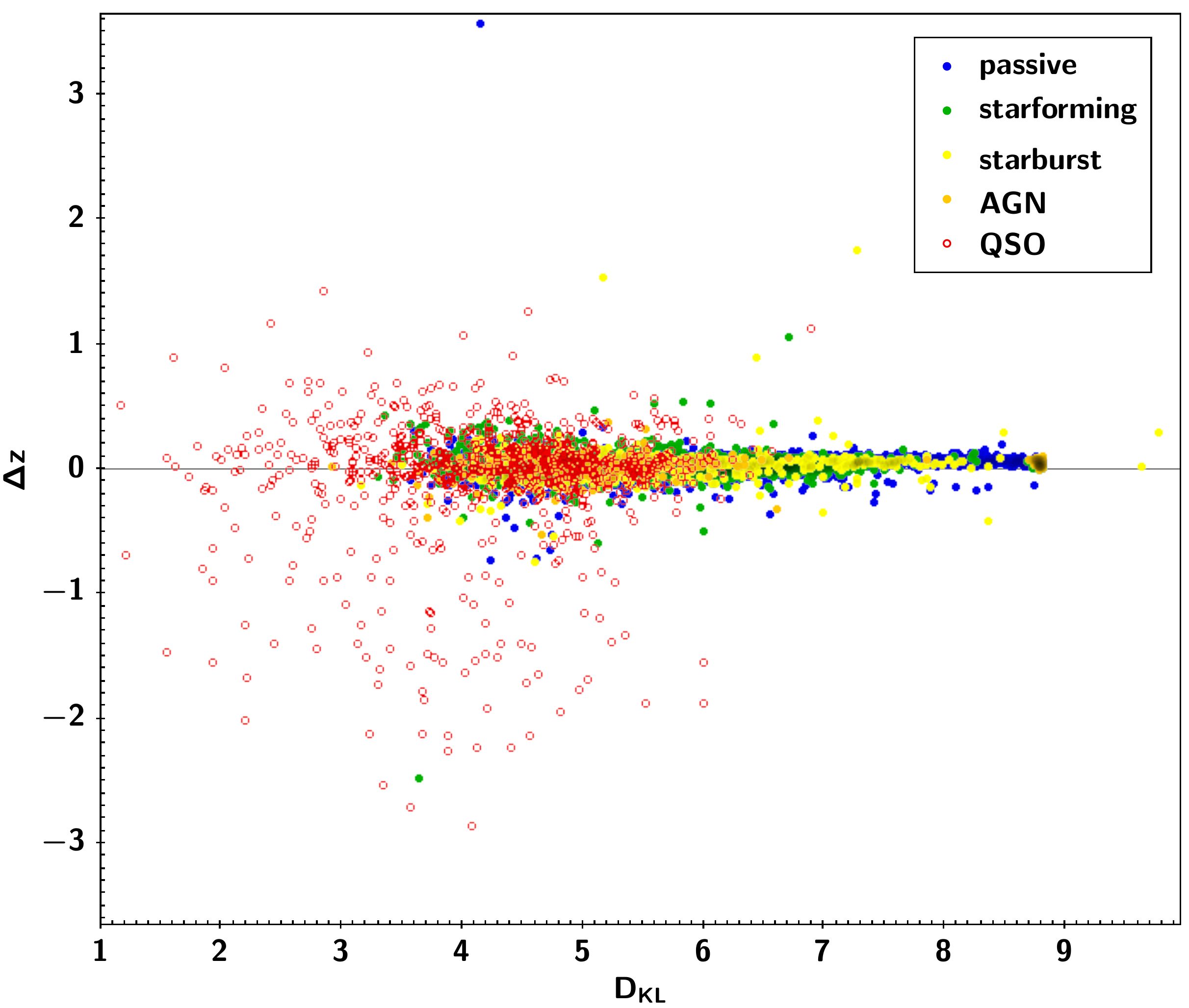}\\
\end{tabular}
 \caption{Left: Y-band magnitude vs the Kullback-Leibler divergence, $D_{KL}$. The PDF of brighter sources carries more information content, i.e. narrower PDF, compared to faint sources. Right: $\Delta z = z_{phot} - z_{spec}$ as a function of information content split per galaxy class. QSO PDFs have the distinctly lower information content compare to normal galaxies and AGN.}
\label{fig:DKL}
\end{figure*}

Template fitting algorithms, in our case specifically LePhare, provide as an output the full probability distribution function (PDF) for the photometric redshift estimate. Figure \ref{fig:PDF} shows a selection of PDFs ranging from very narrow to multimodal to very broad PDFs normalized to unity at the peak of the distribution for demonstration purposes. The solid lines show the photometric redshift assignment, while the dashed lines show the spectroscopic redshift value. The first panel of the first row shows an example of a very good photo-z estimation, where the PDF is narrow and centered on the true redshift value. The second panel of the first row shows an equally narrow solution, but centered on the wrong value. The second and third rows show PDFs that are broad and/or multimodal but still include the true redshift solution.

It is possible to rank the PDFs according to the information gained compared to a flat distribution by means of the Kullback-Leibler divergence \citep{KL1951}:
\begin{equation}
{\rm{D_{KL}(P||Q)}} = \int_{-\infty}^{+\infty}p(x)\log_2\frac{p(x)}{q(x)}dx,
\label{eq:DKL}
\end{equation}
where P and Q are two continuous random variables and p, q their corresponding probability density functions. The $\rm{D_{KL}}$ divergence, a useful diagnostic of information theory, has been used in the literature for quantifying the information gain for example, when performing bayesian analysis of X-ray spectra \citep{Buchner2014} and when modeling the $\rm{5-10\;keV}$ AGN luminosity function \citep{Fotopoulou2016a}. Further information can be found in \citet{Bishop2006}. 

Here we consider as P the photometric redshift PDF and as Q the extremely agnostic case in which we only know that a source must be between redshift zero and six, assuming a flat distribution. The information gain is measured in bits since the logarithm with base 2 is used.\footnote{One bit corresponds to the reduction of the standard deviation of a Gaussian distribution by a factor of three.}
In Fig. \ref{fig:DKL} (a) we show the trend between the $\rm{D_{KL}}$ and the J band magnitude. As expected, brighter sources (J<20) tend to have higher $\rm{D_{KL}}$ values signifying that the PDF carries more information with respect to a flat distribution, thus signifying a narrow PDF. As we move to fainter objects, the $\rm{D_{KL}}$ is also reduced to lower values reflecting the difficulty of constraining the PDF of a faint object. The same trends hold for all photometric bands. However, we do note a small cloud of sources with $\rm{D_{KL}}<2$. These are sources with broad PDFs similar to the bottom row of Fig. \ref{fig:PDF}. On the other end, we find sources with the highest $\rm{D_{KL}}$ values, aggregated at $\rm{D_{KL}}\sim 8.7$. These correspond to sources with $z_{phot}=0.0-0.01$ with very narrow PDFs, a telling sign of failed photo-z, most of the time due to noisy photometry.

It is also instructive to examine the $\rm{D_{KL}}$ values according to the galaxy classes as determined with classifier B. Panel (b) of Fig. \ref{fig:DKL} shows the difference $\rm{\Delta z = z_{phot}-z_{spec}}$ as a function of the $\rm{D_{KL}}$. The color coding corresponds to the five classes considered here as given in the legend. Most notably, there is no clear-cut $\rm{D_{KL}}$ that can be used to identify good PDFs. However, we note that the QSO class has systematically lower  $\rm{D_{KL}}$ values compared to the normal galaxies and also that the AGNs show similar values as the normal galaxies. This is explained by the discriminating features present in galaxy and AGN SEDs (e.g., Balmer break) contrary to QSO SEDs which are mostly featureless power-laws hence leading to less constrained photometric redshift solutions.

\subsubsection{Final sample}\label{sec:final}

Gathering all information from the previous sections, we select as our final sample the configuration of consolidation B (absolute probability threshold of 40\% for normal galaxies and AGN, 20\% for QSO), imposing rejection of stars (P[star]<50\%) and outliers (P[outlier]>20\%) and rejecting sources that have either too narrow or too broad PDFs ($\rm{D_{KL}}<2$ or $\rm{D_{KL}}>8.5$, 282 sources, 3\% of the sample). The final validation sample consists of 10.655 objects with photometric redshift accuracy $\rm{\sigma_{NMAD}=0.039}$ with 2.3\% outliers.

\begin{figure*}
\centering
\begin{tabular}{cc}
\includegraphics[height=0.85\columnwidth]{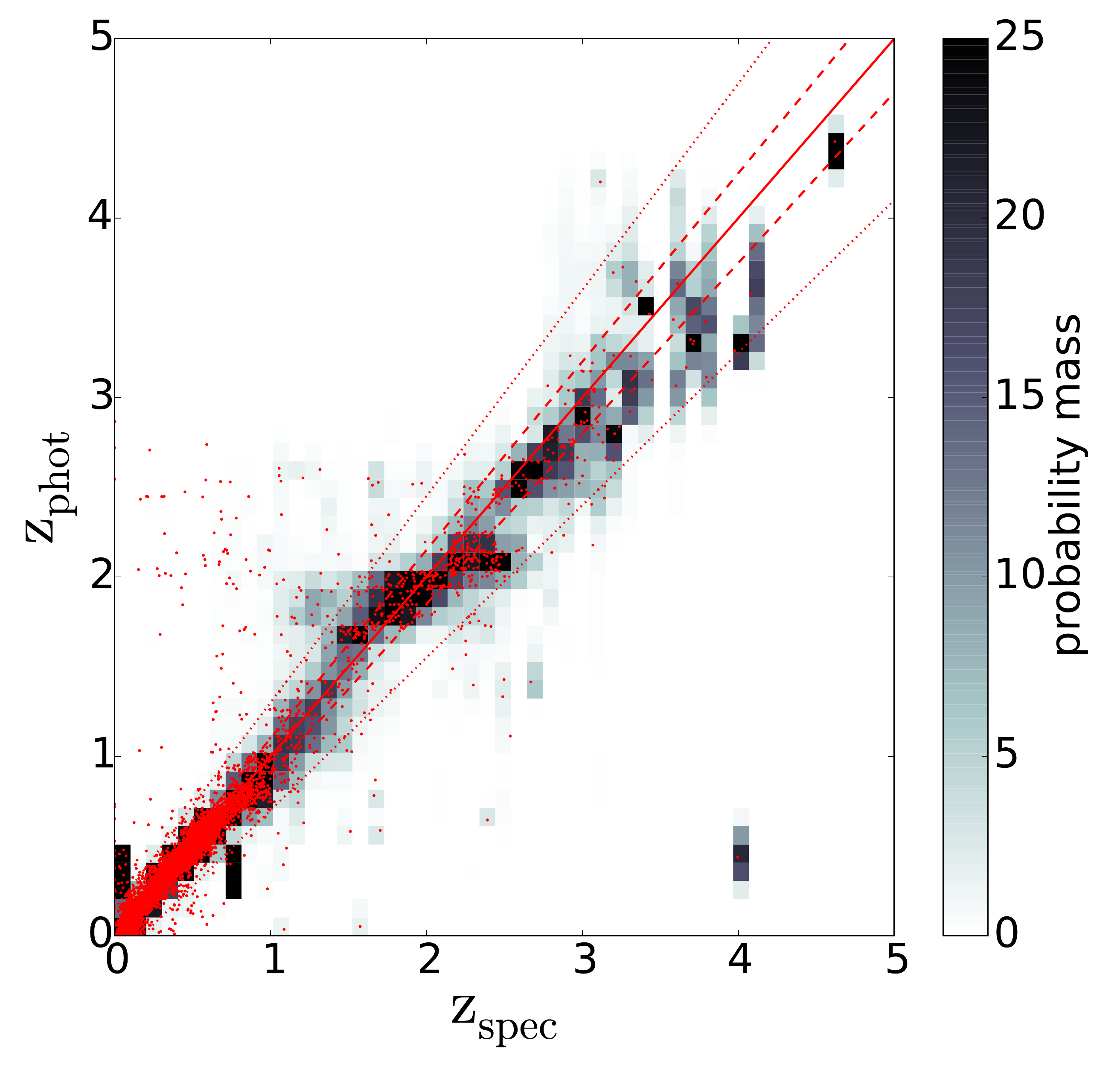}&
\includegraphics[width=\columnwidth]{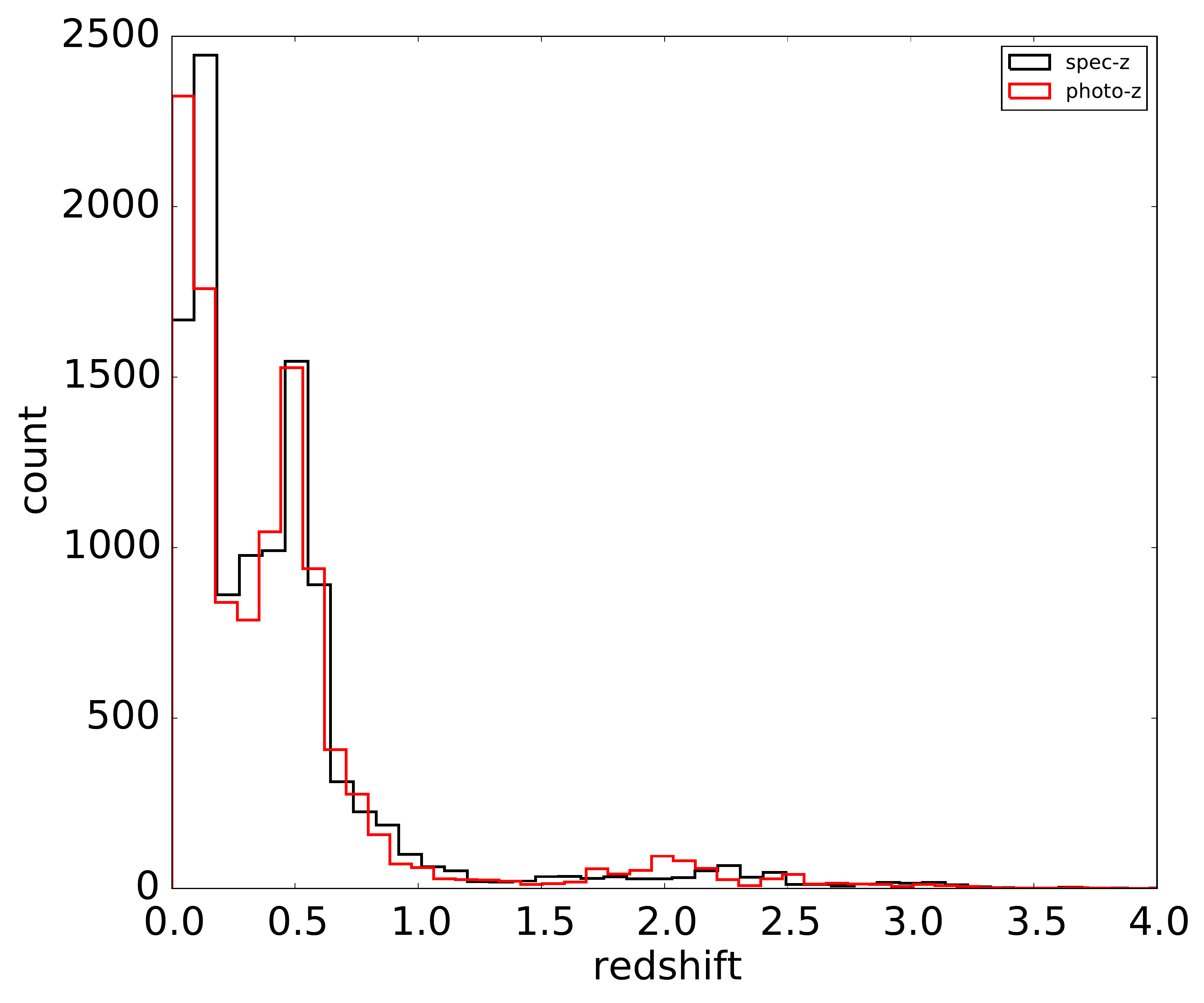}
\end{tabular}
 \caption{Left: histogram of stacked PDFs of the final sample. The solid line shows the diagonal, the dashed and dotted lines mark the $\rm{|z_{phot}-z_{spec}|/(1+z_{spec})}=0.05, 0.15$ respectively. The red dots show the point estimate of the photometric redshifts, here the mode of the PDF. Right: comparison of the photometric (red) and spectroscopic (black) redshift distributions.}
\label{fig:consolidation}
\end{figure*}

On the left hand side of Fig. \ref{fig:consolidation} we show the photometric redshift versus the spectroscopic redshift. The red dots show the mode of the PDF, adopted as point-estimate representation of the photometric redshift estimation. The solid red line is the diagonal while the dashed and dotted lines show the region of $\rm{|z_{phot}-z_{spec}|/(1+z_{spec})}=0.05, 0.15$ respectively. The gray scale shows the 2-D histogram of the stacked PDFs binned at resolution $\rm{\Delta z=0.1}$. The gray scale shows the amount of probability per cell. The black-colored cells show regions containing more than 25\% of the probability mass. The right hand side of the same Figure shows the redshift distribution of the photometric redshift point estimates compared to the spectroscopic redshift values.

\section{Discussion}

\subsection{Galaxy classes}\label{sec:galclass}

In order to identify the optimal photometric redshift libraries, we test four model library setups as shown in Table \ref{tab:SEDfit}. For this comparison we use the validation sample with photometric bands from the u-band to the W2 filter.

In Case 0 we include all galaxy and AGN models in a single library. This is done to test the discriminating power of $\chi^2$. Case 0 is the least optimal photometric redshift estimation, since the AGN and QSO are not treated in a special way and emission lines are not present in the templates ($\sigma=0.06$, $\eta$=9.2\%). However, it has been used previously in the literature mostly by including one or two QSO templates within a galaxy library.

Case I resembles the set-up that was adopted by the COSMOS team as described in \citet{Ilbert2009,Salvato2009,Salvato2011}. This set-up was originally created to optimize the photometric redshift of galaxies and X-ray detected AGN. Briefly, X-ray detected sources are confronted with QSO and AGN hybrid models to account for the combined emission from the QSO and the host galaxy. The templates are empirical and hence they include observed emission lines. The EXTNV and QSOV model libraries consist of the same templates (No. 32-61 of Table \ref{tab:SEDlib}) with the application of different $B$-band absolute magnitude priors: $\rm{-24<M_B<-8}$ for strong AGNs (extended and not varying sources with $\rm{F_x>8\times10^{-15}erg\cdot s^{-1}\cdot cm^{-2}}$) and $\rm{-30<M_B<-20}$ for QSO dominated objects (optical point-like or varying sources). These AGN templates span a range in AGN -- host galaxy combinations ranging from non-active (e.g., CB1, S0) to pure QSO templates (e.g., pl\_QSO, pl\_QSOH, pl\_TQSO1). \citet{Salvato2009} selected these templates as the best representation of the active galaxy population in the COSMOS survey.
The remaining sources that are either not X-ray detected or are X-ray detected but appear extended in the optical, are not varying and have $\rm{F_x<8\times10^{-15}erg\cdot s^{-1}\cdot cm^{-2}}$ are fitted using normal galaxy models and luminosity prior of $\rm{-24<M_B<-8}$ (No. 1 - 31 of Table \ref{tab:SEDlib}).

This method has been successfully applied to other fields such as the Lockman Hole \citet{Fotopoulou2012}, the Chandra Deep Field South \citep{Hsu2014} and AEGIS-X \citep{Nandra2015} and it is the current state of the art, used when X-ray data are available for the whole field in consideration. However, the correct implementation of the method requires splitting the sample into point-like and varying sources and also having an estimate of X-ray flux. This information is not available homogeneously across our sample, therefore we cannot test the result of a COSMOS-like approach directly for this sample. Instead, we used the optimal library set-up identified for the COSMOS field in \citet{Salvato2011} and use a machine-learning classifier to identify which objects are best fitted for each of the three classes. We find that this library setup shows very good performance both in terms of accuracy ($\sigma=0.035$) and catastrophic outliers ($\eta=2.8\%$), achieving an improvement over Case 0 by factor of two on the accuracy and by a factor of three on the fraction of outliers.

Case II is an extension of the previous set up with two main differences. We divide the normal galaxy library into passive (No. 1-8 models in Table \ref{tab:SEDlib}) and starforming systems (No. 9-31 models) and we use the hybrid AGN models (No 36-48 and 50-55, AGN hybrids library) separately from the pure QSO models (No 49, 56-61, pure QSO library) for a total of four model libraries. We have applied the corresponding $B$-band luminosity prior of  $\rm{-24<M_B<-8}$ for normal galaxies and AGN and $\rm{-30<M_B<-20}$ for QSO. 
We find that this setup represents a significant improvement over Case 0 and slightly under-performs compared to Case I ($\sigma_{II}=0.04$, $\eta=2.9\%$).

Finally, Case III is similar to Case II with the extra separation between starforming (No. 9-19 models) and extreme-starforming galaxies (which we refer to as starburst for short, No 20-31 models). We see that Case III shows better performance compared to all previous cases in terms of catastrophic outliers ($\eta=2.3\%$) and achieves slightly worse performance compared to Case I in terms of accuracy ($\sigma=0.039$). We can examine further the accuracy per galaxy class, also given in Table \ref{tab:SEDfit}, keeping in mind that each class contains a different collection of models and a different galaxy population chosen by the machine-learning classification. The performances resemble the expectation for each population, namely passive and spiral galaxies have SEDs reach in discriminating features that can be identified through model fitting showing $\sigma\sim0.04$ and 0.5-1.0\% fraction of catastrophic outliers. The degeneracies present in the starburst and AGN colors creates a higher fraction of catastrophic outliers (2-3\%), however the accuracy remains very good and close to the passive and spiral galaxies ($\sigma\sim0.04-0.05$). Lastly, as expected the QSO population shows the least optimal photometric redshift performance ($\sigma$=0.07 and $\eta~20\%$), mostly due to their featureless SEDs which is particularly problematic when using only broadband photometry.

It is evident that the simple inclusion of AGN templates in a galaxy library leads to the worst performance (Case 0). We recover the good behavior of the setup of \citet{Salvato2009,Salvato2011}, where the split between galaxies and AGN leads to an improvement higher than a factor of two both on accuracy and outlier rate compared to Case 0. More interestingly, the further separation of the galaxy library and the consideration of pure AGN and QSO libraries is beneficial when we opt for pure classes of objects, for example, separating the starforming and starburst galaxies in two distinct classes and AGN from QSO, reducing the outlier rate and while maintaining comparable accuracy.

\begin{table*}
\begin{center}
\caption{Template fitting class set-up and respective performances. The numbers quoted in this table include star and outlier rejection.}
\label{tab:SEDfit}
\begin{tabular}{clcccc|rcccc} \hline \hline
\multirow{2}{*}{case} 	 & \multirow{2}{*}{model library}				 & model  & extinction  & $\rm{M_{abs}}$  & emission &$N_{sample}$&$\sigma_{class}$ &  $\eta_{class}$&$\sigma_{all}$ &  $\eta_{all}$\\
		 &  				 & number 		 & law 		 & prior  & lines & & & (\%)& & (\%)\\\hline
\multirow{2}{*}{0}		 & \multirow{2}{*}{all models} 			 & \multirow{2}{*}{1-61} 			 & Calzetti, 	 & \multirow{2}{*}{-8,-24}  & \multirow{2}{*}{no} & \multirow{2}{*}{13\,856}&\multirow{2}{*}{--} & \multirow{2}{*}{--} &\multirow{2}{*}{0.060} & \multirow{2}{*}{9.2}\\
		            &	   					 & 				 &  Prevot 	 & 		   & &  &\\ \hline
\multirow{3}{*}{I} 	& galaxy 				 & 1-31			 &  Calzetti & -8,-24  & yes & 9~703 & 0.038 & 2.7\\
		            & EXTNV 				 & 32-61		 &  Prevot 	 & -8,-24  & empirical & 575	&  0.044 & 4.2 & 0.036 & 2.8\\
		            & QSOV 					 & 32-61		 &  Prevot 	 & -20,-30 & empirical & 2~882 & 0.029 & 3.2\\ \hline
\multirow{4}{*}{II}	& passive 				 & 1-8			 &  -- 		 & -8,-24  & yes & 4~143 & 0.041 & 0.8\\
		            & starforming 			 & 9-31 		 &  Calzetti & -8,-24  & yes & 7~095 & 0.037 & 1.4 & \multirow{2}{*}{0.041} & \multirow{2}{*}{2.9}\\
		            & AGN hybrids 			& 36-48, 50-55	 &  Prevot 	 & -8,-24  & empirical& 682 & 0.058 & 3.52\\
		            & pure QSO 				& 49,56-61		 &  Prevot 	 & -20,-30 & empirical& 896 & 0.085 & 24.7\\\hline
\multirow{5}{*}{III} & passive 				& 1-8			 &  -- 		 & -8,-24  & yes& 4~897 & 0.040 & 0.7\\
		             & starforming 			& 9-19 		     &  Calzetti & -8,-24  & yes & 4~102 & 0.032 & 1.0\\
		             & extreme starforming	& 20-31 		 &  Calzetti & -8,-24  & yes & 874 & 0.047 & 2.1 & 0.039 & 2.3\\
		             & AGN hybrids 			& 36-48, 50-55	 &  Prevot 	 & -8,-24  & empirical& 272 & 0.053 & 2.6 \\
		             & pure QSO 			& 49,56-61		 &  Prevot 	 & -20,-30 & empirical& 792 & 0.077 & 19.3\\\hline\hline
\end{tabular}
\end{center}
\end{table*}

\subsection{Feature Importance}\label{sec:importance}
One of the attractive features of Random Forest is the relative ranking of the discriminating power of the input attributes. In Table \ref{tab:features} we list the top 10 most important features identified by the Random Forest for each of our three classifiers. The subscript $3$ refers to magnitudes estimated within $3''$ aperture diameter. For Classifier A, the star-galaxy separator, the infrared colors carry the dominant discriminating power, especially the WISE bands. The top three most important features are the colors $\rm{J_3-W1}$, $\rm{Y_3-W1}$, $\rm{J_3-W2}$. The colors identified by the Random Forest correspond to the color-color plots presented in this paper and also found in similar forms with the literature. 

The clear separation between the star and galaxy population as seen in Fig. \ref{fig:test_star}, at least for the population with $\rm{g-J>2}$ leaves little ambiguity. However, the population with $\rm{g-J<2}$ is a locus occupied both by galaxies and stars. In this area a machine learning classifier based on a multiwavelength identification has a clear advantage over color selection methods, since the introduction of morphological attributes and additional colors can lift this degeneracy. Furthermore, Random Forest in particular provides the probability for a source to be a star which allows for the tuning of the completeness and purity of the final sample as demonstrated in \S \ref{sec:consolidation}.

Similarly, for Classifier B, the galaxy-class separator, the top three features are the near-infrared $\rm{H}$ and $\rm{K}$ colors with the WISE $\rm{W2}$ bands. However, they are immediately followed by the $\rm{W1-W2}$ color and a combination of optical and near-infrared colors including the $\rm{g}$ and $\rm{i}$ bands. The WISE bands have been already proposed in the literature as selection method for QSO sources \citep{Stern2012}. Lastly, the outliers (Classifier C) are identified primarily through their optical colors ($\rm{r}$, $\rm{z}$, $\rm{i}$ bands) selecting preferentially QSOs and stars (see Fig. \ref{fig:test_outlier}).

Intuitively we could expect that the size of the object would be a powerful discriminatory attribute. The ranking of random forest places the half light radius (depending on run and photometry used) in the following positions: Classifier A [stars]: 40-60, classifier B [galaxies]: 100-200, classifier C [outliers]: 35-50. The relative ranking of shape among star, galaxy, outlier classifiers follows the idea that the shape is more important discriminatory feature for stars and outliers (comprised mostly out of QSOs and stars) and less important for galaxies. However, in all cases the presence of near and mid infrared photometry carries significantly more information to distinguish between the classes. In the absence of near- and mid-infrared photometry the half-light radius enters the top 5 of important attributes.

\begin{table}
\begin{center}
\caption{Feature importance for the three classifiers.}
\label{tab:features}
\begin{tabular}{r@{-}lr@{-}lr@{-}l} \hline \hline
\multicolumn{2}{c}{A}	  &  	\multicolumn{2}{c}{B}	&   \multicolumn{2}{c}{C} 		 \\\hline
J$_3$&W1 & H&W2         &   r&z$_3$ \\
Y$_3$&W1 & K&W2         &   r&i\\
J$_3$&W2 & W1&W2        &   r$_3$&i$_3$\\
H$_3$&W2 & g&J          &   K$_3$&W2\\
Y$_3$&W2 & i&W2         &   r&z\\
z$_3$&W2 & g&K          &   r&Y$_3$\\
K&J$_3$  & g&H          &   H&J$_3$ \\
H$_3$&W1 & i&W1         &   H&W2\\
z$_3$&W1 & r&H          &   i&u$_3$\\
K&H$_3$  & g$_3$&i$_3$  &   K&J$_3$\\\hline\hline
\end{tabular}
\end{center}
\end{table}

\subsection{Dependence on photometric depth}
\begin{figure*}
    \centering
    \begin{tabular}{cc}
    \includegraphics[width=0.5\linewidth]{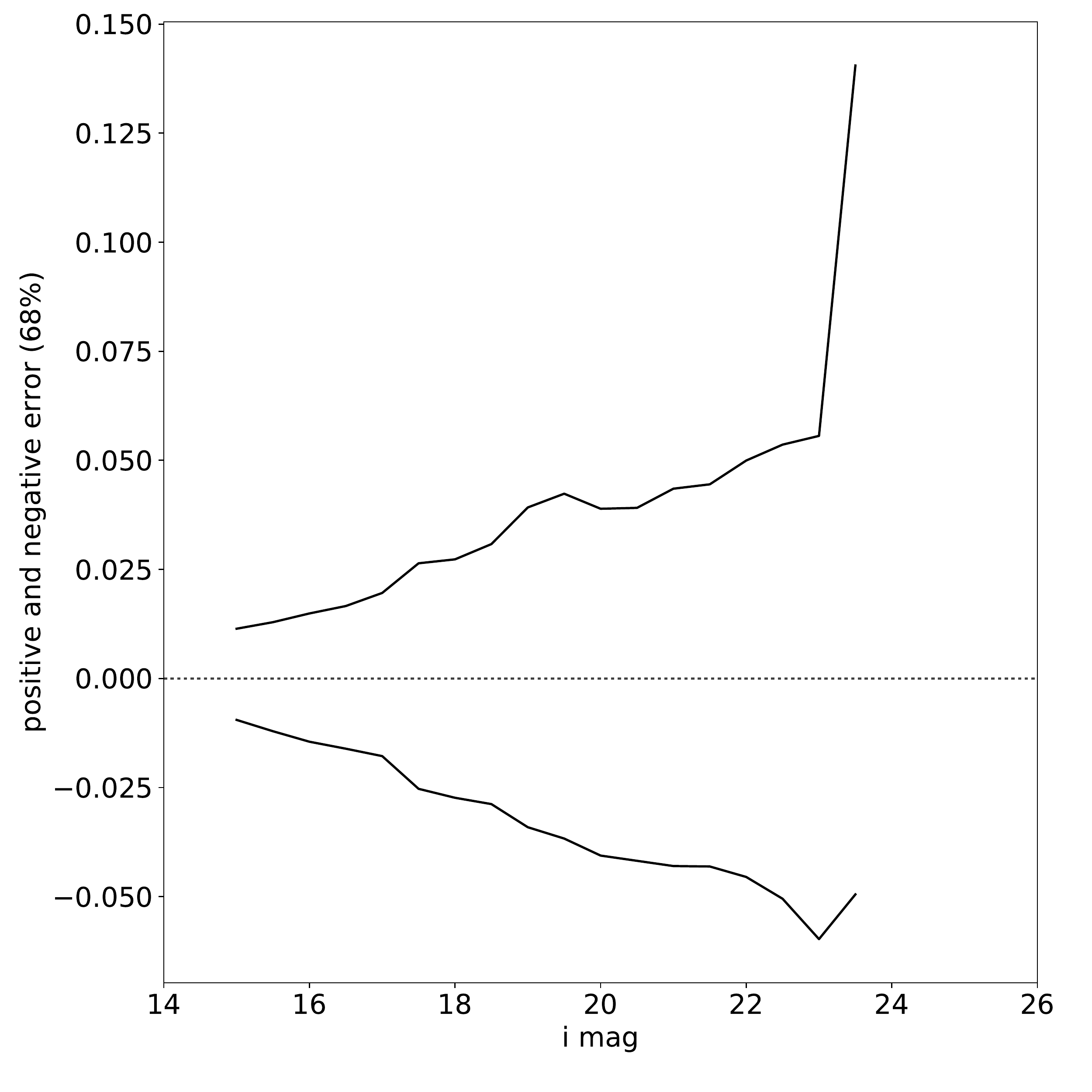}&\includegraphics[width=0.5\linewidth]{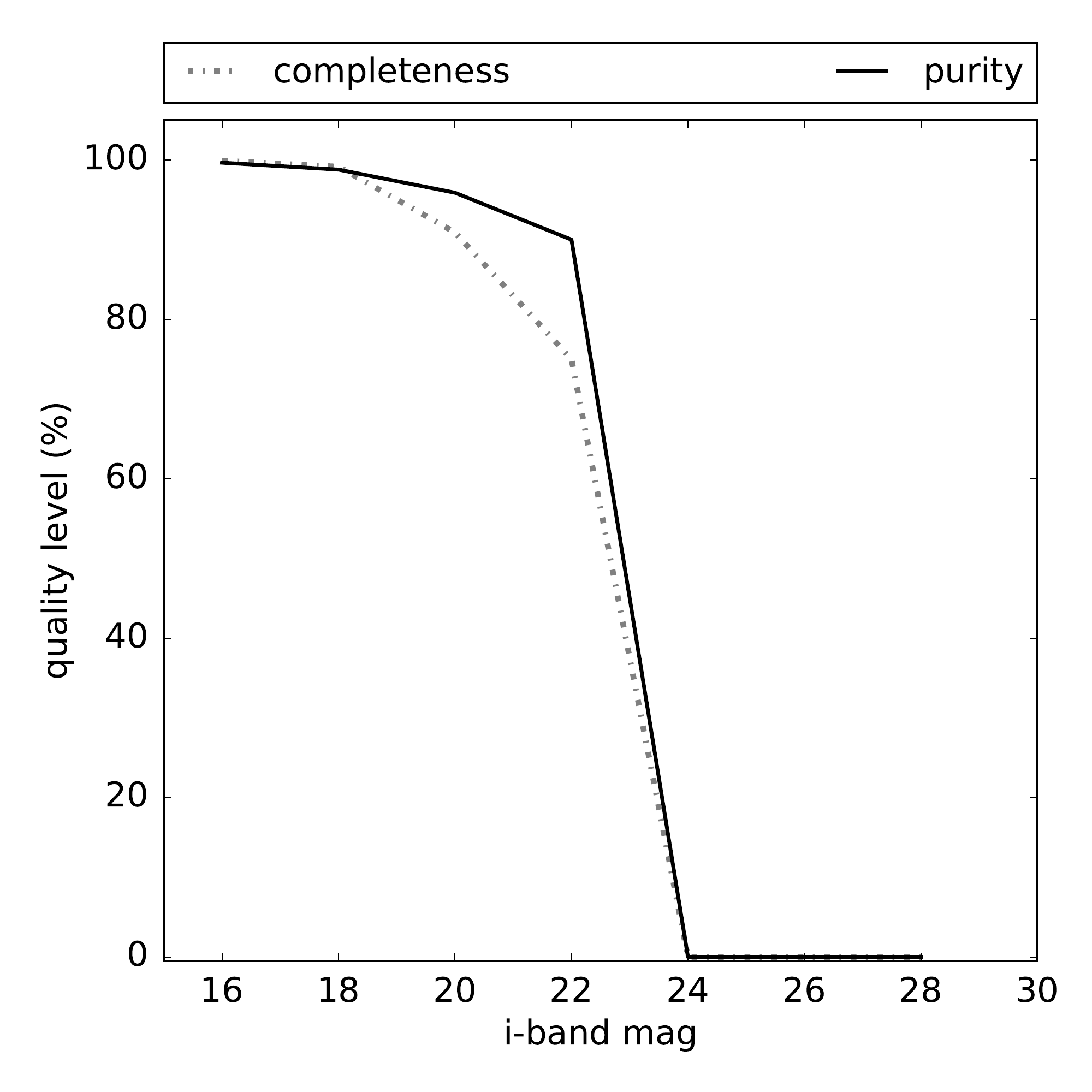}
    \end{tabular}
    \caption{Dependence of photometric redshift uncertainty (left)  and star classification performance (right) as a function of $i$-band magnitude. The left hand side figure shows the median 1-$\sigma$ interval per 0.5 magnitude bin. The right hand side shows the purity and completeness of the star classifier per magnitude bin. The purity of the classification remains high (>80\%) up to $i_{AB}=23$ and decreases sharply closer to the detection limit (see also Fig. \ref{fig:reddist} (b) ). Similar behavior is found for all classification categories.}
    \label{fig:starperformance}
\end{figure*}

Previous photometric redshift studies have demonstrated the impact of magnitude on photometric redshift accuracy \citep[for example Fig. 12 and Fig.10 in][respectively]{Ilbert2009,Fotopoulou2012}. As expected, fainter objects tend to display photometric redshift of lesser quality due to the larger uncertainties associated to photometric measurements.
    \newline
    The CPz method shows a similar trend in performance with magnitude. Fainter objects have larger photometric uncertainties associated to them, hence resulting in broader PDFs. Figure \ref{fig:starperformance} shows the median 1-$\sigma$ interval per $i$-band magnitude bin (left-hand side, magnitude step 0.5). At the same time, fainter objects will also have less accurate classification due to the inherent noise of the magnitude measurement itself. The right hand side of Figure \ref{fig:starperformance} shows the completeness and purity of the star classification (classifier A) as a function of the $i$-band magnitude (right hand side). The classifier shows good quality (>80\%) even up to $i_{AB}=23$, while it quickly declines approaching the magnitude limit.
   
\subsection{Dependence on photometric filters}\label{sec:depfilt}

Existing and future surveys will provide a view of the sky from the ultra-violet (GALEX), to the optical (SDSS, LSST), near-infrared (Euclid) and mid-infrared (WISE). With the exception of the VISTA Hemisphere Survey (VHS, PI McMahon) covering the southern hemisphere at $\rm{K_{AB}=20}$ magnitude, there are no foreseen plans for a deep all-sky K-band survey. The same holds for the ultra-violet and mid-infrared where the GALEX and WISE observations remain the state of the art respectively. We discuss here the impact of the inclusion of these datasets in the training of the classifier and the SED fitting.

The impact of the exclusion of a photometric band is twofold on our method, affecting both the classification and the SED fitting.
As seen on Table \ref{tab:features} the K-band combined with the WISE filter and other near-infrared bands offer strong discriminating information for galaxy classification (Classifier B) and outlier detection (Classifier C). As discussed in the previous section, the ability of CPz to pre-classify an object in order to use a limited number of models for the determination of photometric redshift through SED fitting leads to better accuracy. In the absence of K-band photometry the highest-ranked features by the Random Forest contain again infrared colors where the K-band is substituted by near- (J, H) or mid-infrared bands (W1, W2). However, the classification score remains the same (~60\%). Table \ref{tab:A} shows the performance of classifier A as a function of the photometry used. Similar conclusions hold for all other classifiers. We find that the inclusion of WISE W1 and W2 bands improves both the accuracy of all classifiers (A, B, and C) and the performance of photometric redshift.

On the other hand, due to the flux limit of the surveys, the inclusion of WISE photometry reduces our initial spectroscopic sample detected in the $u$-$K$ bands from 78\,776 sources to 49\,220 sources (62.5\%), however without sacrificing the high redshift ($z>1$) population. However, the inclusion of GALEX photometry reduces the sample to just 19\%, or 15\,064 sources and is limited to low redshift sources ($z<1$) due to the very bright flux limit of the GALEX wide area survey ($\rm{FUV\sim21 AB}$). Finally, the creation of SEDs from the FUV to mid-IR, including both GALEX and WISE observations, would consist of only 10\,003 sources (13\%).

The absence of K-band has a more prominent effect on the photometric redshift determination since the gap in the wavelength coverage leads to more catastrophic outliers. We find that when considering only photometry from the u-band to the H-band the inclusion of a K-band has the most significant impact on starburst, AGN and QSO improving the accuracy up to 2\% and the catastrophic outlier fraction by 2-10\%. However, if an extension to the mid-infrared is available by including WISE data we find that the accuracy reached is improved by 1-2\% while the fraction of catastrophic outliers drops by 3-10\%. While all galaxy populations benefit from the inclusion of infrared observations, the AGNs show the most prominent improvement by the inclusion of the K-band and WISE data.

Limiting the photometric coverage to optical bands, both classification and photometric redshift suffer a decrease in quality. For example, the star classification score decreases from 98.8\% for the u-K case to 97.5\% and 96.5\% for u-z and g-z respectively. In Table \ref{tab:A} we summarize the performance of each quality measure. Interestingly, in the optical-only case the half-light radius plays a more important role in star-galaxy classification entering the top 5 most important attributes (see also \S \ref{sec:importance}).

\begin{table}
\begin{center}
\caption{Classifier A, accuracy per run of test sample for star identification. The columns are the classification measures of quality defined in Section \ref{sec:quality}}
\label{tab:A}
\begin{tabular}{r@{-}l|cccccccc} \hline \hline
\multicolumn{2}{c}{Run} 	&  ACC    & P  & R  & F1  & F  \\\hline
g&z  &  0.976  &  0.949  &  0.923  &  0.936  &  0.012 \\
u&z  &  0.981  &  0.960  &  0.940  &  0.950  &  0.009 \\
\hline
g&H  &  0.980  &  0.959  &  0.931  &  0.945  &  0.009 \\
u&H  &  0.982  &  0.964  &  0.946  &  0.955  &  0.009 \\\hline
u&K  &  0.989  &  0.972  &  0.968  &  0.970  &  0.006 \\
u&K--IR  &  0.997  &  0.991  &  0.989  &  0.99  &  0.002 \\
UV--u&K  &  0.994  &  0.986  &  0.894  &  0.938  &  0.001 \\
UV--u&K--IR  &  0.997  &  0.972  &  0.929  &  0.950  &  0.001 \\
\hline\hline
\end{tabular}
\end{center}
\end{table}

\subsection{Comparison to the literature}

\paragraph{{\bf Classification}}

\begin{figure*}
\begin{tabular}{ccc}
\includegraphics[height=0.55\columnwidth]{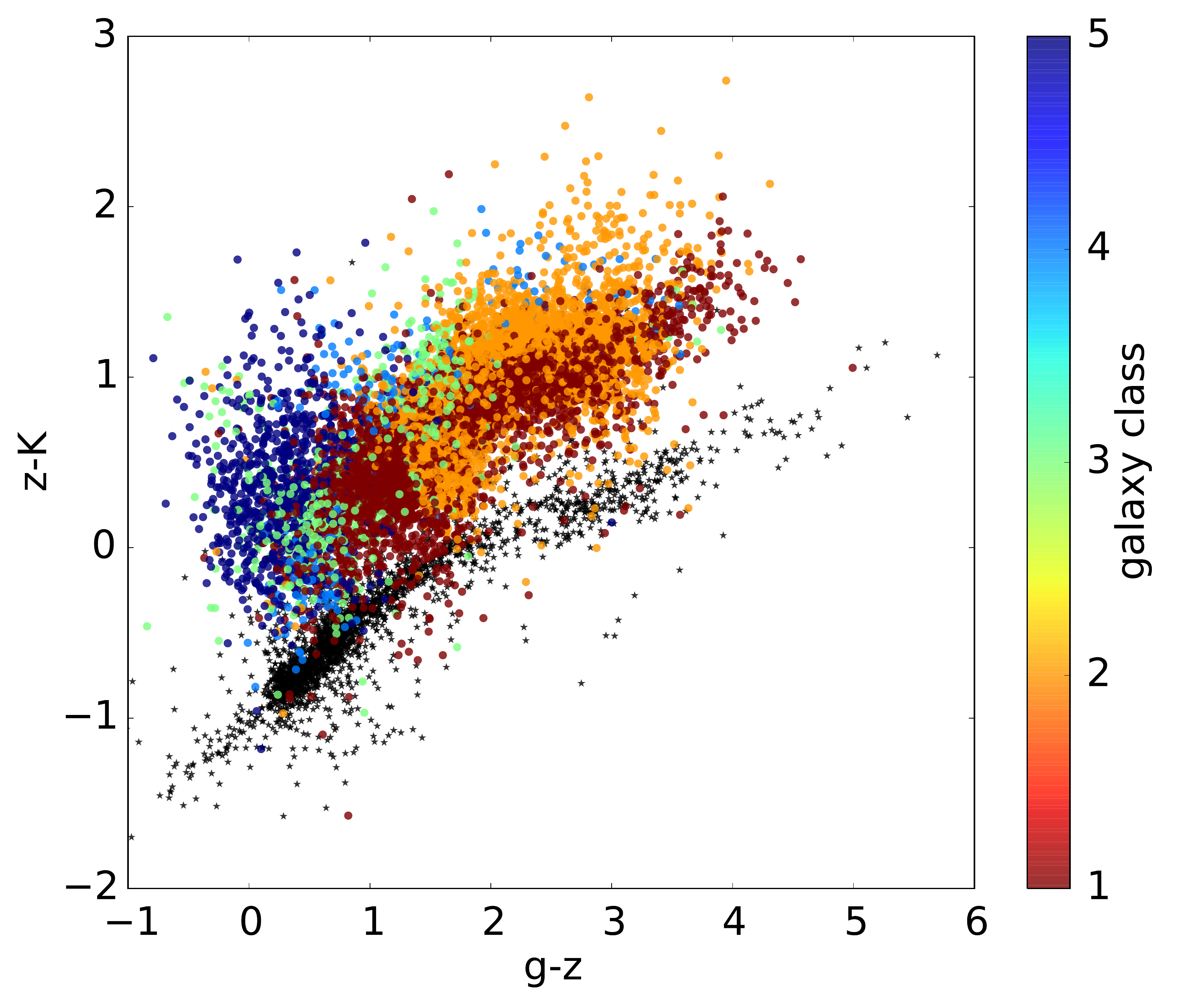} & \includegraphics[height=0.55\columnwidth]{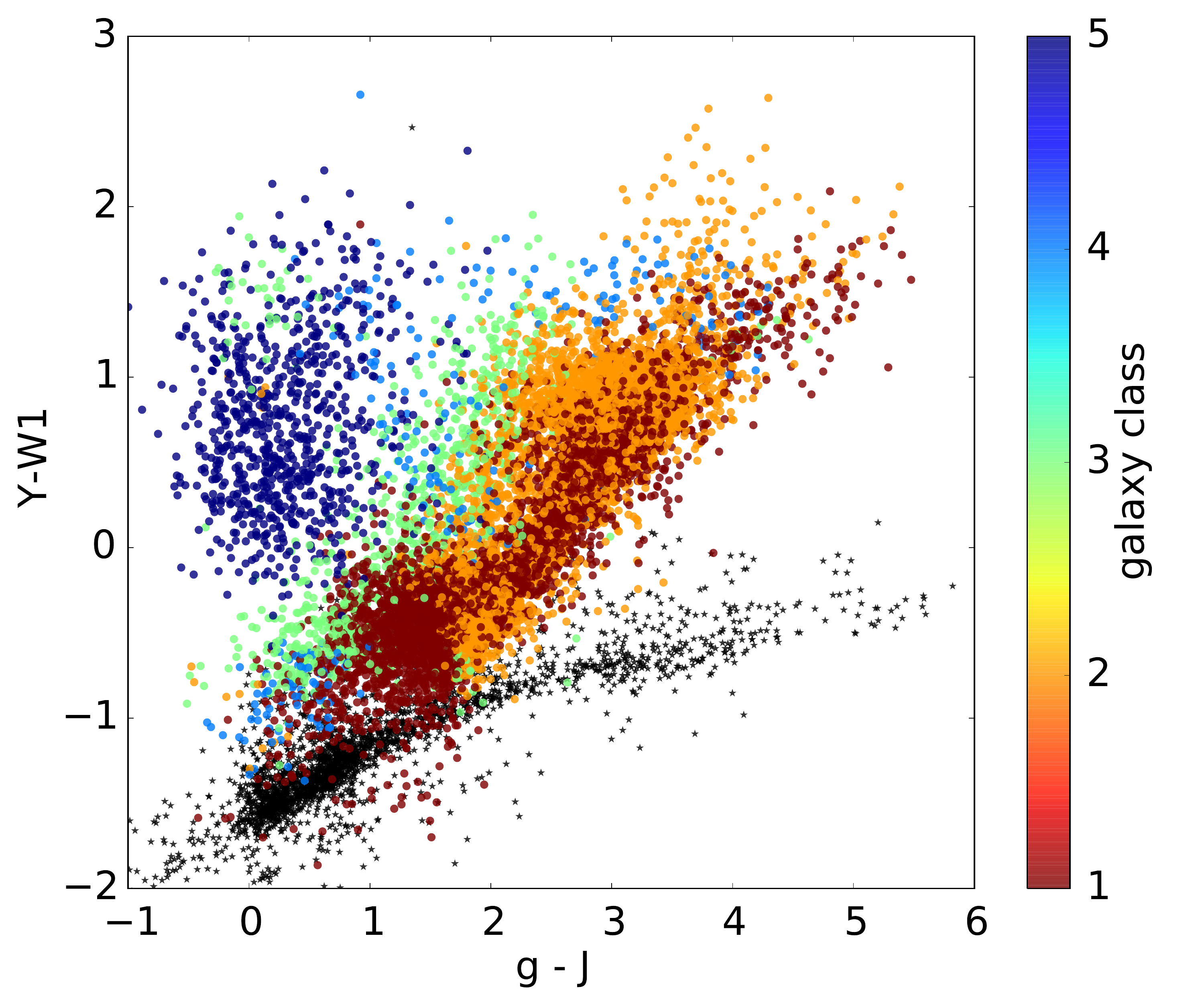} 
& \includegraphics[height=0.55\columnwidth]{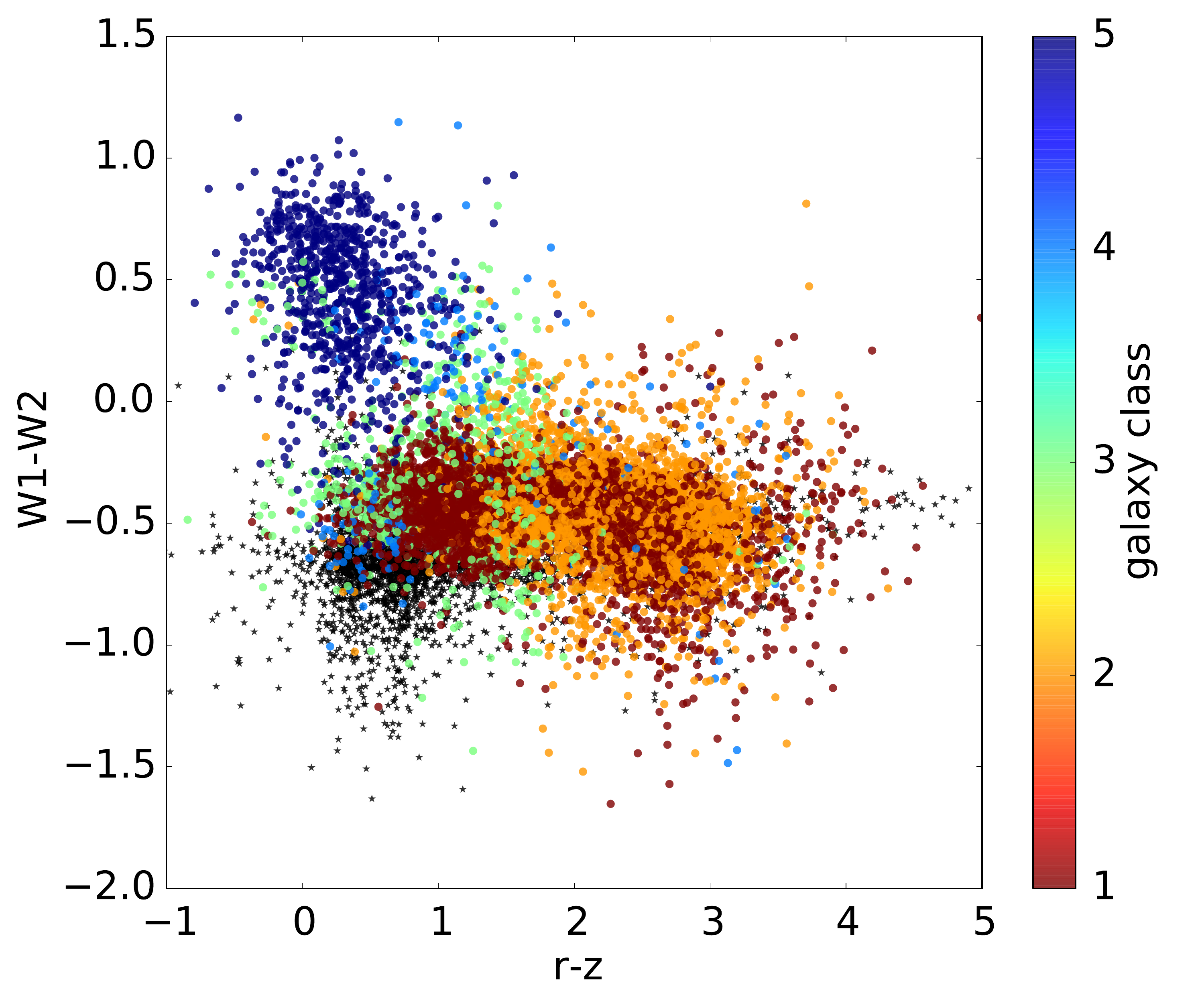} \\
(a)  & (b) & (c)\\
\end{tabular}
\caption{Comparison of the CPz classification (colorbar) to literature color-color plots.}
\label{fig:class_literature}
\end{figure*}

Several color plots have been used in the literature over the years to identify the nature of galaxies, mostly splitting between passive and starforming and AGN versus normal galaxies. In Fig. \ref{fig:class_literature} we compare our classification\footnote{Our galaxy classification is tuned to identify the optimal photometric redshift model library, which might differ from the morphological, or spectroscopic classification due to for example aperture effects. However, these effects will be less evident as we move to higher redshifts.} denoted with the colorbar with some commonly used color plots. In all plots, the smaller black dots denote stars identified with classifier A. Panel (a) is the equivalent of the BzK plot using our filter set using the colors z-K vs g-z. Despite the large overlap of the galaxy populations the general separation between passive and starforming galaxies is on average reproduced. However, in panel (b) we show a modified version which leads to a better distinction between the classes as identified by the Random Forest. By using longer wavelengths (Y-W1 vs g-J) both stars and QSOs are separated more distinctly which showcases the ability of the Random forest to identify automatically colors with high discriminating power.
The bottom panels of Fig. \ref{fig:class_literature} show two AGN classification plots. The left-hand side shows the W1-W2 plot of \citet{Stern2012}. The authors introduced a selection criterion of QSO using W1-W2>0.18 (in AB). Our classifier is in agreement with this criterion however with our probabilistic class assignment we can identify the nature of the sources in the transition area between the normal galaxies and the QSO and effectively separate stars from passive galaxies which are largely overlapping in this plot.

\paragraph{{\bf Photometric redshifts}}

We cross-matched within $0.7''$ our final consolidated sample of $\sim$10\,000 sources (\S \ref{sec:final}) with the template fitting method in the MLS-VIPERS survey of \citet{Moutard2016} which provide photometric redshifts in the CFHTLS-W1 field yielding 2\,399 sources in common. Only 10 sources in our sample are not present in the MLS catalog located at 0.5 < z$_{spec}$<1.28. Out of those CPz classified one source as QSO (class 5) and the remaining are passive (class 1) or spiral (class 2). We also cross-matched our results with the pure machine learning estimates of SDSS DR12 \citep{Alam2015} yielding 7\,954 sources. A total of 2\,983 sources present in our sample were missing from the SDSS photometric redshift catalog. About 41\% of the missing sources (1\,227 sources) have $g<21$, formally above the detection limit of SDSS. These sources span the redshift range of z=0-4 and 41\% are classified by CPz as normal galaxies (504 sources) while 59\% as AGN or QSO (723 sources). The remaining 1\,756 sources that are too faint to be detected by SDSS are mostly located at z<1 (93\%) and the majority is classified as normal galaxies (92\%). There are only 272 sources that are in common in all three datasets, mostly located at z<1 as dictated by the SDSS sources.

In Table \ref{tab:cases} we give a comparison of the performance for each method and sample. We see that the CPz method performs better in terms of outlier fraction than the pure template fitting method of the MLS estimation showing comparable accuracy $\sigma$=0.03 but less catastrophic outliers ($\eta_{CPz}$=2.0\% compared to $\eta_{MLS}$=3.4\%). The same behavior is consistent across all galaxy classes as identified by CPz. MLS shows comparable or slightly better accuracy, but higher fraction of outliers. The class that stands out the most is the QSO which perform significantly better with the CPz method ($\sigma_{CPz}$=0.126, $\eta_{CPz}$=34.6\% compared to $\sigma_{MLS}$=0.517, $\eta_{MLS}$=71.6\%). Comparing the SDSS photometric redshifts, which is a pure machine-learning method we see that for galaxy classes well represented at $z<1$, SDSS outperforms CPz. However, this is not true for the QSO class, where CPz shows better accuracy approximately by a factor of two. Lastly, we also present for completeness the comparison of the 272 objects that in common among the three classes. Even though the limited number of objects is prohibitive to make any strong claims, we see that all methods perform well since the majority of the objects are low redshift ($z<1$) normal galaxies. Once again, CPz shows distinct advantage in treating properly AGN and QSO within a unified framework of photometric redshift estimation.

\begin{figure*}
\begin{tabular}{cc}
\includegraphics[width=0.95\columnwidth]{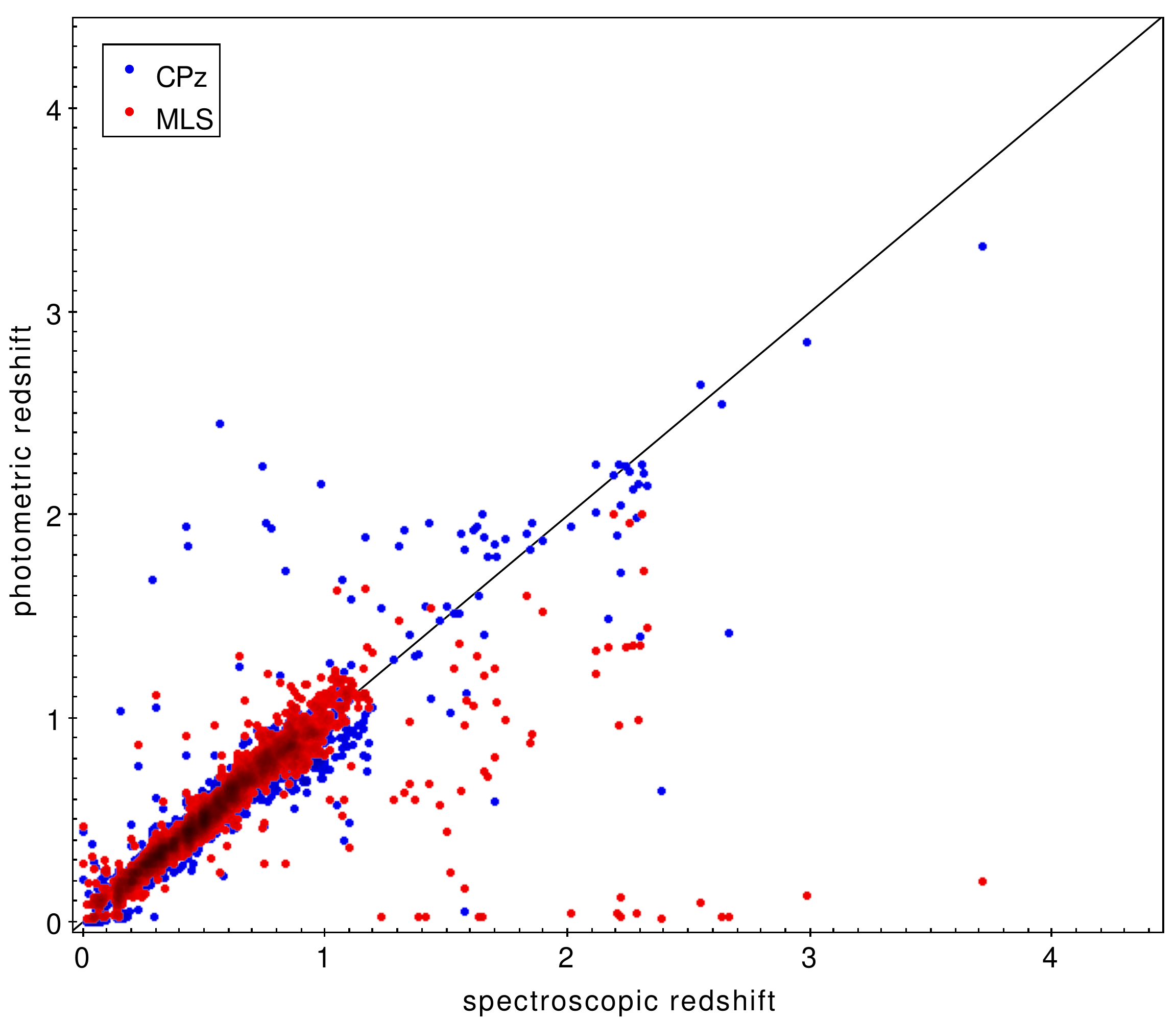} 
& 
\includegraphics[width=0.95\columnwidth]{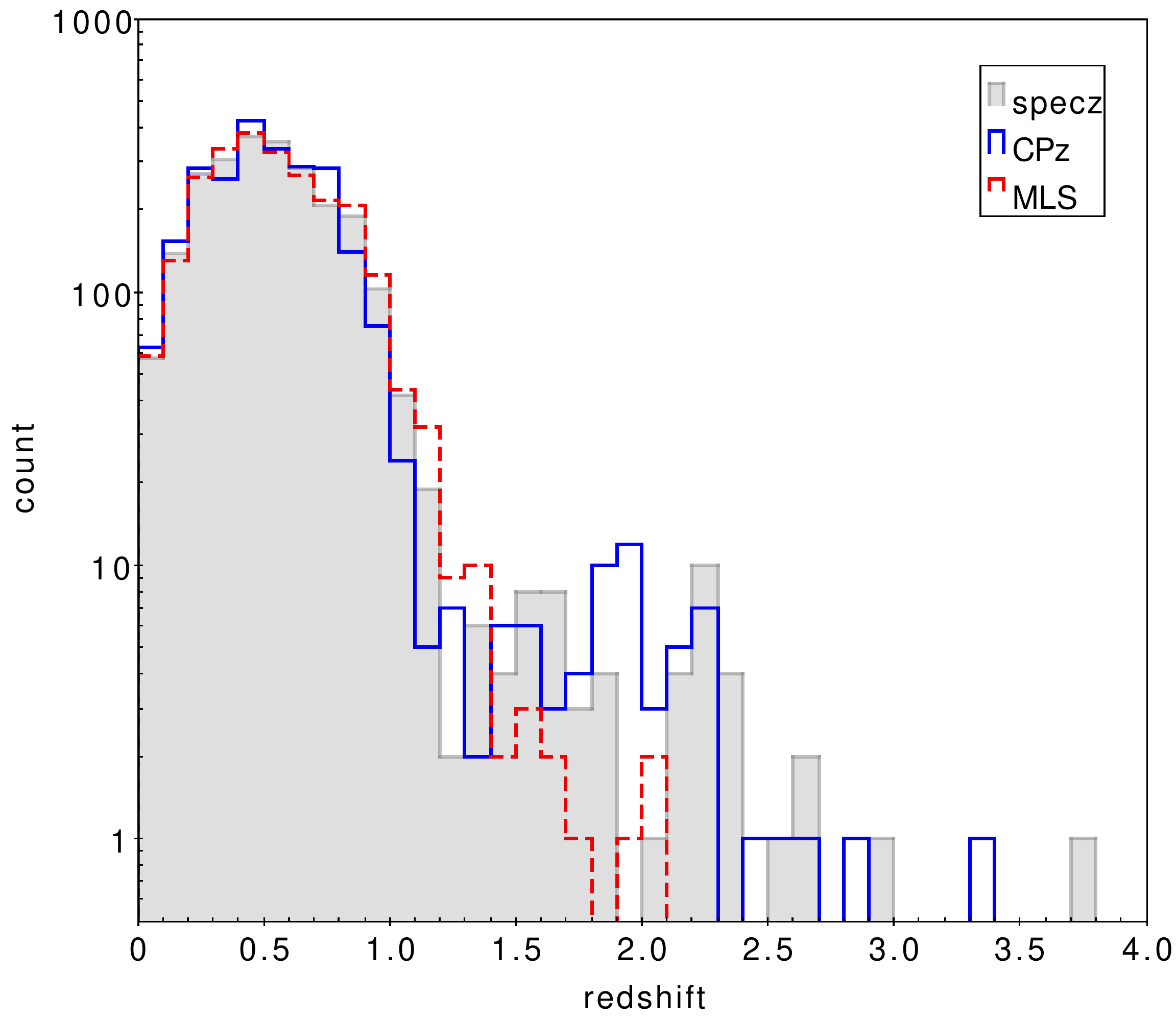}\\
(a)  & (b) \\
\includegraphics[width=0.95\columnwidth]{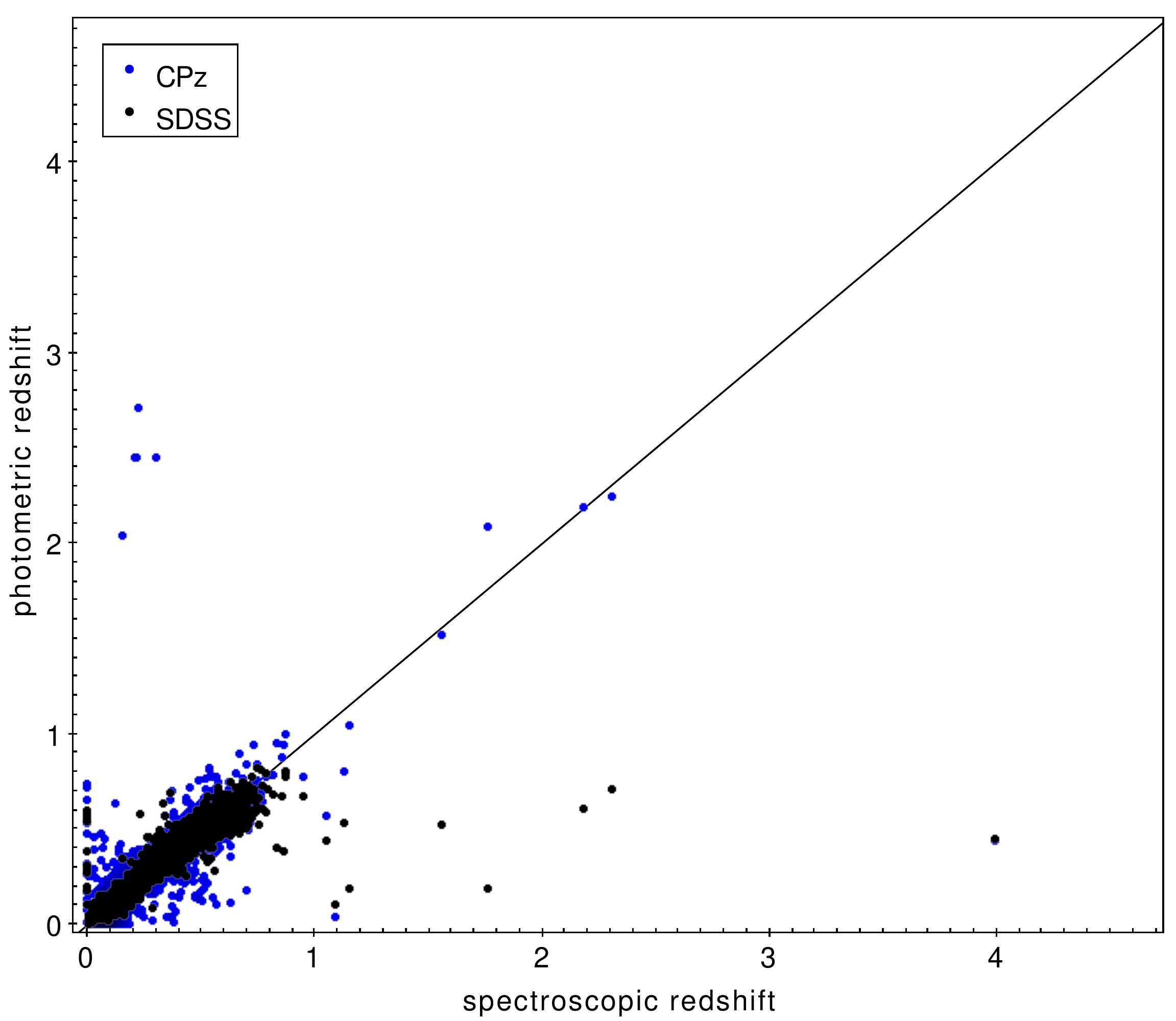} & 
\includegraphics[width=0.95\columnwidth]{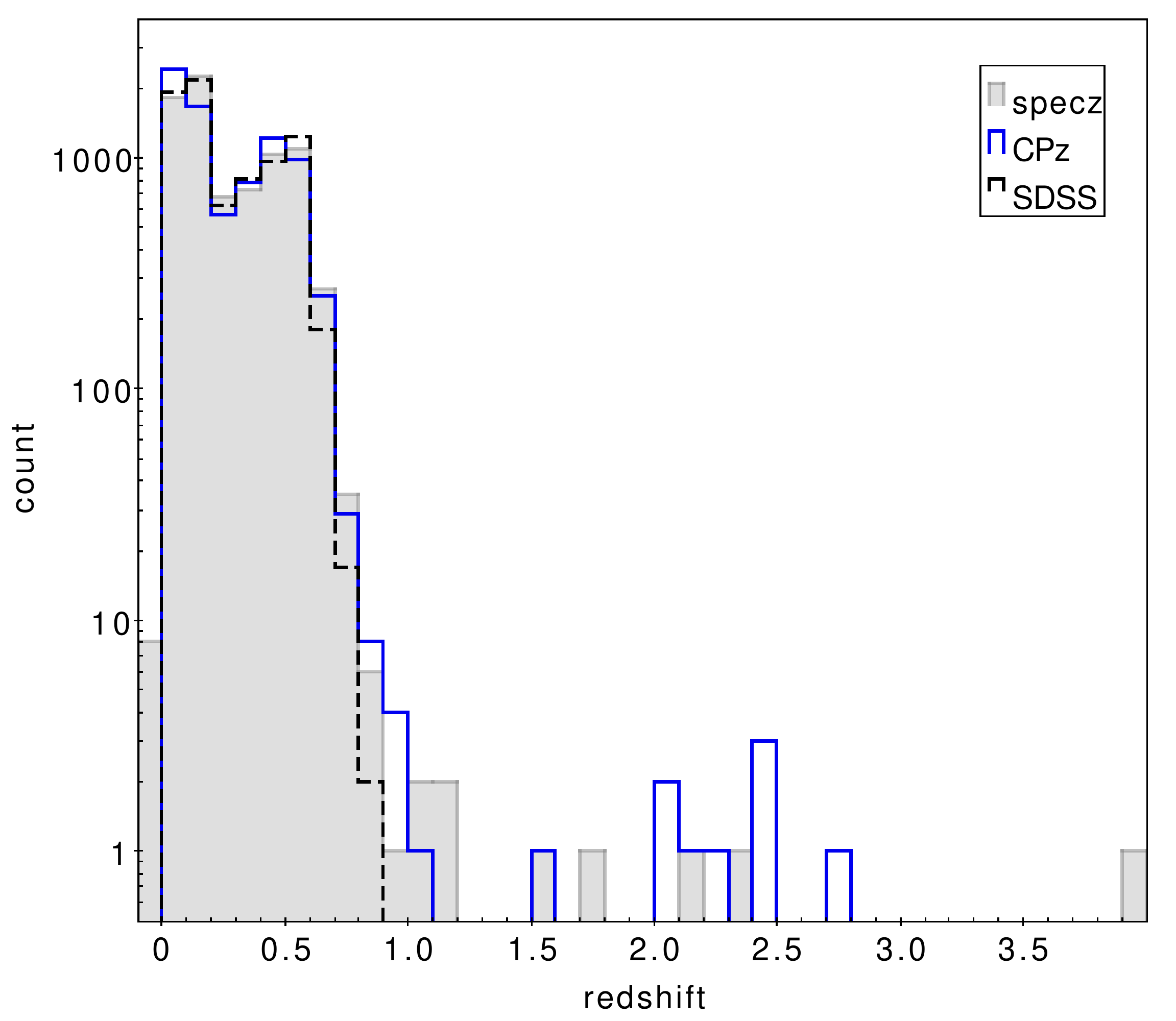} \\
(c)  & (d) \\
\includegraphics[width=0.95\columnwidth]{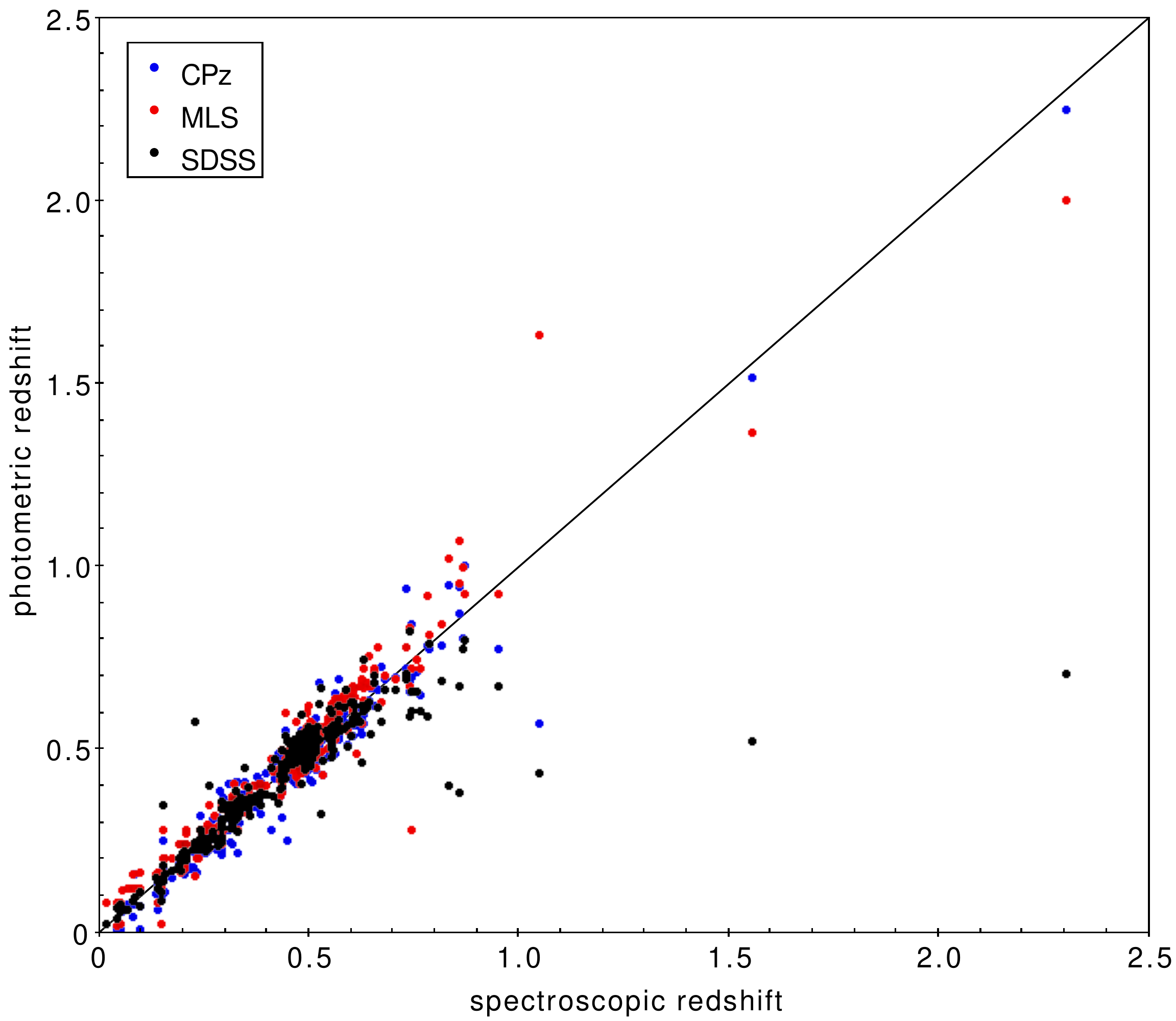} &
 \includegraphics[width=0.95\columnwidth]{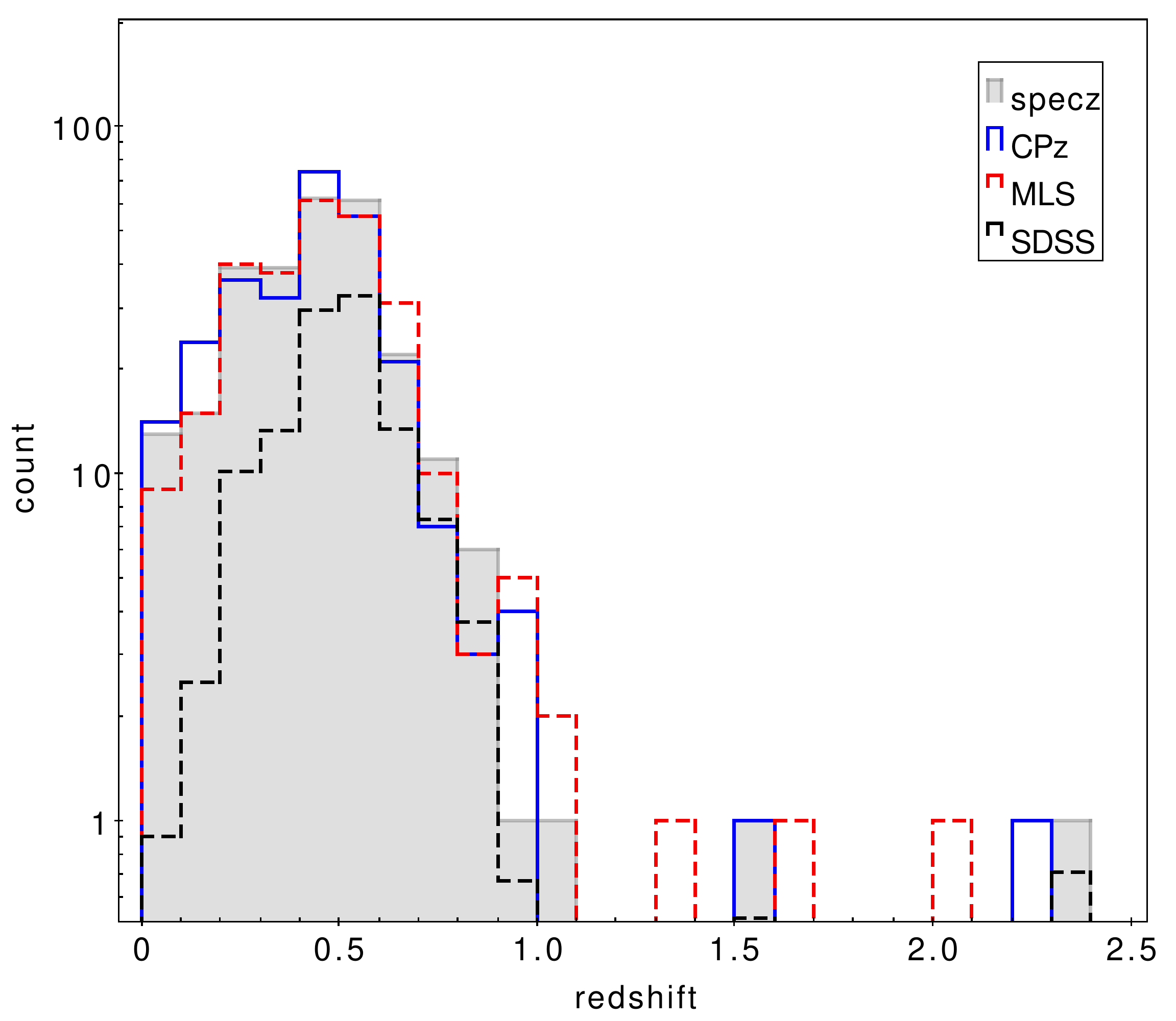} \\
  (e) &(f)\\
\end{tabular}
\caption{Comparison of the CPz photometric redshifts (blue) to (a) 2\,399 objects with pure template fitting method (MLS, red) and (c) 7\.954 objects with pure machine-learning estimate (SDSS, black) (e) 272 objects in common in all three methods.}
\label{fig:photoz_literature}
\end{figure*}

However, the summary of the performance given by $\sigma$ and $\eta$ does not encapsulate all the systematic effects that are commonly present in photometric redshift estimation. In Fig. \ref{fig:photoz_literature} we plot the $\rm{z_{phot}}$ vs $\rm{z_{spec}}$ for each of the three matched samples. It is evident that both the MLS and SDSS photometric redshift estimates have been optimized for the low redshift population ($z<1$). Even though for template fitting methods this is largely due to the decisions on the template set optimization and absolute luminosity prior application, for machine-learning methods the limitation is intrinsic to the method. If a machine-learning training sample does not contain high-redshift objects, the algorithm will never predict a high redshift solution. However with our approach of pre-classifying the sources and using template fitting methods we are able to retrieve high-redshift solutions without systematic problems at least up to redshift of four (see Fig. \ref{fig:consolidation}). 

The present work has focused on training and applying the CPz method on a spectroscopic sample in order to demonstrate its validity. In order to have a deeper understanding of the classification performance and explore further applications, we must apply this method on a flux limited sample. With the application of the CPz methodology to the XXL-Survey, we will release the photometric catalogs and the CPz outcome including the CFHTLS-W1 field used in this work. With a flux limited sample of about eight million objects we will be able to discuss in detail the impact of the relative galaxy populations in the training sample and any trends with redshift on the classification and photometric redshift performance.

\begin{table*}
\begin{center}
\caption{Photometric redshift performance of the final CPz sample compare to the MLS and SDSS photometric redshifts.}
\label{tab:cases}
\scriptsize
\begin{tabular}{@{\extracolsep{2pt}}lccc|ccccccccccccccc} \hline \hline
	&   \multicolumn{3}{c|}{combined sample} 		& \multicolumn{3}{c}{passive}	&   \multicolumn{3}{c}{starforming} &\multicolumn{3}{c}{starburst}	&   \multicolumn{3}{c}{AGN} &\multicolumn{3}{c}{QSO}	 \\ \cline{2-4} \cline{5-7} \cline{8-10} \cline{11-13} \cline{14-16} \cline{17-19} 
	& N & $\sigma$	&   $\eta$ & N & $\sigma$	&   $\eta$ & N & $\sigma$	&   $\eta$		& N & $\sigma$	&   $\eta$ & N & $\sigma$	&   $\eta$ & N & $\sigma$	&   $\eta$	 \\\hline
CPz & \multirow{2}{*}{2~399} & 0.035 & 2.0& \multirow{2}{*}{723}  & 0.034 & 0.4 & \multirow{2}{*}{1~014} & 0.033 & 1.1 & \multirow{2}{*}{514} & 0.036 & 1.2 & \multirow{2}{*}{67} & 0.049 & 1.5 & \multirow{2}{*}{81} & 0.126 & 34.6 \\
MLS &                       & 0.031 & 3.4&                       & 0.028 & 0.6 &                        & 0.032 & 1.0 &                      & 0.028 & 1.6 &                     & 0.042 & 3.0 &                     & 0.517 & 71.6 \\ \hline	
CPz & \multirow{2}{*}{7~954} & 0.038 & 0.9& \multirow{2}{*}{4~240}  & 0.041 & 0.6 & \multirow{2}{*}{3~174} & 0.032 & 0.9 & \multirow{2}{*}{336} & 0.071 & 1.8 & \multirow{2}{*}{194} & 0.056 & 2.6 & \multirow{2}{*}{10} & 0.345 & 50.0 \\
SDSS &                       & 0.016 & 0.4&                        & 0.016 & 0.4 &                        & 0.016 & 0.3 &                      & 0.016 & 0.9 &                     & 0.021 & 0.0 &                     & 0.600 & 60.0 \\ \hline	
CPz & \multirow{3}{*}{272} & 0.033 & 0.4& \multirow{3}{*}{122}  & 0.034 & 0.0 & \multirow{3}{*}{122} & 0.032 & 0.0 & \multirow{2}{*}{17} & 0.051 & 0.0 & \multirow{3}{*}{8} & 0.044 & 0.0 & \multirow{3}{*}{3} & 0.027 & 33.3 \\
MLS &                      & 0.034 & 0.7&                       & 0.032 & 0.0 &                      & 0.034 & 0.0 &                     & 0.045 & 0.0 &                    & 0.047 & 12.5 &                   & 0.143 & 33.3 \\ 
SDSS &                    & 0.023 & 2.6&                      & 0.019 & 0.0 &                      & 0.022 & 1.6 &                     & 0.044 & 11.8 &                   & 0.124 & 0.0 &                    & 0.600 & 100.0 \\
\hline\hline
\end{tabular}
\end{center}
\end{table*}

\section{Conclusions}

We introduce the classification-aided photometric redshift method (CPz), an automatic method to identify stars, estimate optimal photometric redshifts for all galaxy populations including AGNs and QSO and identify photometric redshift outliers. The method consists of three stages. In the first stage, we fit star and galaxy, AGN, and QSO models to all the observations. In the second stage, we create all color combinations and pre-process the data by normalizing and whitening them. Next, we train three classifiers using a machine-learning algorithm to identify 1) stars 2) the optimal photometric redshift library setup 3) photometric redshift outliers. The final stage consists of the consolidation of the results, where the selected probability thresholds tailored to the specific science case are applied.

We have shown that:
\begin{itemize}
\item Using a restricted set of attributes, expected to be widely available with the scheduled large surveys we successfully classify objects as stars versus galaxies.
\item The best photometric redshift results are obtained when the sample is split in passive, starforming, starburst, AGN and QSO, without overlap between the classes.
\item Inclusion of $W1$ and $W2$ filters of WISE photometry combined with the $u$-$K$ filters of Euclid and LSST bring a significant improvement both in accuracy and number of outliers and in the identification of stars.
\item Most importantly, we able to identify AGN and QSO based on their broadband colors.
\end{itemize}

The sample used for this work was restricted only to sources with spectroscopic redshift information. Therefore, the classifier scores presented here should be considered only indicative since the performance achieved for a given survey will depend on the available photometry and training sample used. Thus, we refrain from making any analysis for example, on the number of objects identified per class. We defer this discussion to a future publication of the application of the method on the XXL Survey (Fotopoulou et al., in prep). Preliminary results of the CPz application on the XXL-1000-AGN sample, the 1000 brightest X-ray sources in the XXL survey, can be found in \citet{Fotopoulou2016b}.

\begin{acknowledgements}
We thank the anonymous referee for the useful comments that helped improve our manuscript. SF acknowledges financial support from the Swiss National Science Foundation.

Based on data products from observations made with ESO Telescopes at the La Silla or Paranal Observatories under ESO programme IDs 179.A-2006 and 179.A-2004.
Funding for SDSS-III has been provided by the Alfred P. Sloan Foundation, the Participating Institutions, the National Science Foundation, and the U.S. Department of Energy Office of Science. The SDSS-III web site is http://www.sdss3.org/.
SDSS-III is managed by the Astrophysical Research Consortium for the Participating Institutions of the SDSS-III Collaboration including the University of Arizona, the Brazilian Participation Group, Brookhaven National Laboratory, Carnegie Mellon University, University of Florida, the French Participation Group, the German Participation Group, Harvard University, the Instituto de Astrofisica de Canarias, the Michigan State/Notre Dame/JINA Participation Group, Johns Hopkins University, Lawrence Berkeley National Laboratory, Max Planck Institute for Astrophysics, Max Planck Institute for Extraterrestrial Physics, New Mexico State University, New York University, Ohio State University, Pennsylvania State University, University of Portsmouth, Princeton University, the Spanish Participation Group, University of Tokyo, University of Utah, Vanderbilt University, University of Virginia, University of Washington, and Yale University.
Based on data products from observations made with ESO Telescopes at the La Silla Paranal Observatory under programme IDs 177.A-3016, 177.A-3017 and 177.A-3018, and on data products produced by Target/OmegaCEN, INAF-OACN, INAF-OAPD and the KiDS production team, on behalf of the KiDS consortium. OmegaCEN and the KiDS production team acknowledge support by NOVA and NWO-M grants. Members of INAF-OAPD and INAF-OACN also acknowledge the support from the Department of Physics \& Astronomy of the University of Padova, and of the Department of Physics of Univ. Federico II (Naples).
This publication makes use of data products from the Wide-field Infrared Survey Explorer, which is a joint project of the University of California, Los Angeles, and the Jet Propulsion Laboratory/California Institute of Technology, funded by the National Aeronautics and Space Administration.
Based on observations obtained with MegaPrime/MegaCam, a joint project of CFHT and CEA/IRFU, at the Canada-France-Hawaii Telescope (CFHT) which is operated by the National Research Council (NRC) of Canada, the Institut National des Science de l'Univers of the Centre National de la Recherche Scientifique (CNRS) of France, and the University of Hawaii. This work is based in part on data products produced at Terapix available at the Canadian Astronomy Data Centre as part of the Canada-France-Hawaii Telescope Legacy Survey, a collaborative project of NRC and CNRS. 
 This research has made use of the ASPIC database, operated at CeSAM/LAM, Marseille, France. 
This paper uses data from the VIMOS Public Extragalactic Redshift Survey (VIPERS). VIPERS has been performed using the ESO Very Large Telescope, under the "Large Programme" 182.A-0886. The participating institutions and funding agencies are listed at http://vipers.inaf.it

GAMA is a joint European-Australasian project based around a spectroscopic campaign using the Anglo-Australian Telescope. The GAMA input catalog is based on data taken from the Sloan Digital Sky Survey and the UKIRT Infrared Deep Sky Survey. Complementary imaging of the GAMA regions is being obtained by a number of independent survey programmes including GALEX MIS, VST KiDS, VISTA VIKING, WISE, Herschel-ATLAS, GMRT and ASKAP providing UV to radio coverage. GAMA is funded by the STFC (UK), the ARC (Australia), the AAO and the participating institutions. The GAMA website is http://www.gama-survey.org/ . 
This research uses data from the VIMOS VLT Deep Survey, obtained from the VVDS database operated by Cesam, Laboratoire d'Astrophysique de Marseille, France.
Funding for PRIMUS is provided by NSF (AST-0607701, AST-0908246, AST-0908442, AST-0908354) and NASA (Spitzer-1356708, 08-ADP08-0019, NNX09AC95G).

\end{acknowledgements}

\appendix
\section{Accompanying catalog}

The input and output of Run uk---IR discussed in detail in this paper is available in electronic form at the CDS via anonymous ftp to  to cdsarc.u-strasbg.fr (130.79.128.5) or via http://cdsweb.u-strasbg.fr/cgi-bin/qcat?J/A+A/.

Specifically, the contents of the catalog are:
\begin{itemize}
	\item Col 1-: Spectroscopic redshift ID.
	\item Col 2,3: Spectroscopic redshift coordinates.
	\item Col 4: Spectroscopic redshift value.
	\item Col 5: Spectroscopic redshift classification (-1=unknown, 0=star, 1=normal galaxy, 2=AGN, 3=QSO).
	\item Col 6: Spectroscopic redshift origin (PRIMUS, GAMA, SDSS, VIPERS, VVDS, 6dF).
	\item Col 7-46: SDSS coordinates, flux radius, 3'' aperture and total magnitude with associated errors in each of the u, g, r, i, z bands.
	\item Col 47-86: CFHTLS-Wide identifier, coordinates, flux radius, 3'' aperture and total magnitude with associated errors in each of the u, g, r, i, z bands.
	\item Col 87-118: KiDS identifier, coordinates, flux radius, 3'' aperture and total magnitude with associated errors in each of the u, g, r, i bands.
	\item Col 119-158: VIKING identifier, coordinates, flux radius, 3'' aperture and total magnitude with associated errors in each of the z, Y, J, H, K bands.
	\item Col 159-197: VIDEO identifier, coordinates, flux radius, 3'' aperture and total magnitudes with associated errors in each of the z, Y, J, H, K bands.
	\item Col 198-208: GALEX identifier and coordinates, FUV and NUV 3'' aperture and total magnitudes with associated errors.
	\item Col 209-219: ALLWISE identifier and coordinates, W1, W2, W3, W4 total magnitudes with associated errors.
	\item Col 220-234: consolidated FUV, NUV, u, g, r, i, z, Y, J, H, K, W1, W2, W3, W4 total photometry for Random Forest input.
	\item Col 235-245: consolidated FUV, NUV, u, g, r, i, z, Y, J, H, K, 3'' aperture photometry for Random Forest input.
	\item Col 246-260: consolidated total photometry errors.
	\item Col 261-271: consolidated 3'' aperture photometry errors.
	\item Col 272-280: Half light radius in u, g, r, i, z, Y, H, K bands.
	\item Col 281: $\chi^2$ of best fit star model.
	\item Col 282-295: Case 0 -- All galaxy models: Photometric redshift ID, best fit photometric redshift, best redshift  68\% lower bound,best redshift  68\% higher bound, best model $\chi^2$, best fit model number as given in Table \ref{tab:SEDlib}, best extinction law, best extinction value, scaling of best model, distance modulus, number of bands used, secondary photometry redshift solution, secondary solution $\chi^2$, secondary solution best model.  
	\item Col 296-308: same as before for Case I -- Galaxies.
	\item Col 309-321: same as before for Case I -- EXTNV.
	\item Col 322-334: same as before for Case I -- QSOV.
	\item Col 335-347: same as before for Case II/III -- Passive.
	\item Col 348-360: same as before for Case II -- Staforming and starburst models.
	\item Col 361-373: same as before for Case II/III -- AGN. 
	\item Col 374-386: same as before for Case II/III -- QSO.
	\item Col 387-399: same as before for Case III -- Starforming.
	\item Col 400-412: same as before for Case III -- Extreme Starforming (Starburst).
	\item Col 413: Source used for training the Random Forest (1=train, 2=test, 3=validate).
	\item Col 414: Classifier A: probability to be a star.

	\item Col 415-417: Classifier B -- Case I: probability to be galaxy, EXTNV, QSOV.
	\item Col 418: Classifier C -- Case I: probability to be an outlier.
	
	\item Col 419-422: Case I: optimal redshift, consolidation type A\footnote{Consolidation type A assigns as optimal class the class with highest probability}, a) no rejection b) rejection of stars c) rejection of outliers d) rejection of stars and outliers. 
	\item Col 423-426: Case I; Same as before for consolidation type B\footnote{Consolidation type B requires a minimum probability threshold of 40\% for galaxies and AGN and 20\% for QSO}.
	\item Col 427-430: Case I: Optimal classification for consolidation type A (-2=outlier, -1=star, 1=galaxy, 2=EXTNV, 3=QSOV).
	\item Col 431-434: Case I: Optimal classification for consolidation type B (-2=outlier, -1=star, 1=galaxy, 2=EXTNV, 3=QSOV).
	
	\item Col 435-438: Classifier B -- Case II: probability to be passive, starforming, AGN, QSO
	\item Col 439: Classifier C -- Case II: probability to be an outlier.
	\item Col 440-443: Case II: optimal redshift, consolidation type A a) no rejection b) rejection of stars c) rejection of outliers d) rejection of stars and outliers.
	\item Col 444-447: Case II: same as before for consolidation type B.
	\item Col 448-451: Case II: optimal classes for consolidation type A (-2=outlier, -1=star, 1=passive, 2=starforming, 3=AGN, 4=QSO).
	\item Col 452-455: Case II: optimal classes for consolidation type B (-2=outlier, -1=star, 1=passive, 2=starforming, 3=AGN, 4=QSO).
	\item Col 456-460: Classifier B -- Case III: probability to be passive, starforming, starburst, AGN, QSO.
	\item Col 461: Classifer C -- Case III: probability to be an outlier.
	\item Col 462-465: Case III: optimal redshift, consolidation type A a) no rejection b) rejection of stars c) rejection of outliers d) rejection of stars and outliers.
	\item Col 466-469: Case III: optimal redshift for Case III, consolidation type B.
	\item Col 470-473: Case III: optimal classes for consolidation type A (-2=outlier, -1=star, 1=passive, 2=starforming, 3=starburst 4=AGN, 5=QSO).
	\item Col 474-477: Case III: optimal classes for consolidation type B (-2=outlier, -1=star, 1=passive, 2=starforming, 3=starburst 4=AGN, 5=QSO).
	
\end{itemize}

\end{document}